%% file: main.tex
\documentclass[12pt,twoside]{mitthesis}
\usepackage{lgrind}
\usepackage{amsmath}
\usepackage{amssymb}
\usepackage{amsthm}
\usepackage{graphicx}
\usepackage{color}
\usepackage{type1cm}
\usepackage{lettrine}
\usepackage{appendix}
\usepackage{rotating}
\usepackage{subfigure}
\usepackage[breaklinks=true]{hyperref}

\newtheorem{Thm}{Theorem}[chapter]

\newtheorem{Prop}{Proposition}[chapter]

\newcommand{\argmin}{\mathop{\mathrm{arg\min}}}

\newcommand{\packet}{u}
\newcommand{\flow}{x}
\newcommand{\invord}{\mu}

\makeatletter
\def\cleardoublepage{\clearpage\if@twoside \ifodd\c@page\else
  \hbox{}
  \thispagestyle{empty}
  \newpage
  \if@twocolumn\hbox{}\newpage\fi\fi\fi}
\makeatother

\begin{document}
\frontmatter
\include{cover}
\pagestyle{headings}
\include{preface}
\include{contents}
\mainmatter
\include{chap1}
\include{chap2}
\include{chap3}

\include{chap4}

\include{chap5}
\include{biblio}
\end{document}

%% file: cover.tex
%
%
%
%
%
%
%
\title{Efficient Operation of Coded Packet Networks}

\author{Desmond S. Lun}
\prevdegrees{B.Sc.,~University of Melbourne (2001) \\
  B.E. (Hons.),~University of Melbourne (2001) \\
  S.M., Massachusetts Institute of Technology (2002)}
\department{Department of Electrical Engineering and Computer Science}
\degree{Doctor of Philosophy in Electrical Engineering and Computer
Science}
\degreemonth{June}
\degreeyear{2006}
\thesisdate{\today}


\supervisor{Muriel M\'edard}{Esther and Harold Edgerton Associate Professor of Electrical Engineering}

\chairman{Arthur C. Smith}{Chairman, Department Committee on Graduate Students}

\maketitle



\cleardoublepage
\pagestyle{empty}
\begin{abstractpage}
\input{abstract}
\end{abstractpage}



%


%% file: abstract.tex
%
%
%

A fundamental problem faced in the design of almost all packet networks
is that of efficient operation---of reliably communicating given
messages among nodes at minimum cost in resource usage.  We present a
solution to the efficient operation problem for coded packet net\-works,
i.e.,\  packet networks where the contents of outgoing packets are
arbitrary, causal functions of the contents of received packets.  Such
networks are in contrast to conventional, routed packet networks, where
outgoing packets are restricted to being copies of received packets and
where reliability is provided by the use of retransmissions.

This thesis introduces four considerations to
coded packet networks:
\begin{enumerate}
\item efficiency,
\item the lack of synchronization in packet networks,
\item the possibility of broadcast links, and
\item packet loss.
\end{enumerate}
We take these considerations and give a prescription for operation that
is novel and general, yet simple, useful, and extensible.

We separate the efficient operation problem into two smaller problems,
which we call network coding---the problem of deciding what coding
operation each node should perform given the rates at which packets are
injected on each link---and subgraph selection---the problem of deciding
those rates.  Our main contribution for the network coding problem is to
give a scheme that achieves the maximum rate of a multicast connection
under the given injection rates.  As a consequence, the separation of
network coding and subgraph selection results in no loss of optimality
provided that we are constrained to only coding packets within a single
connection.  Our main contribution for the subgraph selection problem is
to give distributed algorithms that optimally solve the
single-connection problem under certain assumptions.
Since the scheme we propose for network coding can easily be implemented
in a distributed manner, we obtain, by combining the solutions for each
of the smaller problems, a distributed approach to the efficient
operation problem.

We assess the performance of our solution for three problems:
minimum-trans\-mission wireless unicast, minimum-weight wireline
multicast, and minimum-energy wireless multicast.  
We find that our solution has the potential to offer significant
efficiency improvements over existing techniques in routed packet
networks, particularly for multi-hop wireless networks.

%% file: preface.tex
\chapter{Preface}

\enlargethispage{-3\baselineskip}

Vladimir Nabokov once opined, ``My loathings are simple: stupidity,
oppression, crime, cruelty, soft music.  My pleasures are the most
intense known to man:  writing and butterfly hunting.'' I share all of
Nabokov's loathings, but only one of his pleasures---and that began only
recently.  Of course, the lepidoptera I've been involved with are none
that Nabokov would recognize or, I imagine, much revere.  Nevertheless,
the butterflies to which I refer---from the butterfly network of
Ahlswede et al.\  (see Figure 7 of \cite{acl00}) to its wireless
counterpart (see Figure 1 of \cite{lmk05-efficient}) to further
generalizations---have certainly given me a great deal of pleasure since
I began investigating network coding in the spring of 2003.

This thesis represents the culmination of my work over the last three
years, which began with the simple question, how would all this actually
work?  I was intrigued by network coding.  But I couldn't quite reconcile
it with the way that I understood data networks to operate.  So I
thought to take the basic premise of network coding and put it in a
model that, at least to me, was more satisfying.  The following pages
lay out a view of coded packet networks that, while certainly not the
only one possible, is one that I believe is simple, relevant, and
extensible---I can only hope that it is sufficiently so to be truly
useful.


Various parts of the work in this thesis appear in various published
papers \cite{deh05, lme04, lmh04, lmk05, lmk05-efficient, lmk06,
lmk05-further, lpf06, lrk05} and various as yet unpublished papers
\cite{lmk, lrm}.  A brief glance at the author lists of the these
papers, and it is evident that I cannot claim sole credit for this
work---many others are involved.

My adviser, Professor Muriel M\'edard, is foremost among them.  I would
like to thank her for all that she has taught me and all that she has
done to aid my development---both professional and personal.  The way
that she manages the multitude of demands on her time continues to amaze
and inspire me.  I would like to thank also my thesis readers,
Professors Michelle Effros, Ralf Koetter, and John Tsitsiklis.  All have
contributed helpful discussions and advice.  I would like to thank Ralf
in particular, as he has served almost as a second adviser to me.  His
insight and enthusiasm have been invaluable.

Various others have contributed to various parts of the work, and I wish
to acknowledge them for it: Niranjan Ratnakar
(Section~\ref{sec:distributed_algorithms}), Dr.\  Payam Pakzad
(Section~\ref{sec:finite-memory}), Dr.\  Christina Fragouli
(Section~\ref{sec:finite-memory}), Professor David Karger
(Section~\ref{sec:dynamic_multicast}), Professor Tracey Ho
(Section~\ref{sec:problem_formulation}), Ebad Ahmed
(Sections~\ref{sec:min_weight} and~\ref{sec:min_energy}), Fang Zhao
(Sections~\ref{sec:min_weight} and~\ref{sec:min_energy}), and Hyunjoo
Lee (Section~\ref{sec:min_weight}).  All have been a delight to work
with.  I am grateful also to Guy Weichenberg and Ed Schofield for their
helpful comments on early drafts of the manuscript.

On a personal level, there are many to thank, but I will keep it brief.
I am aware of Friedrich Nietzsche's maxim, ``Ein Mensch mit Genie ist
unausstehlich, wenn er nicht mindestens noch zweierlei dazu besitzt:
Dankbarkeit und Reinlichkeit.''  [A man with spirit is unbearable if he
does not have at least two other things: gratitude and cleanliness.]
And, while I shan't discuss my cleanliness, I certainly don't wish any
of my friends or family to feel that I am not grateful for the favor
they have shown me.  I am.  But I want to keep this to those to whom I
am really indebted the most: Mum, Dad, Guy, and Katie.  I love you all.

\begin{flushright}
\emph{Desmond S. Lun \\
Cambridge, Mass. \\
April 2006}
\end{flushright}

\emph{This research was supported by the National Science Foundation
under grant nos.\  CCR-0093349 and CCR-0325496;
by the Army Research Office through University of California
subaward no.\  S0176938; and
by the Office of Naval Research under
grant no.\  N00014-05-1-0197.}

%% file: contents.tex
\tableofcontents
\newpage
\listoffigures
\newpage
\listoftables

%% file: chap1.tex
\chapter{Introduction}
\label{sec:introduction}

\lettrine{A}{fundamental} problem faced in the design of almost all
packet networks is that of efficient operation---of reliably
communicating given messages among nodes at minimum cost in resource
usage.  At present, the problem is generally addressed in the following
way: messages admitted into the network are put into packets that are
routed hop-by-hop toward their destinations according to paths chosen to
meet the goal of efficiency, e.g.,\  to achieve low energy consumption,
to achieve low latency, or, more generally, to incur low cost of any
sort.  As packets travel along these paths, they are occasionally lost
because of various reasons, which include buffer overflow, link outage,
and collision; so, to ensure reliability, retransmissions of
unacknowledged packets are sent either on a link-by-link basis, an
end-to-end basis, or both.  This mode of operation crudely characterizes
the operation of the internet and has held sway since at least its
advent.

But much has changed about packet networks since the advent of the
internet.  The underlying communications technologies have changed, as
have the types of services demanded, and, under these changes, the mode
of operation described above has met with difficulties.  We give two
examples.  First, while wireline communications were once dominant in
packet networks, wireless communications involving nodes on the ground,
in the air, in space, and even underwater are now increasingly
prevalent.  In networks where such wireless links are present, this mode
of operation can certainly be made to work, but we encounter
problems---most notably with the use of retransmissions.  Wireless links
are highly unreliable compared to wireline ones and are sometimes
associated with large propagation delays, which means that, not only are
more retransmissions required, but packet acknowledgments are themselves
sometimes lost or subject to large delay, leading to substantial
inefficiencies from the retransmission of unacknowledged packets.  
Moreover, hop-by-hop routing fails to exploit the
inherent broadcast nature often present in wireless links, leading to
further inefficiencies.  

Second, while unicast services were once the norm, multicast services
are now required for applications such as file distribution and
video-conferencing.  For multicast services, hop-by-hop routing means
routing over a tree, which is difficult to do efficiently---finding the
minimum-cost tree that spans a multicast group equates to solving the
Steiner tree problem, which is a well-known \textsc{np}-complete problem
\cite{bhj83, wax88}.  Moreover, if there are many receivers, many
retransmitted packets may be needed, placing an unnecessary load on the
network and possibly overwhelming the source.  Even if the source
manages, packets that are retransmitted are often useful only to a
subset of the receivers and redundant to the remainder.

The problems we mentioned can and generally have been resolved to some
degree by ad hoc methods and heuristics.  But that is hardly
satisfactory---not only from an intellectual standpoint, since ad hoc
solutions do little for our understanding of the fundamental problem,
but also from a practical standpoint, since they tend to lead to
complex, inefficient designs that are more art than science.  Indeed, as
Robert G. Gallager has commented, ``much of the network field is an art
[rather than a science]'' \cite{gal01}.  And while it is evident that
engineering real-world systems is an activity that will always lie
between an art and a science, it is also evident that the more we base
our designs on scientific principles, the better they will generally be.

In this thesis, therefore, we eschew such ``routed'' packet networks
altogether in favor of a new approach: we consider coded packet
networks---generalizations of routed packet networks where the contents
of outgoing packets are arbitrary, causal functions of the contents of
received packets.  In this context, we consider the same fundamental
problem, i.e.,\  we ask, how do we operate coded packet networks
efficiently?

We present a prescription for the operation of coded packet networks
that, in certain scenarios (e.g.,\  in multi-hop wireless networks),
yields significant efficiency improvements over what is achievable in
routed packet networks.  We begin, in
Section~\ref{sec:coded_packet_networks}, by discussing coded packet
networks in more detail and by clarifying the position of our work,
then, in Section~\ref{sec:network_model}, we describe our network model.
We outline the body of the thesis in Section~\ref{sec:thesis_outline}.

\section{Coded packet networks\label{sec:coded_packet_networks}}

The basic notion of network coding, of performing coding operations on
the contents of packets throughout a network, is generally attributed to
Ahlswede et al.\ \cite{acl00}.  Ahlswede et al.\  never explicitly
mentioned the term ``packet'' in \cite{acl00}, but their network model,
which consists of nodes interconnected by error-free point-to-point
links, implies that the coding they consider occurs above channel coding
and, in a data network, is presumably applied to the contents of
packets.

Still, their work is not the first to consider coding in such a network
model.  Earlier instances of work with such a network model include
those by Han \cite{han80} and Tsitsiklis \cite{tsi93}.  But the work of
Ahlswede et al.\  is distinct in two ways: First, Ahlswede et al.\
consider a new problem---multicast.  (The earlier work considers the
problem of transmitting multiple, correlated sources from a number of
nodes to a single node.)  Second, and more importantly, the work of
Ahlswede et al.\  was quickly followed by other work, by Li et al.\
\cite{lyc03} and by Koetter and M\'edard \cite{kom03}, that showed that
codes with a simple, linear structure were sufficient to achieve
capacity in the multicast problem.  This result put structure on the
codes and gave hope that practicable capacity-achieving codes could be
found.  


The subsequent growth in network coding was explosive.  Practicable
capacity-achieving codes were quickly proposed by Jaggi et al.\
\cite{jsc05}, Ho et al.\  \cite{hmk}, and Fragouli and Soljanin
\cite{frs04}.  Applications to network management \cite{hmk05},
network tomography \cite{frm05, hlc05},
overlay networks \cite{gkr05, jlc05, zlg04}, 
and wireless networks 
\cite{gdp04, khs05-multirelay, sae05-crosslayer, wck05-information, wck05}
were studied; 
capacity in random networks \cite{rsw05},
undirected networks \cite{lil05, llj05}, 
and Aref networks \cite{rak} was studied;
security aspects
were studied \cite{bhn05, cay02, fms04, hlk04, jlh05}; 
the extension to non-multicast problems 
was studied \cite{dfz05, kkh05, mek03, ral03, rtk06, rii04};
and further code
constructions based on convolutional codes and other notions were
proposed 
\cite{cwj03, erf04, erf05-efficient, frs04-connection, hkm05, lem05}.
Most notoriously, network coding has been adopted as a core technology
of Microsoft's Avalanche project \cite{gkr05}---a research
project that aims to develop a peer-to-peer file distribution system,
which may be in competition with existing systems such as BitTorrent. 

Of the various work on network coding, we draw particular attention to
the code construction by Ho et al.\  \cite{hmk}.  Their construction is
very simple: they proposed that every node construct its linear code
randomly and independently of all other nodes, and, while random linear
codes were not new (the study of random linear codes dates as early as
the work of Elias \cite{eli56} in the 1950s), the application of such
codes to the network multicast problem was.  Some years earlier, Luby
\cite{lub02} searched for codes for the transmission of packets over a
lossy link and discovered random linear codes, constructed according to
a particular distribution, with remarkable complexity properties.  This
work, combined with that of Ho et al.,\ led to a resurgence of interest
in random linear codes (see, e.g.,\ \cite{adm05, cwj03, dem, may02,
pfs05, sho04}) and to the recognition of a powerful technique that we
shall exploit extensively: random linear coding on packets.

The work we have described has generally focused on coding and
capacity---growing, as it has, from coding theory and information
theory---and has been removed from networking theory, which generally
focuses on notions such as efficiency and quality of service.  While it
is adequate, and indeed appropriate, to start in this way, it is clear
that, with network coding being concerned with communication networks,
topics under the purview of networking theory must eventually be
broached.

This thesis makes an attempt.  It introduces four considerations absent
from the original work of Ahlswede et al.: First, we consider efficiency
by defining a cost for inefficiency.  This is a standard framework in
networking theory, which is used, e.g.,\  in the optimal routing problem
(see, e.g.,\  \cite[Sections 5.4--5.7]{beg92}).  Second, we consider the
lack of synchronization in packet networks, i.e.,\  we allow packet
injections and receptions on separate links to occur at completely
different rates with arbitrary degrees of correlation.  Third, we
consider the possibility of broadcast links, i.e.,\  we allow links in
the network to reach more than one node, capturing one of the key
characteristics of wireless networks.  Fourth, we consider packet loss,
i.e.,\  we allow for the possibility that packets are not received at
the end or ends of the link into which they are injected.  

Some of these considerations are present in other, concurrent work.  For
example, efficiency is also considered in \cite{cxn04, wck05}; and the
possibility of broadcast links and packet loss are also considered in
\cite{gdp04, khs05-multirelay}.  These papers offer alternative
solutions to special cases of the problem that we tackle.  We take all
four considerations and give a prescription for operation that is novel
and general, yet simple, useful, and extensible.

\section{Network model\label{sec:network_model}}

We set out, in this section, to present our network model.  The intent
of the model is to capture heterogeneous networks composed of wireline
and wireless links that may or may not be subject to packet losses.
Thus, the model captures a wide variety of networks, affording us a
great degree of generality.  

But that is not to say that we believe that coding should be applied to
all networks.  There is a common concern about the wisdom of doing
coding in packet networks since coding, being a more complicated
operation than routing, increases the computational load on nodes, which
are often already overtaxed in this regard.  Indeed, in high-speed
optical networks, bottlenecks are caused almost exclusively by
processing at nodes rather than by transmission along links \cite{chm98,
ppp03}.  But high-speed optical networks are certainly not the only type
of network of interest, and there are others where coding seems more
immediately applicable.  Two such types of networks are
application-level overlay networks and multi-hop wireless networks---in
both cases, having coding capability at nodes is feasible, and we expect
bottlenecks to originate from links rather than nodes.

We represent the topology of the network with a directed hypergraph
$\mathcal{H} = (\mathcal{N}, \mathcal{A})$, where $\mathcal{N}$ is the
set of nodes and $\mathcal{A}$ is the set of hyperarcs. A
\emph{hypergraph} is a generalization of a graph, where, rather than
arcs, we have hyperarcs. A \emph{hyperarc} is a pair $(i,J)$, where $i$,
the start node, is an element of $\mathcal{N}$ and $J$, the set of end
nodes, is a non-empty subset of $\mathcal{N}$.

Each hyperarc $(i,J)$ represents a broadcast link from node $i$ to nodes
in the non-empty set $J$.  In the special case where $J$ consists of a
single element $j$, we have a point-to-point link.  The hyperarc is now
a simple arc and we sometimes write $(i,j)$ instead of $(i,\{j\})$.  The
link represented by hyperarc $(i,J)$ may be lossless or lossy, i.e.,\  it
may or may not be subject to packet erasures.

To establish the desired connection or connections, packets are injected
on hyperarcs.
Let $A_{iJ}$ be the counting process describing the arrival of packets
that are injected on hyperarc $(i,J)$, and let $A_{iJK}$ be the counting
process describing the arrival of packets that are injected on hyperarc
$(i,J)$ and received by exactly the set of nodes $K \subset J$;
i.e.,\  for $\tau \ge 0$, $A_{iJ}(\tau)$ is the total number of packets
that are injected on hyperarc $(i,J)$ between time 0 and time $\tau$,
and $A_{iJK}(\tau)$ is the total number of packets that are injected on
hyperarc $(i,J)$ and received by all nodes in $K$ (and no nodes in
$\mathcal{N}\setminus K$) between time 0 and time $\tau$.
For example, suppose that three packets are injected on hyperarc 
$(1, \{2, 3\})$ between time 0 and time $\tau_0$ and that, of these three
packets, one is received by node 2 only, one is lost entirely, and one
is received by both nodes 2 and 3; then we have
$A_{1(23)}(\tau_0) = 3$, $A_{1(23)\emptyset}(\tau_0) = 1$, 
$A_{1(23)2}(\tau_0) = 1$, $A_{1(23)3}(\tau_0) = 0$, and
$A_{1(23)(23)}(\tau_0) = 1$.  We have $A_{1(23)2}(\tau_0) = 1$
not $A_{1(23)2}(\tau_0) = 2$ because, while two packets are received by
node 2, only one is received by exactly node 2 and no other nodes.  
Similarly, we have $A_{1(23)3}(\tau_0) = 0$ not 
$A_{1(23)3}(\tau_0) = 1$ because, while one packet is received by node
3, none are received by exactly node 3 and no other nodes.

We assume that $A_{iJ}$ has an average rate $z_{iJ}$ and that $A_{iJK}$
has an average rate $z_{iJK}$; more precisely, we assume that
\[
\lim_{\tau \rightarrow \infty} \frac{A_{iJ}(\tau)}{\tau} = z_{iJ}
\]
and that
\[
\lim_{\tau \rightarrow \infty} \frac{A_{iJK}(\tau)}{\tau} = z_{iJK}
\]
almost surely.  
Hence, we have $z_{iJ} = \sum_{K \subset J} z_{iJK}$ and,
if the link is lossless, we have $z_{iJK} = 0$ for all
$K \subsetneq J$.

The vector $z$, consisting of $z_{iJ}$, $(i,J) \in \mathcal{A}$, defines
the rate at which packets are injected on all hyperarcs in the network,
and we assume that it must lie within some constraint set $Z$.  Thus,
the pair $(\mathcal{H}, Z)$ defines a capacitated graph that represents
the network at our disposal, which may be a full, physical network or a
subnetwork of a physical network.  The vector $z$, then, can be thought
of as a subset of this capacitated graph---it is the portion actually
under use---and we call it the \emph{coding subgraph} for the desired
connection or connections.  For the time being, we make no assumptions
about $Z$ except that it is a convex subset of the positive orthant
containing the origin.  This assumption leaves room for $Z$ to take
complicated forms; and indeed it does, particularly when the underlying
physical network is a wireless network, where transmissions on one link
generally interfere with those on others.  For examples of forms that
$Z$ may take in wireless networks, see \cite{crs03, jpp03, jxb03, kon05,
wck05, xjb04}.

We associate with the network a convex cost function $f$ that maps
feasible coding subgraphs to real numbers and that we seek to minimize.
This cost function might represent, e.g.,\  energy consumption, average
latency, monetary cost, or a combination of these considerations.
We assume convexity primarily for simplicity and
tractability.  Certainly, cases where $f$ is non-convex may still be
tractable, but proving general results is difficult.  We expect, at any
rate, that most cost functions of interest will indeed be convex, and
this is generally true of
cost functions representing the considerations that we have mentioned.

With this set-up, the objective of the \emph{efficient operation
problem} is to establish a set of desired connections at specified rates
at minimum cost.  This is the problem we address.

As the following example will illustrate, the problem we have defined is
certainly non-trivial.  Nevertheless, its scope is limited:
we consider rate, or throughput, to be the sole factor that
is explicitly important in determining the quality of a connection, and
we consider the rates of packet injections on hyperarcs (i.e.,\  the
coding subgraph) to be the sole factor that contributes to its cost.
Rate is frequently the most important factor under consideration, but
there are others.  For example, memory usage, computational load, and
delay are often also important factors.  At present, we unfortunately do
not have a clean way to consider such factors.  We discuss the issue
further in Section~\ref{sec:finite-memory} and
Chapter~\ref{chap:conclusion}.

\subsection{An example}
\label{sec:example1}

We refer to this example as the \emph{slotted Aloha relay channel}, and
we shall return to it throughout the thesis.  This example serves to
illustrate some of the capabilities of our approach, especially as they
relate to the issues of broadcast and interference in multi-hop wireless 
networks.

One of most important issues in multi-hop wireless networks is medium
access, i.e.,\ determining how radio nodes share the wireless medium.  A
simple, yet popular, method for medium access control is slotted Aloha
(see, e.g.,\  \cite[Section 4.2]{beg92}), where nodes with packets to
send follow simple random rules to determine when they transmit.  In
this example, we consider a multi-hop wireless network using slotted
Aloha for medium access control.

\begin{figure}
\centering
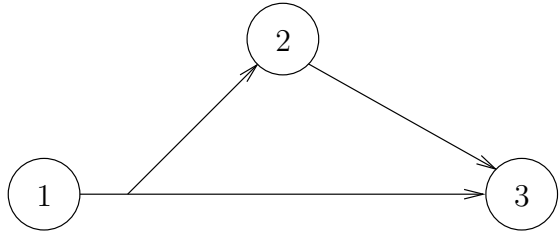
\caption{The slotted Aloha relay channel.}
\label{fig:packet_relay}
\end{figure}

We suppose that the network has the simple topology shown in
Figure~\ref{fig:packet_relay} and that, in this network, we wish to
establish a single unicast connection of rate $R$ from node 1 to node 3.
The random rule we take for transmission is that the two transmitting
nodes, node 1 and node 2, each transmit packets independently in a given
time slot with some fixed probability.  In coded packet networks, nodes
are never ``unbacklogged'' as they are in regular, routed slotted Aloha
networks---nodes can transmit coded packets whenever they are given the
opportunity.
Hence $z_{1(23)}$, the rate of packet injection on hyperarc 
$(1, \{2,3\})$, is the probability that node 1 transmits a packet in a
given time slot, and likewise $z_{23}$, the rate of packet injection
on hyperarc $(2, 3)$, is the probability that node 2 transmits a packet
in a given time slot.
Therefore, $Z = [0,1]^2$, i.e.,\  $0 \le z_{1(23)} \le 1$
and $0 \le z_{23} \le 1$.

If node 1 transmits a packet and node 2 does not, then the packet is
received at node 2 with probability $p_{1(23)2}$, at node 3 with
probability $p_{1(23)3}$, and at both nodes 2 and 3 with probability 
$p_{1(23)(23)}$ (it is lost entirely with probability
$1 - p_{1(23)2} - p_{1(23)3} - p_{1(23)(23)}$).
If node 2 transmits a packet and node 1 does not, then the packet is
received at node 3 with probability $p_{233}$ (it is lost entirely with
probability $1-p_{233}$).
If both nodes 1 and 2 each transmit a packet, then the packets collide
and neither of the packets is received successfully anywhere.

It is possible that simultaneous transmission does not necessarily
result in collision, with one or more packets being received.  This
phenomenon is referred to as multipacket reception capability
\cite{gvs88} and is decided by lower-layer implementation details.  In
this example, however, we simply assume that simultaneous transmission
results in collision.  

Hence, we have 
\begin{align}
z_{1(23)2} &= z_{1(23)}(1-z_{23})p_{1(23)2}, \label{eqn:1100} \\
z_{1(23)3} &= z_{1(23)}(1-z_{23})p_{1(23)3}, \\
z_{1(23)(23)} &= z_{1(23)}(1-z_{23})p_{1(23)(23)},
\end{align}
and
\begin{equation}
z_{233} = (1-z_{1(23)})z_{23}p_{233}. \label{eqn:1200}
\end{equation}

We suppose that our objective is to set up the desired connection while
minimizing the total number of packet transmissions for
each message packet, perhaps for
the sake of energy conservation or conservation of the wireless medium
(to allow it to be used for other purposes, such as other connections).
Therefore 
\[ 
f(z_{1(23)}, z_{23}) = z_{1(23)} + z_{23}.  
\]

The slotted Aloha relay channel is very similar to the relay channel
introduced by van der Meulen \cite{meu71}, and determining the capacity
of the latter is one of the famous, long-standing, open problems of
information theory.  The slotted Aloha relay channel is related to the
relay channel (hence its name), but different.  While the relay channel
relates to the physical layer, we are concerned with higher layers, and
our problem is ultimately soluble.  Whether our solution has any bearing
on the relay channel is an interesting issue that remains to be
explored.

We return to the slotted Aloha relay channel in
Sections~\ref{sec:example2} and~\ref{sec:example3}.

\section{Thesis outline\label{sec:thesis_outline}}

The main contribution of this thesis is to lay out, for coded packet
networks conforming to our model, a solution to the efficient operation
problem that we have posed, namely, the problem of establishing a set of
desired connections at specified rates at minimum cost.  This solution
is contained in Chapters~\ref{chap:network_coding}
and~\ref{chap:subgraph_selection}.  

Chapter~\ref{chap:network_coding} looks at the problem of determining
what coding operation each node should perform given the coding
subgraph.  We propose using a particular random linear coding scheme
that we show can establish a single multicast connection at rates
arbitrarily close to its capacity in a given coding subgraph.  This
means that, at least for establishing a single multicast connection,
there is no loss of optimality in using this coding scheme and
determining the coding subgraph independently.  The optimality to which
we refer is with respect to the efficient operation problem that we have
defined, which, as we have mentioned, does not explicitly consider
factors such as memory usage, computational load, and delay.  In
Section~\ref{sec:finite-memory}, we include memory usage as a factor
under explicit consideration.  We modify the coding scheme to reduce the
memory usage of intermediate nodes and assess, by analysis and computer
simulation, the effect of this modification on various performance
factors.

Chapter~\ref{chap:subgraph_selection}, on the other hand, looks at the
problem of determining the coding subgraph.  We argue that, even when we
wish to establish multiple connections, it suffices, in many instances,
simply to use the coding scheme described in
Chapter~\ref{chap:network_coding} and to determine the coding subgraph
independently.  Thus, this problem, of determining the coding subgraph,
can be written as a mathematical programming problem, and, under
particular assumptions, we find distributed algorithms for performing
the optimization.  We believe that these algorithms may eventually form
the basis for protocols used in practice.


In Chapter~\ref{chap:performance_evaluation}, we evaluate, by computer
simulation, the performance of the solution we laid out and compare it
to the performance of existing techniques for routed packet networks.
We find that our solution has the potential to offer significant
efficiency improvements, particularly for multi-hop wireless networks.
For some readers, this chapter may be the one to read first.  It can be
understood more or less independently of
Chapters~\ref{chap:network_coding} and~\ref{chap:subgraph_selection} and
is, in a sense, ``the bottom line''---at least in so far as we have
managed to elucidate it.  The interested reader may then proceed to
Chapters~\ref{chap:network_coding} and~\ref{chap:subgraph_selection} to
understand the solution we propose.

Our conclusion, in Chapter~\ref{chap:conclusion}, gives a final
perspective on our work and discusses the road ahead.  

%% file: packet_relay.pdftex_t
\begin{picture}(0,0)%
\includegraphics{packet_relay}%
\end{picture}%
\setlength{\unitlength}{3947sp}%
\begingroup\makeatletter\ifx\SetFigFont\undefined%
\gdef\SetFigFont#1#2#3#4#5{%
  \reset@font\fontsize{#1}{#2pt}%
  \fontfamily{#3}\fontseries{#4}\fontshape{#5}%
  \selectfont}%
\fi\endgroup%
\begin{picture}(3466,1441)(1868,-1794)
\put(3601,-661){\makebox(0,0)[b]{\smash{{\SetFigFont{12}{14.4}{\familydefault}{\mddefault}{\updefault}{\color[rgb]{0,0,0}2}%
}}}}
\put(2101,-1636){\makebox(0,0)[b]{\smash{{\SetFigFont{12}{14.4}{\familydefault}{\mddefault}{\updefault}{\color[rgb]{0,0,0}1}%
}}}}
\put(5101,-1636){\makebox(0,0)[b]{\smash{{\SetFigFont{12}{14.4}{\familydefault}{\mddefault}{\updefault}{\color[rgb]{0,0,0}3}%
}}}}
\end{picture}%

%% file: chap2.tex
\chapter{Network Coding}
\label{chap:network_coding}

\lettrine{T}{his} chapter deals with what we call the network coding
part of the efficient operation problem.  We assume that the coding
subgraph $z$ is given, and we set out to determine what coding operation
each node should perform.  We propose using a particular random linear
coding scheme that we show can establish a single multicast connection
at rates arbitrarily close to its capacity in $z$.  More precisely, for
a given coding subgraph $z$, which gives rise to a particular set of
rates $\{z_{iJK}\}$ at which packets are received, the coding scheme we
study achieves (within an arbitrarily small factor) the maximum possible
throughput when run for a sufficiently long period of time.  Exactly how
the injection rates defined by $z$ relates to the reception rates
$\{z_{iJK}\}$ and how the losses, which establishes this relationship,
are caused is immaterial for our result---thus, losses may be due to
collisions, link outage, buffer overflow, or any other process that
gives rise to losses.  The only condition that we require the losses to
satisfy is that they give rise to packet receptions where the average
rates $\{z_{iJK}\}$ exist, as our network model specifies (see
Section~\ref{sec:network_model}).

As a consequence of the result, in establishing a single multicast
connection in a network, there is no loss of optimality in the efficient
operation problem from separating subgraph selection and network coding.
We deal with subgraph selection in
Chapter~\ref{chap:subgraph_selection}.

We begin, in Section~\ref{sec:coding_scheme}, by precisely specifying
the coding scheme we consider then, in
Section~\ref{sec:coding_theorems}, we give our main result: that this
scheme can establish a single multicast connection at rates arbitrarily
close to its capacity in $z$.  In Section~\ref{sec:error_exponents}, we
strengthen these results in the special case of Poisson traffic with
i.i.d.\  losses by giving error exponents.  These error exponents allow
us to quantify the rate of decay of the probability of error with coding
delay and to determine the parameters of importance in this decay.

In both these sections, we consider rate, or throughput, of the desired
connection to be the sole factor of explicit importance.  In
Section~\ref{sec:finite-memory}, we include memory usage as a factor of
explicit importance.  We modify the coding scheme to reduce the memory
usage of intermediate nodes, and we study the effect of this
modification.

\section{Coding scheme}
\label{sec:coding_scheme}

The specific coding scheme we consider is as follows.
We suppose that, at the source node, we have $K$ message packets
$w_1, w_2, \ldots, w_K$, which are vectors of length $\lambda$ over some
finite field $\mathbb{F}_q$.  (If the packet length is $b$ bits, then we
take $\lambda = \lceil b / \log_2 q \rceil$.)  The message packets are
initially present in the memory of the source node. 

The coding operation performed by each node is simple to describe and is
the same for every node: received packets are stored into the node's
memory, and packets are formed for injection with random
linear combinations of its memory contents 
whenever a packet injection occurs on an
outgoing link.  The coefficients of the combination are drawn uniformly
from $\mathbb{F}_q$.  

Since all coding is linear, we can write any
packet $\packet$ in the network as a linear combination of 
$w_1, w_2, \ldots, w_K$, namely,
$\packet = \sum_{k=1}^K \gamma_k w_k$. 
We call $\gamma$ the \emph{global
encoding vector} of $\packet$, and we assume that it is sent along with
$\packet$,
as side information in its header. 
The overhead this incurs (namely, $K \log_2 q$ bits)
is negligible if packets are sufficiently large.

Nodes are assumed to have unlimited memory.  The scheme can be modified
so that received packets are stored into memory only if their global
encoding vectors are linearly-independent of those already stored.  This
modification keeps our results unchanged while ensuring that nodes never
need to store more than $K$ packets.
The case where nodes can only store fewer than $K$ packets is
discussed in Section~\ref{sec:finite-memory}.

A sink node collects packets and, if it has $K$ packets with
linearly-independent global encoding vectors, it is able to recover the
message packets.  Decoding can be done by Gaussian elimination.  The
scheme can be run either for a predetermined duration or, in the case
of rateless operation, until successful decoding at the sink nodes.  We
summarize the scheme in Figure~\ref{fig:summary_RLC}.

\begin{figure}
\centering
\framebox{
\begin{minipage}{0.88\textwidth}
\noindent\textbf{Initialization:}
\begin{itemize}
\item The source node stores the message packets $w_1, w_2, \ldots, w_K$ in its memory.
\end{itemize}
\noindent\textbf{Operation:}
\begin{itemize}
\item When a packet is received by a node,
\begin{itemize}
\item the node stores the packet in its memory.
\end{itemize}
\item When a packet injection occurs on an outgoing link of a node,
\begin{itemize}
\item the node forms the packet from a random linear combination of 
the packets in its memory.  Suppose the node has $L$ packets $\packet_1,
\packet_2, \ldots, \packet_L$ in its memory.  Then the packet formed is
\[
\packet_0 := \sum_{l=1}^L \alpha_l \packet_l,
\]
where $\alpha_l$ is chosen according to a uniform distribution over the
elements of $\mathbb{F}_q$.  
The packet's global encoding vector $\gamma$, which satisfies
$\packet_0 = \sum_{k=1}^K \gamma_k w_k$, is placed in its header.
\end{itemize}
\end{itemize}
\noindent\textbf{Decoding:}
\begin{itemize}
\item Each sink node performs Gaussian elimination on the set of global
encoding vectors from the packets in its memory.  If it is able to find
an inverse, it applies the inverse to the packets to obtain $w_1,
w_2, \ldots, w_K$; otherwise, a decoding error occurs.
\end{itemize}
\end{minipage} }
\caption{Summary of the random linear coding scheme we consider.}
\label{fig:summary_RLC}
\end{figure}

The scheme is carried out for a single block of $K$
message packets at the source.  If the source has more packets to send,
then the scheme is repeated with all nodes flushed of their memory
contents.


Related random linear coding schemes are described in \cite{cwj03, hmk}
for the application of multicast over lossless wireline packet networks,
in \cite{dem} for data dissemination, and in \cite{adm05} for data
storage.  Other coding schemes for lossy packet networks are described
in \cite{gdp04} and \cite{khs05-multirelay}; the scheme described in the
former requires placing in the packet headers side information that
grows with the size of the network, while that described in the latter
requires no side information at all, but achieves lower rates in
general.  Both of these coding schemes, moreover, operate in a
block-by-block manner, where coded packets are sent by intermediate
nodes only after decoding a block of received packets---a strategy that
generally incurs more delay than the scheme we describe, where
intermediate nodes perform additional coding yet do not decode
\cite{pfs05}.  

\section{Coding theorems}
\label{sec:coding_theorems}

In this section, we specify achievable rate intervals for the coding
scheme in various scenarios.  
The fact that the intervals we specify are the largest possible
(i.e.,\  that the scheme is capacity-achieving) 
can be seen by simply noting that the rate of a connection must be
limited by the rate at which distinct packets are being received over
any cut between the source and the sink.  
A formal converse can be obtained
using the cut-set bound for multi-terminal networks 
(see \cite[Section 14.10]{cot91}).

\subsection{Unicast connections}

\begin{figure}
\centering
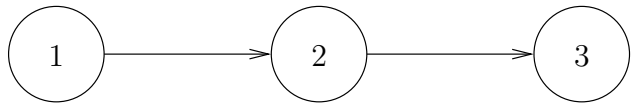
\caption{A network consisting of two point-to-point links in tandem.}
\label{fig:two_links}
\end{figure}

We develop our general result for unicast connections by extending
from some special cases.  We begin with the simplest non-trivial
case: that of two point-to-point links in tandem (see Figure~\ref{fig:two_links}).

Suppose we wish to establish a connection of rate arbitrarily close to
$R$ packets per unit time from node 1 to node 3.  Suppose further
that the coding scheme is run for a total time $\Delta$, from time 0
until time $\Delta$, and that, in this time, a total of $N$ packets is
received by node 2.  We call these packets $v_1, v_2, \ldots, v_N$.

Any packet $\packet$ received by a node is a linear
combination of $v_1, v_2, \ldots, v_N$, so we can write 
\[
\packet = \sum_{n=1}^N \beta_n v_n.
\]
Now, since $v_n$ is formed by a random linear combination of the message
packets $w_1, w_2, \ldots, w_K$, we have
\[
v_n = \sum_{k=1}^K \alpha_{nk} w_k
\]
for $n = 1, 2, \ldots, N$.  Hence
\[
\packet = \sum_{k=1}^K \left( \sum_{n=1}^N \beta_n \alpha_{nk} \right) w_k,
\]
and it follows that the $k$th component of the global encoding vector of
$\packet$ is given by
\[
\gamma_k = \sum_{n=1}^N \beta_n \alpha_{nk}.
\]
We call the vector $\beta$ associated with $\packet$ the \emph{auxiliary
encoding vector} of $\packet$, and we see that any node that receives 
$\lfloor K(1+\varepsilon) \rfloor$ or more packets with 
linearly-independent auxiliary encoding vectors has 
$\lfloor K(1+\varepsilon) \rfloor$ packets whose global encoding vectors
collectively form a random $\lfloor K(1+\varepsilon) \rfloor \times K$
matrix over $\mathbb{F}_q$, with all entries chosen uniformly.  If this
matrix has rank $K$, then node 3 is able to recover the message
packets.  The probability that a random 
$\lfloor K(1+\varepsilon) \rfloor \times K$ matrix has rank $K$ is, by a
simple counting argument,
$\prod_{k=1+\lfloor K(1+\varepsilon) \rfloor -K}^{\lfloor
K(1+\varepsilon) \rfloor} (1 - 1/q^k)$, which can be made arbitrarily
close to 1 by taking $K$ arbitrarily large.  Therefore, to determine
whether node 3 can recover the message packets, we essentially need only
to determine whether it receives $\lfloor K(1+\varepsilon) \rfloor$ or
more packets with linearly-independent auxiliary encoding vectors.

Our proof is based on tracking the propagation of what we call
\emph{innovative} packets.  Such packets are innovative in the sense
that they carry new, as yet unknown, information about $v_1, v_2,
\ldots, v_N$ to a node.  It turns out that the propagation of innovative
packets through a network follows 
the propagation of jobs through a queueing network, 
for which fluid flow models give good approximations.
We present the
following argument in terms of this fluid analogy and defer the formal
argument to Appendix~\ref{app:formal_two-link_tandem} at the end of this
chapter.

Since the packets being received by node 2 are the packets $v_1, v_2,
\ldots, v_N$ themselves, it is clear that every packet being received by
node 2 is innovative.  Thus, innovative packets arrive at node 2 at a
rate of $z_{122}$, and this can be approximated by fluid flowing in at
rate $z_{122}$.  These innovative packets are stored in node 2's memory,
so the fluid that flows in is stored in a reservoir.

Packets, now, are being received by node 3 at a rate of $z_{233}$, but
whether these packets are innovative depends on the contents of node 2's
memory.  If node 2 has more information about $v_1, v_2, \ldots, v_N$
than node 3 does, then it is highly likely that new information will be
described to node 3 in the next packet that it receives.  Otherwise, if
node 2 and node 3 have the same degree of information about $v_1, v_2,
\ldots, v_N$, then packets received by node 3 cannot possibly be
innovative.  Thus, the situation is as though fluid flows into node 3's
reservoir at a rate of $z_{233}$, but the level of node 3's reservoir is
restricted from ever exceeding that of node 2's reservoir.  The level of
node 3's reservoir, which is ultimately what we are concerned with, can
equivalently be determined by fluid flowing out of node 2's reservoir at
rate $z_{233}$.

\begin{figure}
\centering
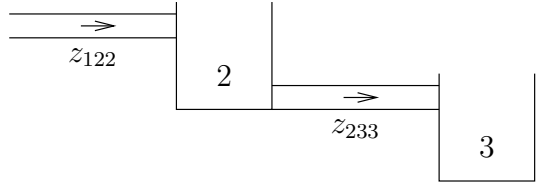
\caption{Fluid flow system corresponding to two-link tandem network.}
\label{fig:two_pipes}
\end{figure}

We therefore see that the two-link tandem network in
Figure~\ref{fig:two_links} maps to the fluid flow system shown in
Figure~\ref{fig:two_pipes}.  It is clear that, in this system, fluid
flows into node 3's reservoir at rate $\min(z_{122}, z_{233})$.  This rate
determines the rate at which packets with new information about $v_1,
v_2, \ldots, v_N$---and, therefore, linearly-independent auxiliary
encoding vectors---arrive at node 3.  Hence the time required for node 3
to receive $\lfloor K(1+\varepsilon) \rfloor$ packets with
linearly-independent auxiliary encoding vectors is, for large $K$,
approximately $K(1 + \varepsilon)/\min(z_{122}, z_{233})$, which implies
that a connection of rate arbitrarily close to $R$ packets per unit time
can be established provided that
\begin{equation}
R \le \min(z_{122}, z_{233}).
\label{eqn:2130}
\end{equation}
The right-hand side of (\ref{eqn:2130}) is indeed the capacity of the
two-link tandem network, and we therefore have the desired
result for this case.

\begin{figure}
\centering
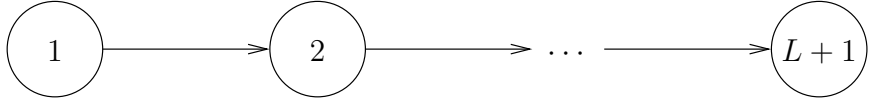
\caption{A network consisting of $L$ point-to-point links in tandem.}
\label{fig:l_links}
\end{figure}

We extend our result to another special case before considering general
unicast connections: we consider the case of a tandem network consisting
of $L$ point-to-point links and $L+1$ nodes (see
Figure~\ref{fig:l_links}).

\begin{figure}
\centering
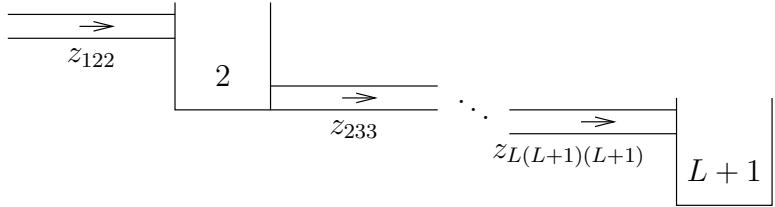
\caption{Fluid flow system corresponding to $L$-link tandem network.}
\label{fig:l_pipes}
\end{figure}

This case is a straightforward extension of that of the two-link tandem
network.  It maps to the fluid flow system shown in
Figure~\ref{fig:l_pipes}.  In this system, it is clear that fluid flows 
into node $(L+1)$'s reservoir at rate 
$\min_{1 \le i \le L}\{z_{i(i+1)(i+1)}\}$.  Hence a connection of rate
arbitrarily close to $R$ packets per unit time from node 1 to node $L+1$
can be established provided that
\begin{equation}
R \le \min_{1 \le i \le L}\{z_{i(i+1)(i+1)}\}.
\label{eqn:2150}
\end{equation}
Since the right-hand
side of (\ref{eqn:2150}) is indeed the capacity of the $L$-link tandem
network, we therefore have the desired result for this
case.  A formal argument is in Appendix~\ref{app:formal_l-link_tandem}.

We now extend our result to general unicast connections.  The strategy
here is simple:  A general unicast connection can be formulated as a
flow, which can be decomposed into a finite number of paths.  Each of
these paths is a tandem network, which is the case that we have just
considered.

Suppose that we wish to establish a connection of rate arbitrarily close
to $R$ packets per unit time from source node $s$ to sink node $t$.
Suppose further that
\[
R \leq \min_{Q \in \mathcal{Q}(s,t)}
  \left\{\sum_{(i, J) \in \Gamma_+(Q)} \sum_{K \not\subset Q} z_{iJK}
  \right\},
\]
where $\mathcal{Q}(s,t)$ is the set of all cuts between $s$ and $t$, and
$\Gamma_+(Q)$ denotes the set of forward hyperarcs of the cut $Q$, i.e.,\  
\[
\Gamma_+(Q) := \{(i, J) \in \mathcal{A} \,|\, i \in Q, J\setminus Q
\neq \emptyset\} .
\]
Therefore, by the max-flow/min-cut theorem (see, e.g.,\ 
\cite[Sections 6.5--6.7]{amo93}, 
\cite[Section 3.1]{ber98}), there exists a
flow vector $\flow$ satisfying
\[
\sum_{\{J | (i,J) \in \mathcal{A}\}} \sum_{j \in J} \flow_{iJj} 
- \sum_{\{j | (j,I) \in \mathcal{A}, i \in I\}} \flow_{jIi} =
\begin{cases}
R & \text{if $i = s$}, \\
-R & \text{if $i = t$}, \\
0 & \text{otherwise},
\end{cases}
\]
for all $i \in \mathcal{N}$, 
\begin{equation}
\sum_{j \in K} \flow_{iJj} \le \sum_{\{L \subset J | L \cap K \neq
\emptyset\}} z_{iJL}
\label{eqn:2600}
\end{equation}
for all $(i,J) \in \mathcal{A}$ and $K \subset J$,
and $\flow_{iJj} \ge 0$
for all $(i,J) \in \mathcal{A}$ and $j \in J$.

Using the conformal realization theorem
(see, e.g.,\  \cite[Section 1.1]{ber98}), we decompose $\flow$ into
a finite set of paths $\{p_1, p_2, \ldots, p_M\}$, 
each carrying positive 
flow $R_{m}$ for $m= 1, 2, \ldots, M$, such that
$\sum_{m=1}^M R_{m} = R$.  
We treat each path $p_m$ as a tandem network and use it to deliver
innovative packets at rate arbitrarily close to $R_m$, 
resulting in an overall rate 
for innovative packets arriving at node $t$
that is arbitrarily close to $R$.
Some care must be take in the interpretation of the flow and its path
decomposition because the same packet may be received by more than one
node.  The details of the interpretation are in
Appendix~\ref{app:formal_general_unicast}

\subsection{Multicast connections}
\label{sec:coding_multicast}

The result for multicast connections 
is, in fact, a straightforward extension of that
for unicast connections.
In this case, rather than a single sink $t$, we have a set
of sinks $T$.  
As in the framework of static broadcasting (see \cite{shu03, shf00}), we
allow sink nodes to operate at different rates.
We suppose that sink $t \in T$ wishes to achieve rate
arbitrarily close to $R_t$, i.e.,\  to recover the $K$ message packets,
sink $t$ wishes to wait for a time $\Delta_t$ that is only marginally
greater than $K/R_t$.  
We further suppose that
\[
R_t \leq \min_{Q \in \mathcal{Q}(s,t)}
  \left\{\sum_{(i, J) \in \Gamma_+(Q)} \sum_{K \not\subset Q} z_{iJK}
  \right\}
\]
for all $t \in T$.  Therefore, by the max-flow/min-cut theorem, there
exists, for each $t \in T$, a flow vector $\flow^{(t)}$ satisfying
\[
\sum_{\{j | (i,J) \in \mathcal{A}\}} \sum_{j \in J} \flow^{(t)}_{iJj} 
- \sum_{\{j | (j,I) \in \mathcal{A}, i \in I\}} \flow^{(t)}_{jIi} =
\begin{cases}
R & \text{if $i = s$}, \\
-R & \text{if $i = t$}, \\
0 & \text{otherwise},
\end{cases}
\]
for all $i \in \mathcal{N}$, 
\begin{equation*}
\sum_{j \in K} \flow^{(t)}_{iJj} \le \sum_{\{L \subset J | L \cap K \neq
\emptyset\}} z_{iJL}
\end{equation*}
for all $(i,J) \in \mathcal{A}$ and $K \subset J$,
and $\flow^{(t)}_{iJj} \ge 0$
for all $(i,J) \in \mathcal{A}$ and $j \in J$.

For each flow vector $\flow^{(t)}$, we go through the same argument as that
for a unicast connection, and we find that the probability of error at
every sink node can be made arbitrarily small by taking $K$
sufficiently large.

We summarize our results with the following
theorem statement.

\begin{Thm}
Consider the coding subgraph $z$.
The random linear coding scheme described in
Section~\ref{sec:coding_scheme} is capacity-achieving for
multicast connections in $z$,
i.e.,\  for $K$ sufficiently large, it can achieve, with
arbitrarily small error probability, a multicast
connection 
from source node $s$ to sink nodes in the set $T$ at rate
arbitrarily close to $R_t$ packets per unit time for each $t \in T$ if
\[
R_t \leq \min_{Q \in \mathcal{Q}(s,t)}
  \left\{\sum_{(i, J) \in \Gamma_+(Q)} \sum_{K \not\subset Q} z_{iJK}
  \right\}
\]
for all $t \in T$.\footnote{In 
earlier versions of this work \cite{lme04, lmk05-further}, we required the
field size $q$ of the coding scheme to approach infinity for
Theorem~\ref{thm:2100} to hold.  This requirement is in fact not
necessary, and the formal arguments in 
Appendix~\ref{app:formal} do not require
it.}
\label{thm:2100}
\end{Thm}

\noindent \emph{Remark.}  The capacity region is determined solely by
the average rates $\{z_{iJK}\}$ at which packets are received.  Thus,
the packet injection and loss processes, which give rise to the packet
reception processes, can in fact take any distribution, exhibiting
arbitrary correlations, as long as these average rates exist.

\subsection{An example}
\label{sec:example2}

We return to the slotted Aloha relay channel described in
Section~\ref{sec:example1}.
Theorem~\ref{thm:2100} implies that the random linear coding scheme
we consider can achieve the desired unicast connection at rates
arbitrarily close to $R$ packets per unit time if
\[
R \le \min(z_{1(23)2} + z_{1(23)3} + z_{1(23)(23)},
  z_{1(23)3} + z_{1(23)(23)} + z_{233}).
\]
Substituting (\ref{eqn:1100})--(\ref{eqn:1200}), we obtain
\begin{multline*}
R \le \min(z_{1(23)}(1-z_{23})(p_{1(23)2} + p_{1(23)3} + p_{1(23)(23)}),
\\
  z_{1(23)}(1-z_{23})(p_{1(23)3} + p_{1(23)(23)}) + (1-z_{1(23)})z_{23}p_{233}).
\end{multline*}
We see that the range of achievable rates is specified completely in
terms of the parameters we control, $z_{1(23)}$ and $z_{23}$, and the
given parameters of the problem, $p_{1(23)2}$, $p_{1(23)3}$, 
$p_{1(23)(23)}$, and $p_{233}$.  It remains only to choose 
$z_{1(23)}$ and $z_{23}$.  This, we deal with in the
next chapter.

\section{Error exponents for Poisson traffic with i.i.d.\  losses}
\label{sec:error_exponents}

We now look at the rate of decay of the probability of
error $p_e$ in the coding delay $\Delta$.  
In contrast to traditional error exponents where coding delay is
measured in symbols, we measure coding delay in time units---time
$\tau = \Delta$ is 
the time at which the sink nodes attempt to decode the
message packets.  The two methods of measuring delay are essentially
equivalent when packets arrive in regular, deterministic intervals.

We specialize to the case of Poisson traffic with i.i.d.\  losses.  
Thus, the process $A_{iJK}$ is a
Poisson process with rate $z_{iJK}$.
Consider the unicast case for now, and
suppose we wish to establish a connection of rate $R$.
Let $C$ be the supremum of all asymptotically-achievable rates.

We begin by deriving an upper bound on the probability of error.
To this end, we take a flow vector $\flow$ from $s$ to $t$ of size $C$
and, following the development in
Appendix~\ref{app:formal}, 
develop a queueing network from it that describes the propagation of
innovative packets for a given innovation order $\mu$.
This queueing network now becomes a Jackson network.
Moreover, as a consequence of Burke's
theorem (see, e.g.,\  \cite[Section 2.1]{kel79}) and the fact that
the queueing network is acyclic, the
arrival and departure processes at all stations are 
Poisson in steady-state.  

Let $\Psi_{t}(m)$ be the arrival time of the $m$th  
innovative packet at $t$, and let $C^\prime := (1-q^{-\mu})C$. 
When the queueing network is in steady-state, the arrival of innovative
packets at $t$ is described by a Poisson process of rate $C^\prime$.
Hence we have
\begin{equation}
\lim_{m \rightarrow \infty} \frac{1}{m}
\log \mathbb{E}[\exp(\theta \Psi_{t}(m))]
= \log \frac{C^\prime}{C^\prime - \theta} 
\label{eqn:2900}
\end{equation}
for $\theta < C^\prime$ \cite{bpt98, pal03}.
If an error occurs, then fewer than $\lceil R\Delta \rceil$
innovative packets are received by $t$ by 
time $\tau = \Delta$, which is
equivalent to
saying that $\Psi_{t}(\lceil R\Delta \rceil) > \Delta$.
Therefore,
\[
p_e \le \Pr(\Psi_{t}(\lceil R\Delta \rceil) > \Delta),
\]
and, using the Chernoff bound, we obtain
\[
p_e \le \min_{0 \le \theta < C^\prime}
\exp\left(
-\theta \Delta + \log \mathbb{E}[\exp(\theta \Psi_{t}(\lceil R\Delta
\rceil) )] 
\right) .
\]
Let $\varepsilon$ be a positive real number.  
Then using equation (\ref{eqn:2900}) we obtain, 
for $\Delta$ sufficiently large,
\[
\begin{split}
p_e &\le \min_{0 \le \theta < C^\prime}
\exp\left(-\theta \Delta
+ R \Delta \left\{\log \frac{C^\prime}{C^\prime-\theta} + \varepsilon \right\} \right)
\\
&= \exp( -\Delta(C^\prime-R-R\log(C^\prime/R)) + R\Delta \varepsilon) .
\end{split}
\]
Hence, we conclude that
\begin{equation}
\lim_{\Delta \rightarrow \infty} \frac{-\log p_e}{\Delta}
\ge C^\prime - R - R\log(C^\prime/R) .
\label{eqn:2910}
\end{equation}

For the lower bound, we examine 
a cut whose flow capacity is $C$.  We take one such cut and denote it by
$Q^*$.  It is
clear that, if fewer than $\lceil R\Delta \rceil$ distinct packets are
received across $Q^*$ in time $\tau = \Delta$, then an error occurs.
The arrival of
distinct packets across $Q^*$ is described by a Poisson
process of rate $C$.  
Thus we have
\[
\begin{split}
p_e &\ge \exp(-C\Delta)
\sum_{l = 0}^{\lceil R\Delta \rceil - 1}
\frac{(C\Delta)^l}{l!}  \\
&\ge \exp(-C \Delta)
\frac{(C\Delta)^{\lceil R\Delta \rceil -1}}
{\Gamma(\lceil R \Delta \rceil)} ,
\end{split}
\]
and, using Stirling's formula, we obtain 
\begin{equation}
\lim_{\Delta \rightarrow \infty} \frac{-\log p_e}{\Delta}
\le C - R - R\log(C/R) .
\label{eqn:2915}
\end{equation}

Since (\ref{eqn:2910}) holds for all positive integers $\mu$, we conclude from
(\ref{eqn:2910}) and (\ref{eqn:2915}) that
\begin{equation}
\lim_{\Delta \rightarrow \infty} \frac{-\log p_e}{\Delta}
= C - R - R\log(C/R) .
\label{eqn:2920}
\end{equation}

Equation (\ref{eqn:2920}) defines the asymptotic rate of decay of the
probability of error in the coding delay $\Delta$.  This asymptotic rate
of decay is determined entirely by $R$ and $C$.  Thus, for a packet
network with Poisson traffic and i.i.d.\  losses employing the coding
scheme described in Section~\ref{sec:coding_scheme}, 
the flow capacity $C$ of the minimum cut of the network is
essentially the sole figure of merit of importance in determining the
effectiveness of the coding scheme for large, but finite, coding delay.
Hence, in deciding how to inject packets to support the desired
connection, a sensible approach is to reduce our attention to this
figure of merit, which is indeed the approach that we take in
Chapter~\ref{chap:subgraph_selection}.

Extending the result from unicast connections to multicast connections
is straight\-forward---we simply obtain (\ref{eqn:2920}) for each sink.

\section{Finite-memory random linear coding}
\label{sec:finite-memory}

The results that we have thus far established about the coding scheme
described in Section~\ref{sec:coding_scheme} show that, from the
perspective of conveying the most information in each packet
transmission, it does very well.  But packet transmissions are not the
only resource with which we are concerned.  Other resources that may be
scarce include memory and computation and, if these resources are as
important or more important than packet transmissions, then a natural
question is whether we can modify the coding scheme of
Section~\ref{sec:coding_scheme} to reduce its memory and computation
requirements, possibly in exchange for more transmissions.

In this section, we study a simple modification.  We take the coding
scheme of Section~\ref{sec:coding_scheme}, and we assume that
intermediate nodes (i.e.,\  nodes that are neither source nor sink nodes)
have memories capable only of storing a fixed, finite number of packets,
irrespective of $K$.  An intermediate node with a memory capable of
storing $M$ packets uses its memory in one of two ways:
\begin{enumerate}
\item as a shift register: 
arriving packets are
stored in memory and, if the memory is already full, the oldest packet
in the memory is discarded; or
\item as an accumulator:
arriving packets are multiplied by a random vector chosen uniformly over
$\mathbb{F}_q^M$, and the product is added to the $M$ memory slots.
\end{enumerate}

We first consider, in Section~\ref{sec:finite-memory_isolation}, the
case of a single intermediate node in isolation.  In this case, the
intermediate node encodes packets and its immediate downstream node
decodes them.  Such a scheme offers an attractive alternative to
comparable reliability schemes for a single link, such as automatic
repeat request (\textsc{arq}) or convolutional coding (see, e.g.,\
\cite{afi05, ayi00}).  In Section~\ref{sec:finite-memory_two-link}, we
consider a network, specifically, the two-link tandem network (see
Figure~\ref{fig:two_links}).  We see that, while limiting the memory of
intermediate nodes certainly results in loss of achievable rate, the
relative rate loss, at least for the two-link tandem network, can be
quantified, and it decays exponentially in $M$.


\subsection{Use in isolation}
\label{sec:finite-memory_isolation}

When used in isolation at a single intermediate node, the encoder takes
an incoming stream of message packets, $u_1, u_2, \ldots$, and forms a
coded stream of packets that is placed on its lossy outgoing link and
decoded on reception.  We assume that the decoder knows, for each
received packet, the linear transformation that 
has been performed on the message packets to yield that packet.
This information can be communicated to the decoder by a variety of
means, which include placing it into the header of each packet as
described in Section~\ref{sec:coding_scheme} (which is certainly viable
when the memory is used as a shift register---the overhead is $M\log_2
q$ bits plus that of a sequence number), and initializing the random
number generators at the encoder and decoder with the same seed.

The task of decoding, then, equates to matrix inversion in
$\mathbb{F}_q$, which can be done straightforwardly by applying Gaussian
elimination to each packet as it is received.  This procedure produces
an approximately-steady stream of decoded packets with an expected delay
that is constant in the length of the input stream.  Moreover, if the
memory is used as a shift register, then the complexity of this decoding
procedure is also constant with the length of the input stream and, on
average, is $O(M^2)$ per packet.

We discretize the
time axis into epochs that correspond to the transmission of an outgoing
packet.  Thus, in each epoch, an outgoing packet is transmitted, which
may be lost, and one or more incoming packets are received.  If
transmission is to be reliable, then the average number of incoming
packets received in each epoch must be at most one.

We make the following assumptions on incoming packet arrivals and
outgoing packet losses, with the understanding that generalizations are
certainly possible.  We assume that, in an epoch, a single packet
arrives independently with probability $r$ and no packets arrive
otherwise, and the transmitted outgoing packet is lost independently
with probability $\varepsilon$ and is received otherwise.  This model is
appropriate when losses and arrivals are steady---and not bursty.

We conduct our analysis in the limit of $q \rightarrow \infty$, i.e.,\
the limit of infinite field size.  We later discuss how the analysis may
be adapted for finite $q$, and quantify by simulation the difference
between the performance in the case of finite $q$ and that of infinite
$q$ in some particular instances.

\begin{figure*}
\centering
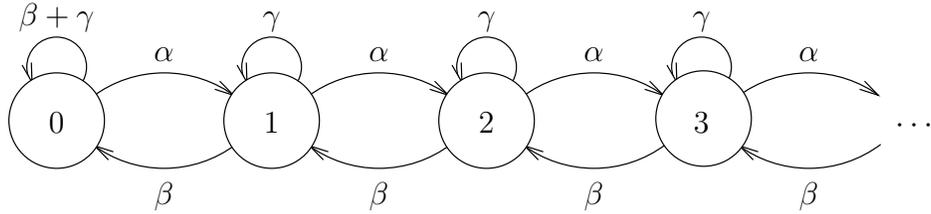
\caption{Markov chain modeling the evolution of the difference between
the number of packets received by the encoder and the number of packets
transmitted and not lost.}
\label{fig:markov_chain}
\end{figure*}

We begin by considering the difference between the number of packets
received by the encoder and the number of packets transmitted and not
lost.  This quantity, we see, evolves according to the infinite-state
Markov chain shown in Figure~\ref{fig:markov_chain}, 
where $\alpha = r\varepsilon$, 
$\beta = (1-r)(1-\varepsilon)$, and
$\gamma = r(1-\varepsilon) + (1-r)\varepsilon$.

At the first epoch, the memory of the encoder is empty and we are in
state 0.  We continue to remain in state 0 in subsequent epochs until
the first packet $u_1$ arrives.  Consider the first outgoing
packet after the arrival of $u_1$.  This packet is either lost or not.
Let us first suppose that it is not lost.  Thus, we remain in state 0,
and the decoder receives a packet that is a random linear combination of
$u_1$, i.e.,\  a random scalar multiple of $u_1$, and, since $q$ is
infinitely large by assumption, this scalar multiple is non-zero with
probability 1; so the decoder can recover $u_1$ from the packet that it
receives.  

Now suppose instead that the first outgoing packet after the arrival of
$u_1$ is lost.  Thus, we move to state 1.  If an outgoing packet is
transmitted and not lost before the next packet arrives, the
decoder again receives a random scalar multiple of $u_1$ and we return
to state 0.  So suppose we are in state 1 and $u_2$ arrives.  Then, the
next outgoing packet is a random linear combination of $u_1$ and $u_2$.
Suppose further that this packet is received by the decoder, so we are
again in state 1.  This packet, currently, is more or less useless to
the decoder; it represents a mixture between $u_1$ and $u_2$ and does
not allow us to determine either.  Nevertheless, it gives, with
probability 1, the decoder some information that it did not previously
know, namely, that $u_1$ and $u_2$ lie in a particular linear subspace
of $\mathbb{F}_q^2$.  
As in Section~\ref{sec:coding_theorems},
we call such an informative packet
$\emph{innovative}$.

\begin{figure*}
\centering
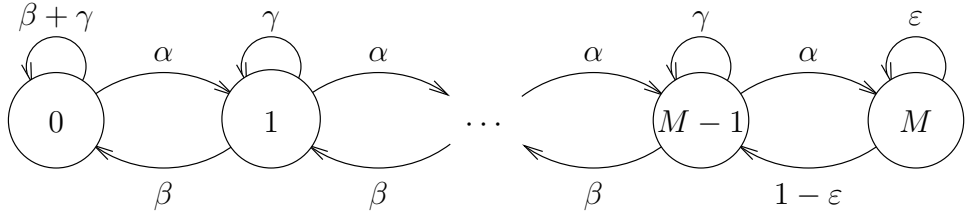
\caption{Markov chain modeling the 
behavior of the coding scheme in the limit of $q \rightarrow \infty$.}
\label{fig:markov_chain3}
\end{figure*}

Any subsequent packet received by the decoder is also innovative with
probability 1.  In particular, if the decoder receives a packet before
the arrival of another packet $u_3$ at the encoder, 
returning us to state 0,
then the decoder is able to recover both $u_1$ and $u_2$.  More
generally, what we see is that, provided that packets arrive only in
states $0, 1, \ldots, M-1$, the decoder is able to recover, at every
return to state 0, the packets that arrived between the current and the
previous return.  If a packet arrives in state $M$, however, loss
occurs.  Information in the encoder's memory is overwritten or
corrupted, and will never be recovered.  The current contents of the
encoder's memory, however, can still be recovered and, from the point of
view of recovering these contents, the coding system behaves as though
we were in state $M$.  Hence, to analyze the performance of the coding
scheme, we modify the Markov chain shown in
Figure~\ref{fig:markov_chain} to that in Figure~\ref{fig:markov_chain3}.
Let $x_t$ be the state of this Markov chain at time $t$.  We can
interpret $x_t$ as the number of innovative packets the encoder has for
sending at time $t$.

We now proceed to derive some quantities that are useful for designing
the parameters of the coding scheme.  We begin with the steady-state
probabilities $\pi_i := \lim_{t \rightarrow \infty} \Pr(x_t = i)$.
Since $\{x_t\}$ is a
birth-death process, its steady-state probabilities are readily
obtained. 
We obtain
\begin{equation}
\pi_i = \frac{\varrho^i(1-\varrho)}{1-\sigma\varrho^M}
\label{eqn:2750}
\end{equation}
for $i = 0, 1, \ldots, M-1$, and
\begin{equation}
\pi_M = \frac{\varepsilon\sigma\varrho^{M-1}(1-\varrho)}{1-\sigma\varrho^M},
\label{eqn:2760}
\end{equation}
where
$\varrho := {\alpha}/{\beta} =
{r\varepsilon}/{(1-r)(1-\varepsilon)}$
and
$\sigma := {r}/{(1-\varepsilon)}$.
We assume $\varrho < 1$, which is equivalent to $r < 1 - \varepsilon$, for,
if not, the capacity of the outgoing link is exceeded, and we cannot
hope for the coding scheme to be effective.

We now derive the probability of packet loss, $p_l$.  Evaluating $p_l$
is not straightforward because, since coded packets depend on each
other, the loss of a packet owing to the encoder exceeding its memory is
usually accompanied by other packet losses.  We derive an upper bound on
the probability of loss.

A packet is successfully recovered by the decoder if the ensuing path
taken in the Markov chain in Figure~\ref{fig:markov_chain3} returns to
state 0 without a packet arrival occurring in state $M$.
Let $q_i$ be the probability that a path, originating in state $i$,
reaches state 0 without a packet arrival occurring in state $M$. 
Our problem is very similar to a random walk, or ruin, problem
(see, e.g.,\  \cite[Chapter XIV]{fel68}).
We obtain
\[
q_i = \frac{1 - \sigma\varrho^{M-i}}{1-\sigma\varrho^M}
\]
for $i = 0, 1, \ldots, M$.

Now, after the coding scheme has been running for some time, a random
arriving packet finds the scheme in state $i$ with probability $\pi_i$
and, with probability $1-\varepsilon$, the scheme returns to state $i$
after the next packet transmission or, with probability $\varepsilon$,
it moves to state $i+1$.  Hence
\[
\begin{split}
1 - p_l &\ge \sum_{i=0}^{M-1}\{(1-\varepsilon)q_i + \varepsilon q_{i+1}\}
\pi_i \\
&= \sum_{i=0}^{M-1} \left\{(1-\varepsilon)
\frac{1-\sigma\varrho^{M-i}}{1-\sigma\varrho^M}
+ \varepsilon \frac{1-\sigma\varrho^{M-i-1}}{1-\sigma\varrho^M}\right\}
\frac{\varrho^i(1-\varrho)}{1-\sigma\varrho^M} \\
&= \frac{1-\varrho}{(1-\sigma\varrho^M)^2}
\left\{
\frac{1-\varrho^M}{1-\varrho}
- (1-\varepsilon)M\sigma\varrho^M 
- \varepsilon M\sigma\varrho^{M-1}
\right\} \\
&= \frac{1}{(1-\sigma\varrho^M)^{2}}
\{1-\varrho^M - (1-2\varepsilon)M\sigma\varrho^M
- \varepsilon M\sigma\varrho^{M-1}
+ (1-\varepsilon)M\sigma\varrho^{M+1}\},
\end{split}
\]
from which we obtain
\begin{equation}
p_l \le \frac{\varrho^{M-1}}{(1-\sigma\varrho^M)^2}
\{\varepsilon M\sigma + (1-2\sigma+M\sigma-2\varepsilon M\sigma)\varrho
-(1-\varepsilon)M\sigma\varrho^2 + \sigma^2\varrho^{M+1} \}.
\label{eqn:2850}
\end{equation}

We have thus far looked at the limit of $q \rightarrow \infty$, while,
in reality, $q$ must be finite.  There are two effects of having finite
$q$:  The first is that, while the encoder may have innovative
information to send to the decoder (i.e.,\  $x_t > 0$), it fails to do so
because the linear combination it chooses is not linearly independent of
the combinations already received by the decoder.  For analysis, we can
consider such non-innovative packets to be equivalent to erasures, and
we find that the effective erasure rate is $\varepsilon (1 - q^{-x_t})$.
The Markov chain in Figure~\ref{fig:markov_chain3} can certainly be
modified to account for this effective erasure rate, but doing so makes
analysis much more tedious.

The second of the effects is that, when a new packet arrives, it may not
increase the level of innovation at the encoder.  When the memory is
used as a shift register, this event arises because a packet is
overwritten before it has participated as a linear factor in any
successfully received packets, i.e.,\  all successfully received packets
have had a coefficient of zero for that packet.  When the memory is used
as an accumulator, this event arises because the random vector chosen to
multiply the new packet is such that the level of innovation remains
constant.
The event of the level of innovation not being
increased by a new packet can be quite disastrous, because it is
effectively equivalent to the encoder exceeding its memory.
Fortunately, the event seems rare; in the accumulator case, we can
quantify the probability of the event exactly as $1 - q^{x_t-M}$.

\begin{figure}
\centering
{\includegraphics{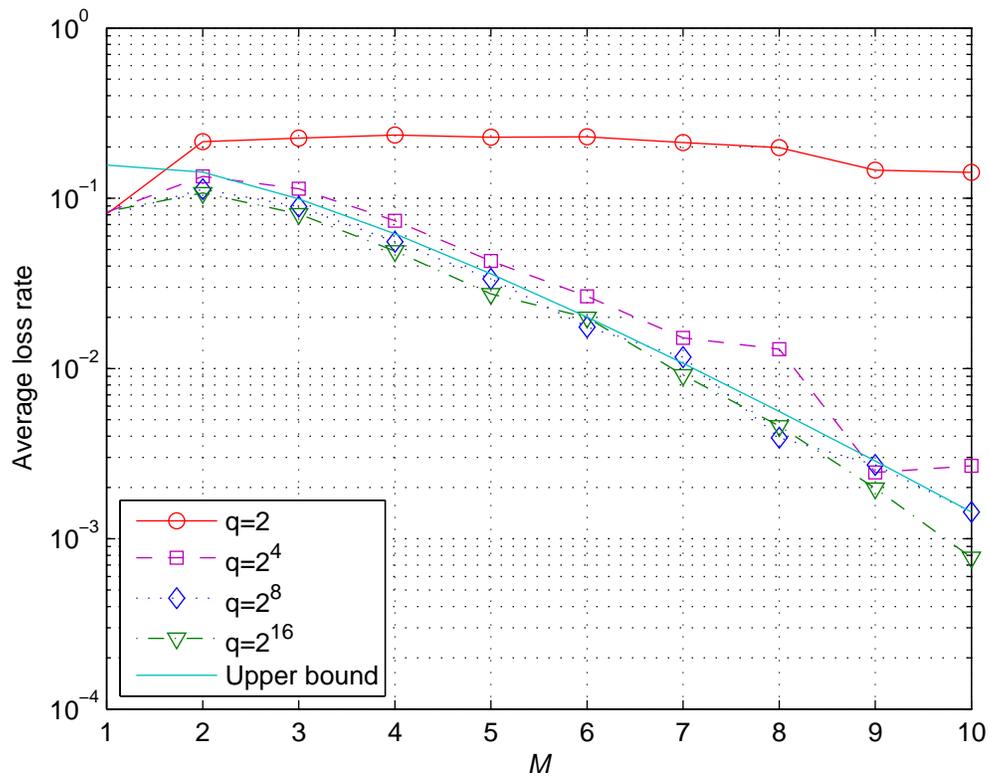}}
\caption[Average loss rate for 200,000 packets 
as a function of memory size $M$ with
$r=0.8$, $\varepsilon = 0.1$, and various coding field sizes $q$.] 
{Average loss rate for 200,000 packets 
as a function of memory size $M$ with
$r=0.8$, $\varepsilon = 0.1$, and various coding field sizes $q$.  The
upper bound on the probability of loss for $q \rightarrow \infty$ is also
drawn.}
\label{fig:lossplot-p0_8-eps0_1}
\end{figure}

\begin{figure}
\centering
{\includegraphics{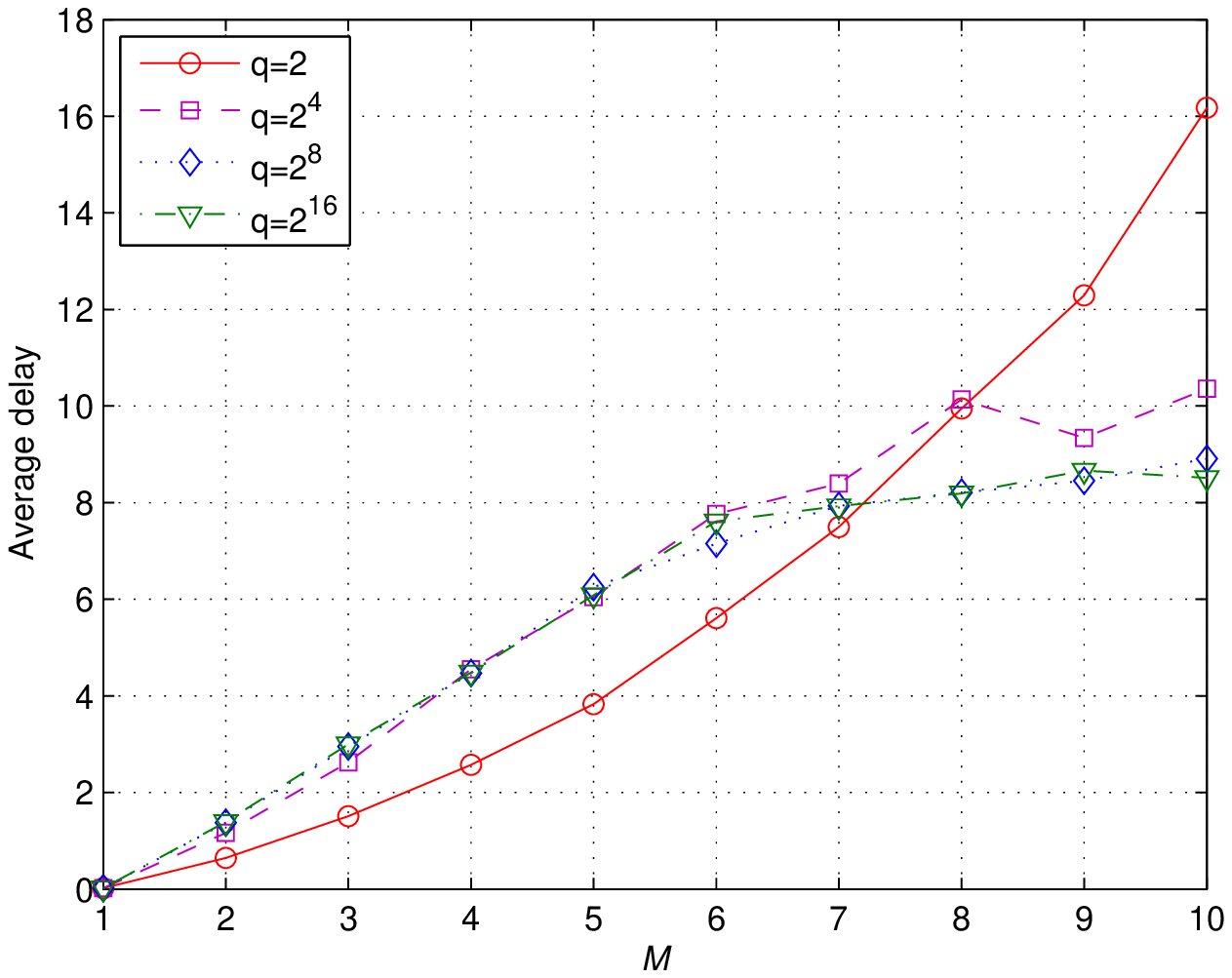}}
\caption{Average delay for 200,000 packets 
as a function of memory size $M$ with
$r=0.8$, $\varepsilon = 0.1$, and various coding field sizes $q$.}
\label{fig:delayplot-p0_8-eps0_1}
\end{figure}

\begin{figure}
\centering
{\includegraphics{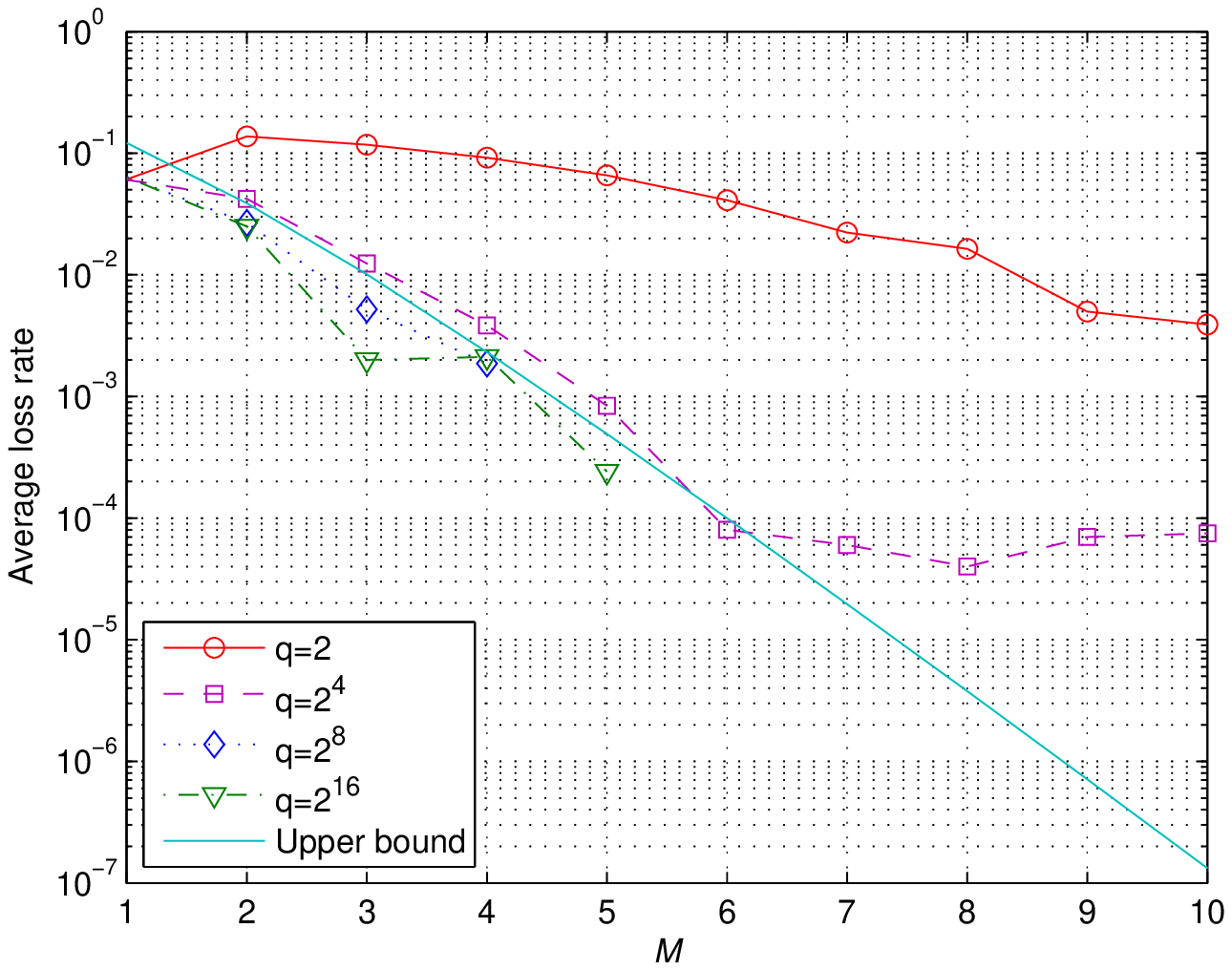}}
\caption[Average loss rate for 200,000 packets 
as a function of memory size $M$ with
$r=0.6$, $\varepsilon = 0.1$, and various coding field sizes $q$.] 
{Average loss rate for 200,000 packets 
as a function of memory size $M$ with
$r=0.6$, $\varepsilon = 0.1$, and various coding field sizes $q$.  The
upper bound on the probability of loss for $q \rightarrow \infty$ is also
drawn.}
\label{fig:lossplot-p0_6-eps0_1}
\end{figure}

\begin{figure}
\centering
{\includegraphics{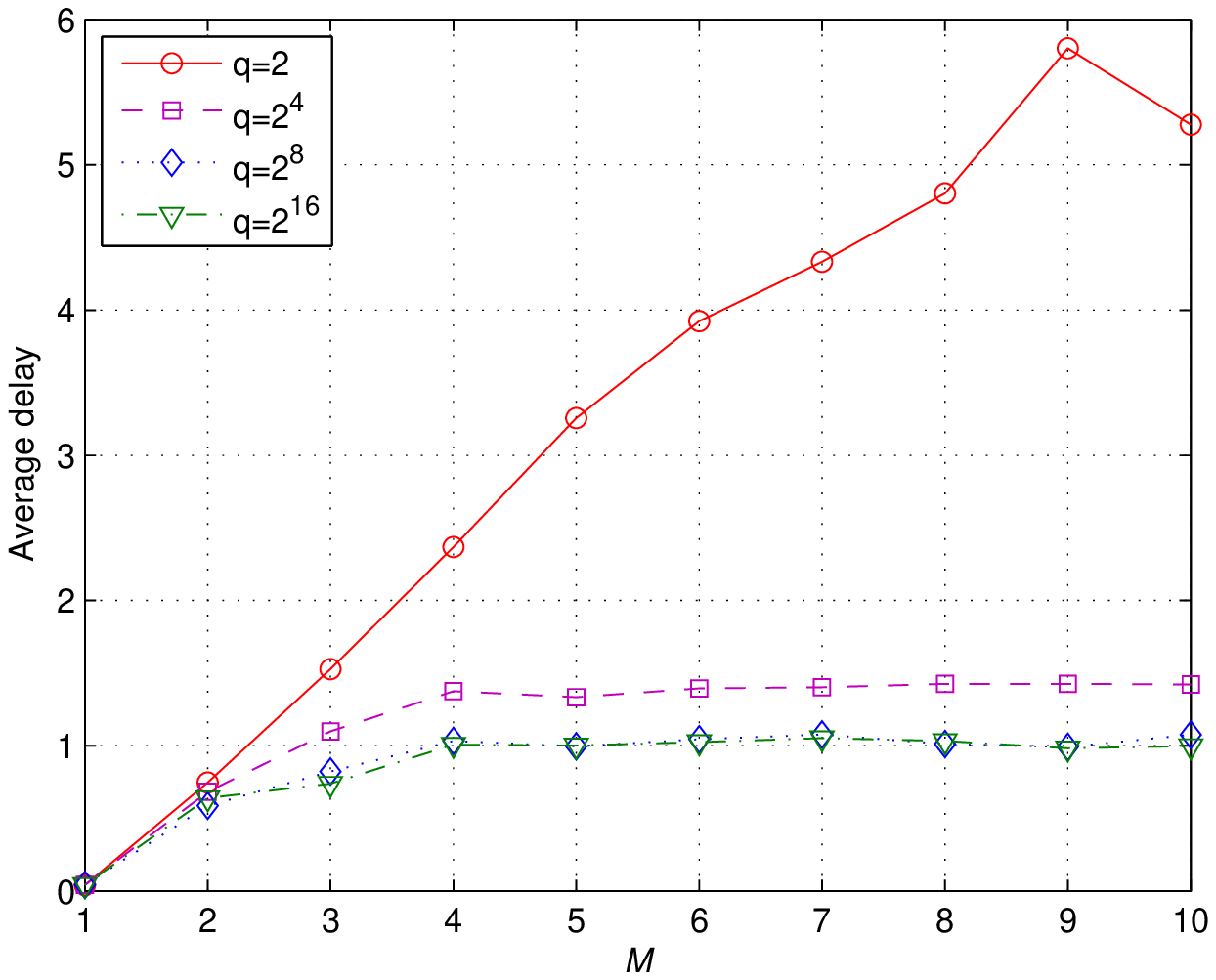}}
\caption{Average delay for 200,000 packets 
as a function of memory size $M$ with
$r=0.6$, $\varepsilon = 0.1$, and various coding field sizes $q$.}
\label{fig:delayplot-p0_6-eps0_1}
\end{figure}

To examine the effect of finite $q$, we chose $\varepsilon =0.1$ and
simulated the performance of the coding scheme for 200,000 packets with
various choices of the parameters $r$, $q$, and $M$ (see
Figures~\ref{fig:lossplot-p0_8-eps0_1}--\ref{fig:delayplot-p0_6-eps0_1}).
We decoded using Gaussian elimination on packets as they were received
and used the encoder's memory as a shift register to keep decoding
complexity constant with the length of the packet stream.  Delay was
evaluated as the number of epochs between a packet's arrival at the
encoder and it being decoded, neglecting transmission delay.  As
expected, we see that average loss rate decreases and average delay
increases with increasing $M$; a larger memory results, in a sense, in
more coding, which gives robustness at the expense of delay.  Moreover,
we see that a field size $q\ge 2^8$ (perhaps even $q \ge 2^4$) is
adequate for attaining loss rates close to the upper bound for infinite
field size.  

\subsection{Use in a two-link tandem network}
\label{sec:finite-memory_two-link}

When finite-memory random linear coding is used in isolation, packets
are sometimes lost because the decoder receives linear combinations
that, although innovative, are not decodable.  For example, suppose the
decoder receives $u_1+u_2$, but is neither able to recover $u_1$ nor
$u_2$ from other packets.  This packet, $u_1 + u_2$, definitely gives
the decoder some information, but, without either $u_1$ or $u_2$, the
packet must be discarded.  This would not be the case, however, if $u_1$
and $u_2$ were themselves coded packets---a trivial example, assuming
that we are not coding over $\mathbb{F}_2$, is if $u_1 = u_2 = w_1$,
where $w_1$ is a message packet for an outer code.

In this section, we consider finite-memory random linear coding in the
context of a larger coded packet network.  We consider the simplest
set-up with an intermediate node: a two-link tandem network (see
Figure~\ref{fig:two_links}) where we wish to establish a unicast
connection from node 1 to node 3.  Node 1 and node 3 use the coding
scheme described in Section~\ref{sec:coding_scheme} without
modification, while node 2 has only $M < K$ memory elements and uses the
modified scheme.  This simple two-link tandem network
serves as a basis for longer tandem networks and more general network
topologies.

We again discretize the time axis.  We assume that, at each epoch,
packets are injected by both nodes 1 and 2 and they are lost
independently with probability $\delta$ and $\varepsilon$, respectively.
Although situations of interest may not have transmissions that are
synchronized in this way, the synchronicity assumption can be relaxed to
an extent by accounting for differences in the packet injection rates
using the loss rates.

We again conduct our analysis in the limit of infinite field size.  The
considerations for finite field size are the same as those mentioned in
Section~\ref{sec:finite-memory_isolation}.  

\begin{figure*}
\centering
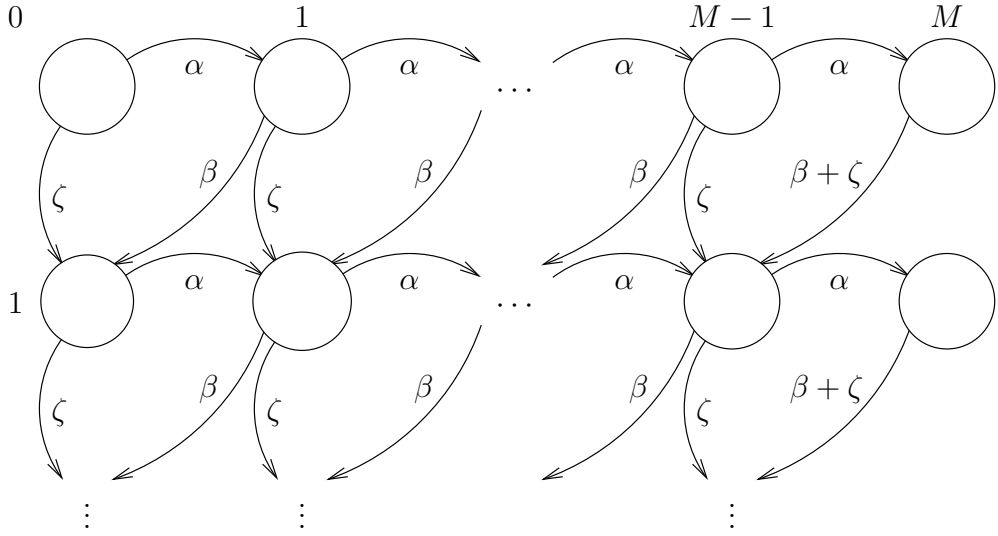
\caption[Markov chain modeling the evolution of $x_t$ and $y_t$.] 
{Markov chain modeling the evolution of $x_t$ and $y_t$.  To
simplify the diagram, we do not show self-transitions.}
\label{fig:markov_chain_2d}
\end{figure*}

Let $x_t$ denote the number of innovative packets (relative to $u_1,
u_2, \ldots, u_N$) node 2 has for sending at time $t$, and let $y_t$ denote
the number of innovative packets received by node 3 at time $t$.  By the
arguments of Section~\ref{sec:finite-memory_isolation}, the following
principles govern the evolution of $x_t$ and $y_t$ over time:
\begin{itemize}
\item As long as $x_t < M$, i.e.,\  the memory does not already have $M$
innovative packets,
node 2 increases the innovation contents of its memory by 1 upon
successful reception of a packet over arc $(1,2)$.
\item As long as $x_t > 0$, i.e.,\  the memory is not completely
redundant, the output of 2 is innovative, so $y_t$ will increase
by 1 provided that transmission over $(2,3)$ is successful.
\end{itemize}
Let $\alpha := (1 - \delta)\varepsilon$, $\beta :=
\delta(1-\varepsilon)$, and
$\zeta := (1-\delta)(1-\varepsilon)$.
Then the evolution of $x_t$ and $y_t$ is modeled by the Markov chain
shown in Figure~\ref{fig:markov_chain_2d}, where the horizontal
coordinate of a state indicates $x_t$, and the vertical coordinate
corresponds to the variable $y_t$.

We see that $\{x_t\}$ evolves as in
Section~\ref{sec:finite-memory_isolation}, so its steady-state
probabilities are given by (\ref{eqn:2750}) and (\ref{eqn:2760}) with
$r=1-\delta$.  Hence,
once the system is sufficiently mixed, the probability that $y_t$
increases at time $t$ is given by 
\[
\begin{split}
\zeta \pi_0 + (1-\varepsilon) \pi_1 + \cdots + (1-\varepsilon) \pi_M
&= (1 - \varepsilon)(1 - \delta \pi_0) \\
&= (1 - \delta)(1 - \pi_M).
\end{split}
\]
Therefore the system can operate at rate 
\begin{equation*}
R = (1-\delta)(1 - \pi_M)
\end{equation*}
with high probability of success.  

Suppose, without loss of generality, that $\delta > \varepsilon$, so
$\varrho < 1$.  Let $R^*$ be the min-cut capacity, or maximum rate, of the
system, which, in this case, is $1-\delta$.  Then the relative rate loss
with respect to the min-cut rate is
\begin{equation}
1 - \frac{R}{R^*} = \pi_M.
\label{eqn:2860}
\end{equation}

As discussed before, our analysis assumes forming linear combinations
over an infinitely large field, resulting in a Markov chain model with
transition probabilities given in Figure~\ref{fig:markov_chain_2d}.  If
on the other hand the field size is finite, we can still find new
expressions for the transition probabilities, although the complete
analysis becomes very complex.  In particular, assume that the memory is
used as an accumulator, so that the contents of the memory at each time
are uniformly random linear combinations, over $\mathbb{F}_q$, of the
received packets at node 2 by that time.  Then, as we have mentioned, if
the innovation content of the memory is $x$ and a new packet arrives at
node 2, the probability that node 2 can increase the innovation of its
memory by $1$ is $(1-q^{x-M})$, independently from all other past
events.  Similarly, the probability that the output of node 2 is
innovative is $(1-q^{-x})$.  

\begin{figure}
\centering
\includegraphics{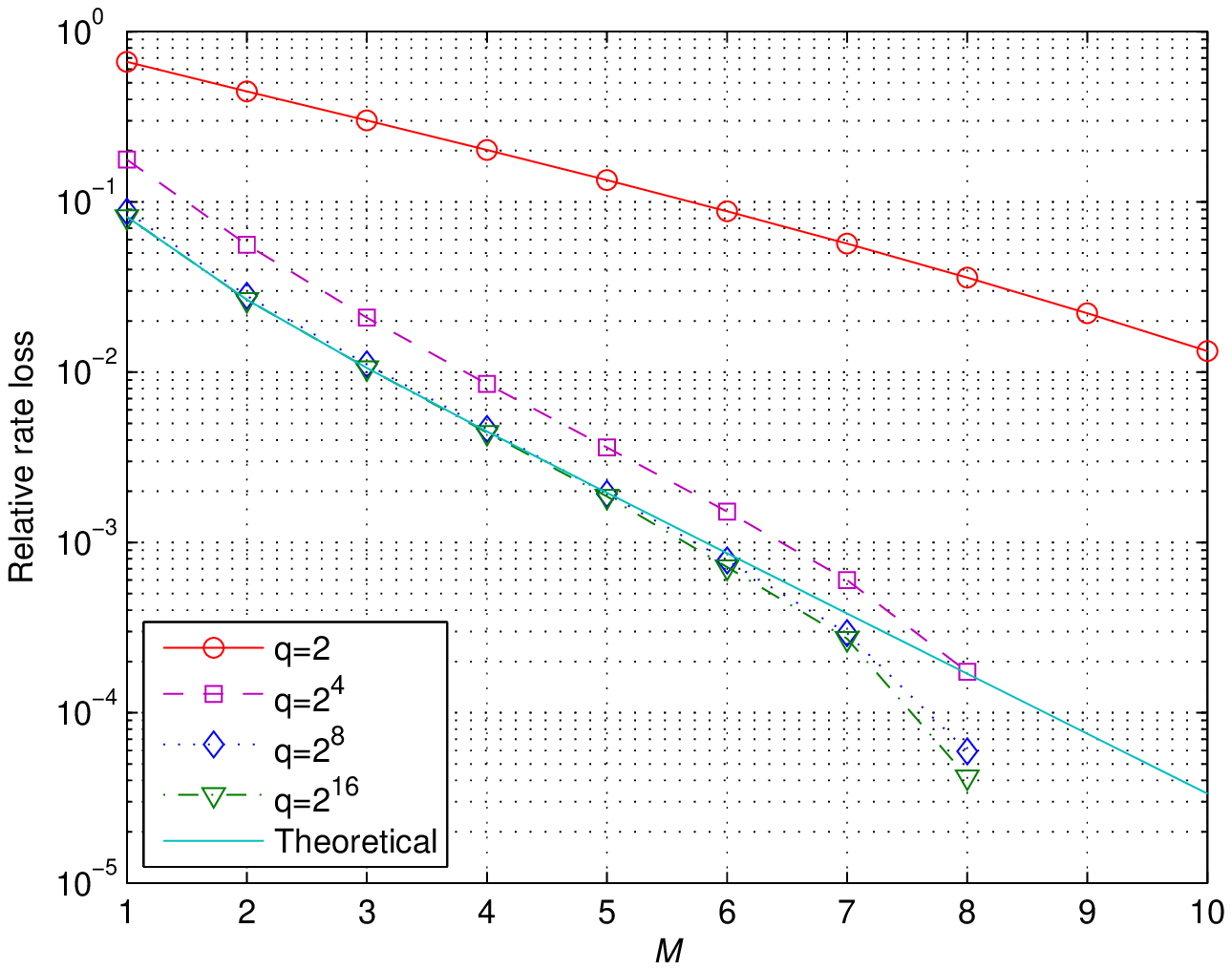}
\caption{Relative rate loss with respect to min-cut rate as a function
of memory size $M$ for $\delta=0.2$, $\varepsilon=0.1$, and various
coding field sizes $q$.}
\label{fig:payamplot3}
\end{figure}

\begin{figure}
\centering
\includegraphics{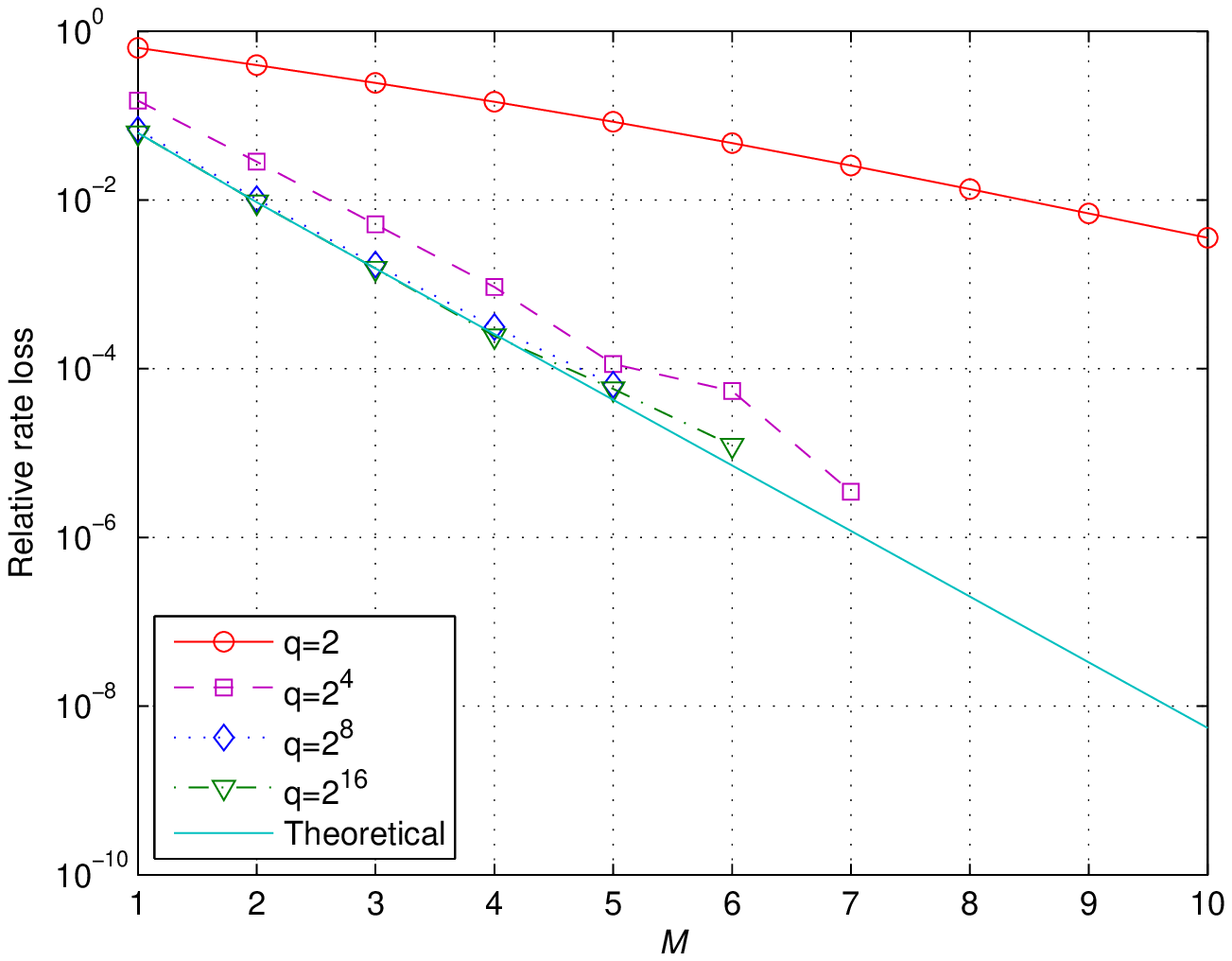}
\caption{Relative rate loss with respect to min-cut rate as a function
of memory size $M$ for $\delta=0.4$, $\varepsilon=0.1$, and various
coding field sizes $q$.}
\label{fig:payamplot4}
\end{figure}

To quantify the effect of operations over a finite field, we simulated
the evolution of this Markov chain for two combinations of $\delta$ and
$\epsilon$ values that were also considered in
Section~\ref{sec:finite-memory_isolation} (see
Figures~\ref{fig:payamplot3} and~\ref{fig:payamplot4}).  The effective
rate is considered to be $R_e:=y_N/N$, where $N$ is the number of packet
transmissions at $A$, and as before, $y_N$ is the number of innovative
packets received at node 3 by time $N$.  We simulated this process for
$N=10^9$ packets.  For different field sizes, we plot the relative rate
loss with respect to the min-cut rate---i.e.,\ $1-R_e/R^*$---as a
function of the memory size.  Also plotted is the theoretical result
from~(\ref{eqn:2860}).  

By comparing Figures~\ref{fig:payamplot3} and~\ref{fig:payamplot4} with
Figures~\ref{fig:lossplot-p0_8-eps0_1}
and~\ref{fig:lossplot-p0_6-eps0_1}, respectively, we see the advantage
that comes from explicitly recognizing that the coding takes place in
the context of a larger coded packet network.
The loss rate in the latter plots essentially equates to
the factor $1-R_e/R^*$ in the former.  Thus, in the limit of infinite
$q$, we compare the probability of loss $p_l$ upper bounded by equation
(\ref{eqn:2850}) and the expression for $1-R/R^*$ given by equation
(\ref{eqn:2860}).  We note that, in both cases, the decay as $M
\rightarrow \infty$ is as $\varrho^M$.  Moreover, it follows from our
discussion that $1-R/R^*$ must be a lower bound for $p_l$, hence $p_l$
itself decays as $\varrho^M$ as $M \rightarrow \infty$.

\begin{subappendices}
\section{Appendix: Formal arguments for main result}
\label{app:formal}

In this appendix, we given formal arguments for Theorem~\ref{thm:2100}.  
Sections~\ref{app:formal_two-link_tandem},
\ref{app:formal_l-link_tandem}, and~\ref{app:formal_general_unicast}
give formal arguments for three special cases of Theorem~\ref{thm:2100}:
the two-link tandem network, the $L$-link tandem network, and general
unicast connections, respectively.

\subsection{Two-link tandem network}
\label{app:formal_two-link_tandem}

All packets received by node 2, namely $v_1,
v_2, \ldots, v_N$, are considered innovative.  We associate with node 2
the set of vectors $U$, which varies with time and is initially empty,
i.e.,\  $U(0) := \emptyset$.  If packet $\packet$ is received by node 2 at time
$\tau$, then its auxiliary encoding vector $\beta$ is added to $U$ at
time $\tau$,
i.e.,\  $U(\tau^+) := \{\beta\} \cup U(\tau)$.

We associate with node 3 the set of vectors $W$, which again varies with
time and is initially empty.  Suppose packet $\packet$, with auxiliary
encoding vector $\beta$, is received by node 3
at time $\tau$.  
Let $\mu$ be a positive integer, which we call the \emph{innovation
order}.
Then we say $\packet$ is innovative 
if $\beta \notin \mathrm{span}(W(\tau))$ and $|U(\tau)| >
|W(\tau)| + \mu - 1$.  If $\packet$ is innovative,
then $\beta$ is added to $W$ at time $\tau$.\footnote{This definition of
innovative differs from merely being informative, which is the sense in
which innovative is used in Section~\ref{sec:finite-memory} and in
\cite{cwj03}.  Indeed, a packet can be informative, in the sense that in
gives a node some new, as yet unknown, 
information about $v_1, v_2, \ldots, v_N$ (or about $w_1, w_2, \ldots,
w_K$), and not satisfy this definition of innovative.  In this appendix,
we have defined innovative so that innovative packets are informative
(with respect to other innovative packets at the node), but not
necessarily conversely.  This allows us to bound, or dominate, the
behavior of the coding scheme, though we cannot describe it exactly.}

The definition of innovative is designed to satisfy two properties:
First, we require
that $W(\Delta)$, the set of vectors in $W$ when the scheme
terminates, is linearly independent.
Second, we require that, when a packet is received by node 3 and 
$|U(\tau)| > |W(\tau)| + \invord - 1$, 
it is innovative with high probability.  The
innovation order $\invord$ is an arbitrary factor that ensures that the
latter property is satisfied.

Suppose $|U(\tau)| > |W(\tau)| + \mu - 1$.
Since $\packet$ is a random linear combination of vectors in $U(\tau)$, it
follows that $\packet$ is innovative with some non-trivial
probability.  More precisely, because $\beta$ is uniformly-distributed
over $q^{|U(\tau)|}$ possibilities, of which at least 
$q^{|U(\tau)|} - q^{|W(\tau)|}$ are not in $\mathrm{span}(W(\tau))$, it
follows that 
\[
\Pr(\beta \notin \mathrm{span}(W(\tau)))
\ge \frac{q^{|U(\tau)|}-q^{|W(\tau)|}}{q^{|U(\tau)|}}
= 1 - q^{|W(\tau)|-|U(\tau)|}
\ge 1 - q^{-\mu}.
\]
Hence $\packet$ is innovative with probability at least
$1-q^{-\mu}$.  Since we can always discard innovative packets, we
assume that the event occurs 
with probability exactly
$1-q^{-\mu}$.  If instead $|U(\tau)| \le |W(\tau)| + \mu - 1$, then we
see that $\packet$ cannot be innovative, and this remains true at
least until another arrival occurs at node 2.  Therefore, for an
innovation order of $\mu$, the
propagation of innovative packets through node 2 is
described by the propagation of jobs through a single-server queueing
station with queue size $(|U(\tau)| - |W(\tau)| - \mu + 1)^+$.

The queueing station is serviced with probability $1-q^{-\mu}$ whenever
the queue is non-empty and a received packet arrives on arc $(2,3)$.  We can
equivalently consider ``candidate'' packets that arrive with probability
$1-q^{-\mu}$ whenever a received packet arrives on arc $(2,3)$ and say that
the queueing station is serviced whenever the queue is non-empty and a
candidate packet arrives on arc $(2,3)$. 
We consider all packets received on arc $(1,2)$ to be candidate packets.

The system we wish to analyze, therefore, is the following simple
queueing system:  Jobs arrive at node 2 according to the arrival of
received packets on arc $(1,2)$ and, with the exception of the first
$\mu - 1$ jobs, enter node 2's queue.  
The jobs in node 2's queue are
serviced by the arrival of candidate packets on arc $(2,3)$ and exit 
after being serviced.  The number of jobs exiting is
a lower bound on the number of packets 
with linearly-independent auxiliary encoding vectors
received by node 3. 

We analyze the queueing system of interest using the fluid
approximation for discrete-flow networks (see, e.g.,\  \cite{chm91,
chy01}).
We do not explicitly account for the fact that the first $\mu-1$
jobs arriving at node 2 do not enter its queue because 
this fact has no effect on job throughput.
Let $B_1$, $B$, and $C$ be the counting processes for the arrival of
received packets on arc $(1,2)$, of innovative packets on
arc $(2,3)$, and of candidate packets on arc $(2,3)$,
respectively.
Let $Q(\tau)$ be the number of jobs queued for service at node 2 at
time $\tau$.
Hence $Q = B_1 - B$.  Let $X := B_1 - C$ and $Y := C - B$.  Then
\begin{equation}
Q = X + Y.
\label{eqn:2100}
\end{equation}
Moreover, we have
\begin{gather}
Q(\tau) dY(\tau) = 0, \\
dY(\tau) \ge 0,
\end{gather}
and
\begin{equation}
Q(\tau) \ge 0
\end{equation}
for all $\tau \ge 0$, and
\begin{equation}
Y(0) = 0.
\label{eqn:2110}
\end{equation}

We observe now that 
equations (\ref{eqn:2100})--(\ref{eqn:2110}) give us
the conditions for a Skorohod problem (see,
e.g.,\  \cite[Section 7.2]{chy01}) and, by the oblique reflection
mapping theorem, there is a well-defined, 
Lipschitz-continuous mapping $\Phi$ such that $Q = \Phi(X)$.  

Let
\begin{gather*}
\bar{C}^{(K)}(\tau) := \frac{C(K\tau)}{K}, \\
\bar{X}^{(K)}(\tau) := \frac{X(K\tau)}{K},
\end{gather*}
and
\[
\bar{Q}^{(K)}(\tau) := \frac{Q(K\tau)}{K}.
\]

Recall that $A_{233}$ is the counting process for the arrival of received
packets on arc $(2,3)$.  Therefore, $C(\tau)$ is the sum of
$A_{233}(\tau)$ Bernoulli-distributed random variables with parameter
$1-q^{-\mu}$.  
Hence
\[
\begin{split}
\bar{C}(\tau) 
&:= 
\lim_{K \rightarrow \infty}
\bar{C}^{(K)}(\tau) \\
& = 
\lim_{K \rightarrow \infty}
(1-q^{-\mu})
\frac{A_{233}(K \tau)}{K}
\qquad \text{a.s.} \\
&=
(1-q^{-\mu})z_{233} \tau
\qquad \text{a.s.},
\end{split}
\]
where the last equality follows by the assumptions of the model.
Therefore
\[
\bar{X}(\tau) :=
\lim_{K \rightarrow \infty} \bar{X}^{(K)}(\tau)
= (z_{122} - (1-q^{-\mu})z_{233}) \tau
\qquad \text{a.s.}
\]
By the Lipschitz-continuity of $\Phi$, then, it follows that
$\bar{Q} := \lim_{K \rightarrow \infty} \bar{Q}^{(K)} 
= \Phi(\bar{X})$,  i.e.,\
$\bar{Q}$ is, almost surely, 
the unique $\bar{Q}$ that satisfies, for some
$\bar{Y}$,
\begin{gather}
\bar{Q}(\tau) =
(z_{122} - (1-q^{-\mu})z_{233})\tau + \bar{Y},
\label{eqn:2200} \\
\bar{Q}(\tau) d\bar{Y}(\tau) = 0, \\
d\bar{Y}(\tau) \ge 0, 
\end{gather}
and
\begin{equation}
\bar{Q}(\tau) \ge 0
\end{equation}
for all $\tau \ge 0$, and
\begin{equation}
\bar{Y}(0) = 0.
\label{eqn:2210}
\end{equation}

A pair $(\bar{Q}, \bar{Y})$ that satisfies
(\ref{eqn:2200})--(\ref{eqn:2210}) is 
\begin{equation}
\bar{Q}(\tau)
= (z_{122} - (1-q^{-\mu})z_{233})^+ \tau
\label{eqn:2220}
\end{equation}
and
\[
\bar{Y}(\tau)
= (z_{122} - (1-q^{-\mu})z_{233})^- \tau,
\]
where, for a real number $x$, $(x)^+ := \max(x, 0)$ and
$(x)^- := \max(-x, 0)$.
Hence $\bar{Q}$ is given by equation (\ref{eqn:2220}).

Recall that node 3 can recover the message packets with high probability
if it receives $\lfloor K(1+\varepsilon) \rfloor$ packets with
linearly-independent auxiliary encoding vectors and that the number of
jobs exiting the queueing system is a lower bound on the number of
packets with linearly-independent auxiliary encoding vectors received by
node 3.  Therefore, node 3 can recover the message packets with high
probability if $\lfloor K(1 + \varepsilon) \rfloor$ or more jobs exit
the queueing system.  Let $\nu$ be the number of jobs that have exited
the queueing system by time $\Delta$.  Then
\[
\nu = B_1(\Delta) - Q(\Delta).
\]
Take $K = \lceil (1-q^{-\mu}) \Delta R_c R / (1 + \varepsilon) \rceil$, 
where $0 < R_c < 1$.  
Then
\[
\begin{split}
\lim_{K \rightarrow \infty} 
\frac{\nu}{\lfloor K(1 + \varepsilon) \rfloor}
&= \lim_{K \rightarrow \infty} 
\frac{B_1(\Delta) - Q(\Delta)}{K (1 + \varepsilon)} \\
&= \frac{z_{122} - (z_{122} - (1 - q^{-\mu}) z_{233})^+}
{(1 - q^{-\mu})R_cR} \\
&= \frac{\min(z_{122}, (1-q^{-\mu}) z_{233})}
{(1 - q^{-\mu})R_cR} \\
&\ge
\frac{1}{R_c} \frac{\min(z_{122}, z_{233})}{R} > 1
\end{split} 
\]
provided that
\begin{equation}
R \le \min(z_{122}, z_{233}).
\label{eqn:2300}
\end{equation}
Hence, for all $R$ satisfying (\ref{eqn:2300}), $\nu \ge \lfloor K(1 +
\varepsilon) \rfloor$ with probability arbitrarily close to 1 for $K$
sufficiently large.  The rate achieved is
\[
\frac{K}{\Delta} 
\ge \frac{(1-q^{-\mu}) R_c}{1 + \varepsilon} R,
\]
which can be made arbitrarily close to $R$ by varying $\mu$, $R_c$, and
$\varepsilon$.

\subsection{$L$-link tandem network}
\label{app:formal_l-link_tandem}

For $i = 2, 3, \ldots, L+1$, we associate with node $i$ the set of
vectors $V_i$, which varies with time and is initially empty. 
We define $U := V_2$ and $W := V_{L+1}$.
As in the case of the two-link tandem, 
all packets received by node 2 are considered
innovative and, if packet $\packet$ is received by
node 2 at time $\tau$, then its auxiliary encoding vector $\beta$ is
added to $U$ at time $\tau$.  
For $i = 3, 4, \ldots, L+1$,
if packet
$\packet$, with auxiliary encoding vector $\beta$, is received by node $i$ at
time $\tau$, then we say $\packet$ is innovative if 
$\beta \notin \mathrm{span}(V_i(\tau))$ and 
$|V_{i-1}(\tau)| > |V_{i}(\tau)| + \mu - 1$.
If $\packet$ is innovative, then $\beta$ is added to
$V_i$ at time $\tau$.  

This definition of innovative is a straightforward extension of that
in Appendix~\ref{app:formal_two-link_tandem}.  
The first property remains the same:
we continue to require
that $W(\Delta)$ is a set of linearly-independent vectors.
We extend the second property so that, when
a packet is received by node $i$ for any $i = 3, 4, \ldots, L+1$ and 
$|V_{i-1}(\tau)| > |V_{i}(\tau)| + \invord - 1$, 
it is innovative with high probability.  

Take some $i \in \{3, 4, \ldots, L+1\}$.  Suppose that packet $\packet$, with
auxiliary encoding vector $\beta$, is received by node $i$ at time
$\tau$ and that $|V_{i-1}(\tau)| > |V_i(\tau)| + \invord - 1$.  Thus, the
auxiliary encoding vector $\beta$ is a random linear combination of
vectors in some set $V_0$ that contains $V_{i-1}(\tau)$.  Hence, because
$\beta$ is uniformly-distributed over $q^{|V_0|}$ possibilities, of
which at least $q^{|V_0|} - q^{|V_i(\tau)|}$ are not in
$\mathrm{span}(V_i(\tau))$, it follows that
\[
\Pr(\beta \notin \mathrm{span}(V_i(\tau))) 
\ge \frac{q^{|V_0|} - q^{|V_i(\tau)|}}{q^{|V_0|}}
= 1 - q^{|V_i(\tau)| - |V_0|}
\ge 1 - q^{|V_i(\tau)| - |V_{i-1}(\tau)|}
\ge 1 - q^{-\invord} .
\]
Therefore $\packet$ is innovative with probability at least $1 - q^{-\invord}$.  

Following the argument in Appendix~\ref{app:formal_two-link_tandem}, we
see, for all $i = 2, 3, \ldots, L$, that
the propagation of innovative packets through node $i$ 
is described by the propagation of
jobs through a single-server queueing station with queue size
$(|V_i(\tau)| - |V_{i+1}(\tau)| - \invord + 1)^+$ and that the queueing
station is serviced with probability $1 - q^{-\invord}$ whenever the queue
is non-empty and a received packet arrives on arc $(i, i+1)$.
We again consider candidate packets that 
arrive with probability $1 - q^{-\invord}$ whenever a received packet
arrives on arc $(i, i+1)$ and say that the queueing station is serviced
whenever the queue is non-empty and a candidate packet arrives on arc
$(i, i+1)$.

The system we wish to analyze in this case is therefore the following
simple queueing network:  Jobs arrive at node 2 according to the arrival
of received packets on arc $(1,2)$ and, with the exception of the first
$\mu - 1$ jobs, enter node 2's queue.  For $i = 2, 3, \ldots, L - 1$,
the jobs in node $i$'s queue are serviced by the arrival of candidate
packets on arc $(i,i+1)$ and, with the exception of the first $\mu - 1$
jobs, enter node $(i+1)$'s queue after being serviced.  The jobs in node
$L$'s queue are serviced by the arrival of candidate packets on arc $(L,
L+1)$ and exit after being serviced.  The number of jobs exiting is a
lower bound on the number of packets with linearly-independent auxiliary
encoding vectors received by node $L+1$.

We again analyze the queueing network of interest using the fluid
approximation for discrete-flow networks, and we again do not explicitly
account for the fact that the first $\mu-1$ jobs arriving at a queueing
node do not enter its queue.  Let $B_1$ be the counting process for the 
arrival of received
packets on arc $(1,2)$.  For $i = 2, 3, \ldots, L$, let 
$B_i$, and $C_i$ be the counting processes for the arrival of 
innovative packets
and candidate packets on arc $(i, i+1)$, respectively.
Let $Q_i(\tau)$ be the number of jobs queued for service at node $i$ at
time $\tau$.  Hence, for $i = 2, 3, \ldots, L$, 
$Q_i = B_{i-1} - B_i$.  Let $X_i := C_{i-1} - C_i$
and $Y_i := C_i - B_i$, where $C_1 := B_1$.
Then, we obtain a Skorohod problem with the following conditions:
For all $i = 2, 3, \ldots, L$,
\[
Q_i = X_i - Y_{i-1} + Y_i.
\]
For all $\tau \ge 0$ and $i = 2, 3, \ldots, L$,
\begin{gather*}
Q_i(\tau) dY_i(\tau) = 0, \\
dY_i(\tau) \ge 0,
\end{gather*}
and
\begin{equation*}
Q_i(\tau) \ge 0.
\end{equation*}
For all $i = 2, 3, \ldots, L$, 
\begin{equation*}
Y_i(0) = 0.
\end{equation*}

Let
\[
\bar{Q}_i^{(K)}(\tau) := \frac{Q_i(K\tau)}{K}
\]
and $\bar{Q}_i := \lim_{K \rightarrow \infty} \bar{Q}^{(K)}_i$
for $i = 2, 3, \ldots, L$.
Then the vector $\bar{Q}$ 
is, almost surely, the unique $\bar{Q}$ that satisfies,
for some $\bar{Y}$, 
\begin{gather}
\bar{Q}_i(\tau) =
\begin{cases}
(z_{122} - (1-q^{-\mu})z_{233}) \tau + \bar{Y}_2(\tau) & \text{if $i=2$}, \\
(1-q^{-\mu})(z_{(i-1)ii} - z_{i(i+1)(i+1)}) \tau + \bar{Y}_i(\tau)
- \bar{Y}_{i-1}(\tau) & \text{otherwise},
\end{cases}
\label{eqn:2400} \\
\bar{Q}_i(\tau) d\bar{Y}_i(\tau) = 0, \\
d\bar{Y}_i(\tau) \ge 0, 
\end{gather}
and
\begin{equation}
\bar{Q}_i(\tau) \ge 0
\end{equation}
for all $\tau \ge 0$ and $i = 2,3,\ldots, L$, and
\begin{equation}
\bar{Y}_i(0) = 0
\label{eqn:2410}
\end{equation}
for all $i = 2,3,\ldots, L$.

A pair $(\bar{Q}, \bar{Y})$ that satisfies
(\ref{eqn:2400})--(\ref{eqn:2410}) is 
\begin{equation}
\bar{Q}_i(\tau)
= (\min(z_{122}, \min_{2 \le j < i} \{
(1-q^{-\mu})z_{j(j+1)(j+1)}
\}) - (1-q^{-\mu}) z_{i(i+1)(i+1)})^+ \tau
\label{eqn:2420}
\end{equation}
and
\[
\bar{Y}_i(\tau)
= (\min(z_{122}, \min_{2 \le j < i} \{
(1-q^{-\mu})z_{j(j+1)(j+1)}
\}) - (1-q^{-\mu}) z_{i(i+1)(i+1)})^- \tau .
\]
Hence $\bar{Q}$ is given by equation (\ref{eqn:2420}).  

The number of jobs that have exited the queueing network by time
$\Delta$ is given by
\[
\nu = B_1(\Delta) - \sum_{i=2}^L Q_i(\Delta).
\]
Take $K = \lceil (1-q^{-\mu}) \Delta R_c R / (1 + \varepsilon) \rceil$, 
where $0 < R_c < 1$.  
Then
\begin{equation}
\begin{split}
\lim_{K \rightarrow \infty} 
\frac{\nu}{\lfloor K(1 + \varepsilon) \rfloor}
&= \lim_{K \rightarrow \infty} 
\frac{B_1(\Delta) - \sum_{i=2}^LQ(\Delta)}{K (1 + \varepsilon)} \\
&=
\frac{\min(z_{122}, \min_{2 \le i \le L}\{(1-q^{-\mu}) z_{i(i+1)(i+1)}\})}
{(1 - q^{-\mu})R_cR} \\
&\ge
\frac{1}{R_c} \frac{\min_{1 \le i \le L}\{z_{i(i+1)(i+1)}\}}{R} > 1
\end{split} 
\label{eqn:2490}
\end{equation}
provided that
\begin{equation}
R \le \min_{1 \le i \le L}\{z_{i(i+1)(i+1)}\}.
\label{eqn:2500}
\end{equation}
Hence, for all $R$ satisfying (\ref{eqn:2500}), $\nu \ge \lfloor K(1 +
\varepsilon) \rfloor$ with probability arbitrarily close to 1 for $K$
sufficiently large.  The rate can again be made arbitrarily close to $R$
by varying $\mu$, $R_c$, and $\varepsilon$.

\subsection{General unicast connection}
\label{app:formal_general_unicast}

Consider a single path $p_m$.  We write $p_m = \{i_1, i_2, \ldots,
i_{L_m}, i_{L_m+1}\}$, where $i_1 = s$ and $i_{L_m+1} = t$.  
For $l = 2, 3, \ldots, L_m+1$, we associate with node $i_l$ the set of
vectors $V^{(p_m)}_l$, which varies with time and is initially empty.
We define $U^{(p_m)} := V^{(p_m)}_2$ and $W^{(p_m)} :=
V^{(p_m)}_{L_m+1}$.

We note that the constraint (\ref{eqn:2600}) can also be written as
\[
\flow_{iJj} \le 
\sum_{\{L \subset J | j \in L\}} \alpha_{iJL}^{(j)} z_{iJL}
\]
for all $(i,J) \in \mathcal{A}$ and $j \in J$, where 
$\sum_{j \in L} \alpha_{iJL}^{(j)} = 1$ for all $(i,J) \in \mathcal{A}$ 
and $L \subset J$, and $\alpha_{iJL}^{(j)} \ge 0$ for all $(i,J) \in
\mathcal{A}$, $L \subset J$, and $j \in L$.
Suppose packet $\packet$, with auxiliary encoding vector $\beta$, is placed on
hyperarc $(i_1, J)$ and received by $K \subset J$, where $K \ni i_2$,
at time $\tau$.  We associate with $\packet$ the independent random variable
$P_\packet$, which takes the value $m$ with probability $R_m
\alpha_{i_1JK}^{(i_2)} / \sum_{\{L \subset J|i_2 \in L\}}
\alpha_{i_1JL}^{(i_2)} z_{iJL}$.  If $P_\packet = m$, then we say $\packet$ is
innovative on path $p_m$, and $\beta$ is added to $U^{(p_m)}$ at time
$\tau$.

Take $l=2,3,\ldots, L_m$.  Now suppose packet $\packet$, with auxiliary
encoding vector $\beta$, is placed on hyperarc $(i_l, J)$ and received
by $K \subset J$, where $K \ni i_{l+1}$, at time $\tau$.  We associate
with $\packet$ the independent random variable $P_\packet$, which takes the value
$m$ with probability $R_m \alpha_{i_lJK}^{(i_{l+1})} / \sum_{\{L \subset
J|i_{l+1} \in L\}} \alpha_{i_lJL}^{(i_{l+1})} z_{iJL}$.  We say $\packet$ is
innovative on path $p_m$ if $P_\packet = m$, $\beta \notin
\mathrm{span}(\cup_{n=1}^{m-1} W^{(p_n)}(\Delta) \cup
V_{l+1}^{(p_m)}(\tau) \cup \cup_{n=m+1}^M U^{(p_n)}(\Delta))$, and
$|V_{l}^{(p_m)}(\tau)| > |V_{l+1}^{(p_m)}(\tau)| + \mu - 1$.

This definition of innovative is somewhat more complicated than that in
Appendices~\ref{app:formal_two-link_tandem}
and~\ref{app:formal_l-link_tandem} because we now have $M$ paths that we
wish to analyze separately.  We have again designed the definition to
satisfy two properties:  First, we require that $\cup_{m=1}^M
W^{(p_m)}(\Delta)$ is linearly-independent.  This is easily verified:
Vectors are added to $W^{(p_1)}(\tau)$ only if they are linearly
independent of existing ones; vectors are added to $W^{(p_2)}(\tau)$
only if they are linearly independent of existing ones and ones in
$W^{(p_1)}(\Delta)$; and so on.  Second, we require that, when a packet
is received by node $i_l$, $P_\packet = m$, and $|V_{l-1}^{(p_m)}(\tau)| >
|V_l^{(p_m)}(\tau)| + \invord - 1$, it is innovative on path $p_m$ with
high probability.  

Take $l \in \{3, 4, \ldots, L_m+1\}$.  Suppose that packet $\packet$, with
auxiliary encoding vector $\beta$, is received by node $i_l$ at time
$\tau$, that $P_\packet = m$, and that $|V_{l-1}^{(p_m)}(\tau)| >
|V_l^{(p_m)}(\tau)| + \invord - 1$.  Thus, the auxiliary encoding vector
$\beta$ is a random linear combination of vectors in some set $V_0$ that
contains $V_{l-1}^{(p_m)}(\tau)$.  Hence $\beta$ is
uniformly-distributed over $q^{|V_0|}$ possibilities, of which at least
$q^{|V_0|} - q^{d}$ are not in $\mathrm{span}(V_l^{(p_m)}(\tau) \cup
\tilde{V}_{\setminus m}$), where $d := {\mathrm{dim}(\mathrm{span}(V_0)
\cap \mathrm{span}(V_l^{(p_m)}(\tau) \cup \tilde{V}_{\setminus m}))}$.
Note that $V_{l-1}^{(p_m)}(\tau)
\cup \tilde{V}_{\setminus m}$ forms a linearly-independent set, so
\[
\begin{split}
d - |V_0| &\le 
\mathrm{dim}(\mathrm{span}(V_{l-1}^{(p_m)}(\tau))
\cap \mathrm{span}(V_l^{(p_m)}(\tau) \cup \tilde{V}_{\setminus m}))
- |V_{l-1}^{(p_m)}(\tau)| \\
&=
\mathrm{dim}(\mathrm{span}(V_{l-1}^{(p_m)}(\tau))
\cap \mathrm{span}(V_l^{(p_m)}(\tau)))
- |V_{l-1}^{(p_m)}(\tau)| \\
&\le |V_l^{(p_m)}(\tau)|
- |V_{l-1}^{(p_m)}(\tau)|
\le -\invord.
\end{split}
\]
Therefore, it follows that
\[
\Pr(\beta \notin \mathrm{span}(V_l^{(p_m)}(\tau) \cup
\tilde{V}_{\setminus m}))
\ge \frac{q^{|V_0|} - q^d}{q^{|V_0|}}
= 1 - q^{d - |V_0|} 
\ge 1 - q^{-\invord}.
\]

We see then that, if we consider only those packets such that $P_\packet = m$,
the conditions that govern the propagation of innovative packets are
exactly those of an $L_m$-link tandem network, which we dealt with in
Appendix~\ref{app:formal_l-link_tandem}.  By recalling the distribution
of $P_\packet$, it follows that the propagation of innovative packets along
path $p_m$ behaves like an $L_m$-link tandem network with average
arrival rate $R_m$ on every link.  Since we have assumed nothing special
about $m$, this statement applies for all $m = 1, 2, \ldots, M$.

Take $K = \lceil (1-q^{-\mu}) \Delta R_c R / (1 + \varepsilon) \rceil$, 
where $0 < R_c < 1$.  
Then, by equation (\ref{eqn:2490}),
\[
\lim_{K \rightarrow \infty} 
\frac{|W^{(p_m)}(\Delta)|}{\lfloor K(1+\varepsilon) \rfloor}
> \frac{R_m}{R}.
\]
Hence
\[
\lim_{K \rightarrow \infty} 
\frac{|\cup_{m=1}^M W^{(p_m)}(\Delta)|}{\lfloor K(1+\varepsilon) \rfloor}
= \sum_{m=1}^M 
\frac{|W^{(p_m)}(\Delta)|}{\lfloor K(1+\varepsilon) \rfloor}
> \sum_{m=1}^M \frac{R_m}{R} = 1.
\]
As before, the rate can be made arbitrarily close to $R$ by varying
$\mu$, $R_c$, and $\varepsilon$.

\end{subappendices}

%% file: two_links.pdftex_t
\begin{picture}(0,0)%
\includegraphics{two_links.eps}%
\end{picture}%
\setlength{\unitlength}{3947sp}%
\begingroup\makeatletter\ifx\SetFigFont\undefined%
\gdef\SetFigFont#1#2#3#4#5{%
  \reset@font\fontsize{#1}{#2pt}%
  \fontfamily{#3}\fontseries{#4}\fontshape{#5}%
  \selectfont}%
\fi\endgroup%
\begin{picture}(3916,614)(1643,-1868)
\put(3601,-1636){\makebox(0,0)[b]{\smash{{\SetFigFont{12}{14.4}{\familydefault}{\mddefault}{\updefault}{\color[rgb]{0,0,0}2}%
}}}}
\put(1951,-1636){\makebox(0,0)[b]{\smash{{\SetFigFont{12}{14.4}{\familydefault}{\mddefault}{\updefault}{\color[rgb]{0,0,0}1}%
}}}}
\put(5251,-1636){\makebox(0,0)[b]{\smash{{\SetFigFont{12}{14.4}{\familydefault}{\mddefault}{\updefault}{\color[rgb]{0,0,0}3}%
}}}}
\end{picture}%

%% file: two_pipes.pdftex_t
\begin{picture}(0,0)%
\includegraphics{two_pipes.eps}%
\end{picture}%
\setlength{\unitlength}{3947sp}%
\begingroup\makeatletter\ifx\SetFigFont\undefined%
\gdef\SetFigFont#1#2#3#4#5{%
  \reset@font\fontsize{#1}{#2pt}%
  \fontfamily{#3}\fontseries{#4}\fontshape{#5}%
  \selectfont}%
\fi\endgroup%
\begin{picture}(3324,1149)(2314,-2473)
\put(3676,-1861){\makebox(0,0)[b]{\smash{{\SetFigFont{12}{14.4}{\familydefault}{\mddefault}{\updefault}{\color[rgb]{0,0,0}2}%
}}}}
\put(5326,-2311){\makebox(0,0)[b]{\smash{{\SetFigFont{12}{14.4}{\familydefault}{\mddefault}{\updefault}{\color[rgb]{0,0,0}3}%
}}}}
\put(2851,-1711){\makebox(0,0)[b]{\smash{{\SetFigFont{12}{14.4}{\familydefault}{\mddefault}{\updefault}{\color[rgb]{0,0,0}$z_{122}$}%
}}}}
\put(4501,-2161){\makebox(0,0)[b]{\smash{{\SetFigFont{12}{14.4}{\familydefault}{\mddefault}{\updefault}{\color[rgb]{0,0,0}$z_{233}$}%
}}}}
\end{picture}%

%% file: l_links.pdftex_t
\begin{picture}(0,0)%
\includegraphics{l_links.eps}%
\end{picture}%
\setlength{\unitlength}{3947sp}%
\begingroup\makeatletter\ifx\SetFigFont\undefined%
\gdef\SetFigFont#1#2#3#4#5{%
  \reset@font\fontsize{#1}{#2pt}%
  \fontfamily{#3}\fontseries{#4}\fontshape{#5}%
  \selectfont}%
\fi\endgroup%
\begin{picture}(5416,614)(1343,-1868)
\put(6451,-1636){\makebox(0,0)[b]{\smash{{\SetFigFont{12}{14.4}{\familydefault}{\mddefault}{\updefault}{\color[rgb]{0,0,0}$L+1$}%
}}}}
\put(3301,-1636){\makebox(0,0)[b]{\smash{{\SetFigFont{12}{14.4}{\familydefault}{\mddefault}{\updefault}{\color[rgb]{0,0,0}2}%
}}}}
\put(1651,-1636){\makebox(0,0)[b]{\smash{{\SetFigFont{12}{14.4}{\familydefault}{\mddefault}{\updefault}{\color[rgb]{0,0,0}1}%
}}}}
\put(4876,-1636){\makebox(0,0)[b]{\smash{{\SetFigFont{12}{14.4}{\familydefault}{\mddefault}{\updefault}{\color[rgb]{0,0,0}$\cdots$}%
}}}}
\end{picture}%

%% file: l_pipes.pdftex_t
\begin{picture}(0,0)%
\includegraphics{l_pipes.eps}%
\end{picture}%
\setlength{\unitlength}{3947sp}%
\begingroup\makeatletter\ifx\SetFigFont\undefined%
\gdef\SetFigFont#1#2#3#4#5{%
  \reset@font\fontsize{#1}{#2pt}%
  \fontfamily{#3}\fontseries{#4}\fontshape{#5}%
  \selectfont}%
\fi\endgroup%
\begin{picture}(4824,1299)(1939,-2473)
\put(6451,-2311){\makebox(0,0)[b]{\smash{{\SetFigFont{12}{14.4}{\familydefault}{\mddefault}{\updefault}{\color[rgb]{0,0,0}$L+1$}%
}}}}
\put(3301,-1711){\makebox(0,0)[b]{\smash{{\SetFigFont{12}{14.4}{\familydefault}{\mddefault}{\updefault}{\color[rgb]{0,0,0}2}%
}}}}
\put(5476,-2161){\makebox(0,0)[b]{\smash{{\SetFigFont{12}{14.4}{\familydefault}{\mddefault}{\updefault}{\color[rgb]{0,0,0}$z_{L(L+1)(L+1)}$}%
}}}}
\put(4126,-2011){\makebox(0,0)[b]{\smash{{\SetFigFont{12}{14.4}{\familydefault}{\mddefault}{\updefault}{\color[rgb]{0,0,0}$z_{233}$}%
}}}}
\put(2476,-1561){\makebox(0,0)[b]{\smash{{\SetFigFont{12}{14.4}{\familydefault}{\mddefault}{\updefault}{\color[rgb]{0,0,0}$z_{122}$}%
}}}}
\put(4876,-1936){\makebox(0,0)[b]{\smash{{\SetFigFont{12}{14.4}{\familydefault}{\mddefault}{\updefault}{\color[rgb]{0,0,0}$\ddots$}%
}}}}
\end{picture}%

%% file: markov_chain.pdftex_t
\begin{picture}(0,0)%
\includegraphics{markov_chain.eps}%
\end{picture}%
\setlength{\unitlength}{3947sp}%
\begingroup\makeatletter\ifx\SetFigFont\undefined%
\gdef\SetFigFont#1#2#3#4#5{%
  \reset@font\fontsize{#1}{#2pt}%
  \fontfamily{#3}\fontseries{#4}\fontshape{#5}%
  \selectfont}%
\fi\endgroup%
\begin{picture}(5708,1330)(1193,-2135)
\put(1501,-1636){\makebox(0,0)[b]{\smash{{\SetFigFont{12}{14.4}{\familydefault}{\mddefault}{\updefault}{\color[rgb]{0,0,0}0}%
}}}}
\put(2851,-1636){\makebox(0,0)[b]{\smash{{\SetFigFont{12}{14.4}{\familydefault}{\mddefault}{\updefault}{\color[rgb]{0,0,0}1}%
}}}}
\put(4201,-1636){\makebox(0,0)[b]{\smash{{\SetFigFont{12}{14.4}{\familydefault}{\mddefault}{\updefault}{\color[rgb]{0,0,0}2}%
}}}}
\put(5551,-1636){\makebox(0,0)[b]{\smash{{\SetFigFont{12}{14.4}{\familydefault}{\mddefault}{\updefault}{\color[rgb]{0,0,0}3}%
}}}}
\put(1501,-961){\makebox(0,0)[b]{\smash{{\SetFigFont{12}{14.4}{\familydefault}{\mddefault}{\updefault}{\color[rgb]{0,0,0}$\beta+\gamma$}%
}}}}
\put(2851,-961){\makebox(0,0)[b]{\smash{{\SetFigFont{12}{14.4}{\familydefault}{\mddefault}{\updefault}{\color[rgb]{0,0,0}$\gamma$}%
}}}}
\put(3526,-2086){\makebox(0,0)[b]{\smash{{\SetFigFont{12}{14.4}{\familydefault}{\mddefault}{\updefault}{\color[rgb]{0,0,0}$\beta$}%
}}}}
\put(3526,-1186){\makebox(0,0)[b]{\smash{{\SetFigFont{12}{14.4}{\familydefault}{\mddefault}{\updefault}{\color[rgb]{0,0,0}$\alpha$}%
}}}}
\put(5551,-961){\makebox(0,0)[b]{\smash{{\SetFigFont{12}{14.4}{\familydefault}{\mddefault}{\updefault}{\color[rgb]{0,0,0}$\gamma$}%
}}}}
\put(4201,-961){\makebox(0,0)[b]{\smash{{\SetFigFont{12}{14.4}{\familydefault}{\mddefault}{\updefault}{\color[rgb]{0,0,0}$\gamma$}%
}}}}
\put(4876,-2086){\makebox(0,0)[b]{\smash{{\SetFigFont{12}{14.4}{\familydefault}{\mddefault}{\updefault}{\color[rgb]{0,0,0}$\beta$}%
}}}}
\put(4876,-1186){\makebox(0,0)[b]{\smash{{\SetFigFont{12}{14.4}{\familydefault}{\mddefault}{\updefault}{\color[rgb]{0,0,0}$\alpha$}%
}}}}
\put(6226,-1186){\makebox(0,0)[b]{\smash{{\SetFigFont{12}{14.4}{\familydefault}{\mddefault}{\updefault}{\color[rgb]{0,0,0}$\alpha$}%
}}}}
\put(6226,-2086){\makebox(0,0)[b]{\smash{{\SetFigFont{12}{14.4}{\familydefault}{\mddefault}{\updefault}{\color[rgb]{0,0,0}$\beta$}%
}}}}
\put(2176,-2086){\makebox(0,0)[b]{\smash{{\SetFigFont{12}{14.4}{\familydefault}{\mddefault}{\updefault}{\color[rgb]{0,0,0}$\beta$}%
}}}}
\put(2176,-1186){\makebox(0,0)[b]{\smash{{\SetFigFont{12}{14.4}{\familydefault}{\mddefault}{\updefault}{\color[rgb]{0,0,0}$\alpha$}%
}}}}
\put(6901,-1636){\makebox(0,0)[b]{\smash{{\SetFigFont{12}{14.4}{\familydefault}{\mddefault}{\updefault}{\color[rgb]{0,0,0}$\cdots$}%
}}}}
\end{picture}%

%% file: markov_chain3.pdftex_t
\begin{picture}(0,0)%
\includegraphics{markov_chain3.eps}%
\end{picture}%
\setlength{\unitlength}{3947sp}%
\begingroup\makeatletter\ifx\SetFigFont\undefined%
\gdef\SetFigFont#1#2#3#4#5{%
  \reset@font\fontsize{#1}{#2pt}%
  \fontfamily{#3}\fontseries{#4}\fontshape{#5}%
  \selectfont}%
\fi\endgroup%
\begin{picture}(6016,1339)(893,-2144)
\put(2551,-1636){\makebox(0,0)[b]{\smash{{\SetFigFont{12}{14.4}{\familydefault}{\mddefault}{\updefault}{\color[rgb]{0,0,0}1}%
}}}}
\put(5251,-1636){\makebox(0,0)[b]{\smash{{\SetFigFont{12}{14.4}{\familydefault}{\mddefault}{\updefault}{\color[rgb]{0,0,0}$M-1$}%
}}}}
\put(6601,-1636){\makebox(0,0)[b]{\smash{{\SetFigFont{12}{14.4}{\familydefault}{\mddefault}{\updefault}{\color[rgb]{0,0,0}$M$}%
}}}}
\put(1201,-1636){\makebox(0,0)[b]{\smash{{\SetFigFont{12}{14.4}{\familydefault}{\mddefault}{\updefault}{\color[rgb]{0,0,0}0}%
}}}}
\put(2551,-961){\makebox(0,0)[b]{\smash{{\SetFigFont{12}{14.4}{\familydefault}{\mddefault}{\updefault}{\color[rgb]{0,0,0}$\gamma$}%
}}}}
\put(3901,-1636){\makebox(0,0)[b]{\smash{{\SetFigFont{12}{14.4}{\familydefault}{\mddefault}{\updefault}{\color[rgb]{0,0,0}$\cdots$}%
}}}}
\put(3226,-1186){\makebox(0,0)[b]{\smash{{\SetFigFont{12}{14.4}{\familydefault}{\mddefault}{\updefault}{\color[rgb]{0,0,0}$\alpha$}%
}}}}
\put(5926,-1186){\makebox(0,0)[b]{\smash{{\SetFigFont{12}{14.4}{\familydefault}{\mddefault}{\updefault}{\color[rgb]{0,0,0}$\alpha$}%
}}}}
\put(4576,-2086){\makebox(0,0)[b]{\smash{{\SetFigFont{12}{14.4}{\familydefault}{\mddefault}{\updefault}{\color[rgb]{0,0,0}$\beta$}%
}}}}
\put(5926,-2086){\makebox(0,0)[b]{\smash{{\SetFigFont{12}{14.4}{\familydefault}{\mddefault}{\updefault}{\color[rgb]{0,0,0}$1-\varepsilon$}%
}}}}
\put(4576,-1186){\makebox(0,0)[b]{\smash{{\SetFigFont{12}{14.4}{\familydefault}{\mddefault}{\updefault}{\color[rgb]{0,0,0}$\alpha$}%
}}}}
\put(5251,-961){\makebox(0,0)[b]{\smash{{\SetFigFont{12}{14.4}{\familydefault}{\mddefault}{\updefault}{\color[rgb]{0,0,0}$\gamma$}%
}}}}
\put(6601,-961){\makebox(0,0)[b]{\smash{{\SetFigFont{12}{14.4}{\familydefault}{\mddefault}{\updefault}{\color[rgb]{0,0,0}$\varepsilon$}%
}}}}
\put(3226,-2086){\makebox(0,0)[b]{\smash{{\SetFigFont{12}{14.4}{\familydefault}{\mddefault}{\updefault}{\color[rgb]{0,0,0}$\beta$}%
}}}}
\put(1876,-1186){\makebox(0,0)[b]{\smash{{\SetFigFont{12}{14.4}{\familydefault}{\mddefault}{\updefault}{\color[rgb]{0,0,0}$\alpha$}%
}}}}
\put(1876,-2086){\makebox(0,0)[b]{\smash{{\SetFigFont{12}{14.4}{\familydefault}{\mddefault}{\updefault}{\color[rgb]{0,0,0}$\beta$}%
}}}}
\put(1201,-961){\makebox(0,0)[b]{\smash{{\SetFigFont{12}{14.4}{\familydefault}{\mddefault}{\updefault}{\color[rgb]{0,0,0}$\beta+\gamma$}%
}}}}
\end{picture}%

%% file: markov_chain_2d.pdftex_t
\begin{picture}(0,0)%
\includegraphics{markov_chain_2d.eps}%
\end{picture}%
\setlength{\unitlength}{3947sp}%
\begingroup\makeatletter\ifx\SetFigFont\undefined%
\gdef\SetFigFont#1#2#3#4#5{%
  \reset@font\fontsize{#1}{#2pt}%
  \fontfamily{#3}\fontseries{#4}\fontshape{#5}%
  \selectfont}%
\fi\endgroup%
\begin{picture}(6205,3334)(1004,-4385)
\put(5551,-4336){\makebox(0,0)[b]{\smash{{\SetFigFont{12}{14.4}{\familydefault}{\mddefault}{\updefault}{\color[rgb]{0,0,0}$\vdots$}%
}}}}
\put(2326,-2161){\makebox(0,0)[rb]{\smash{{\SetFigFont{12}{14.4}{\familydefault}{\mddefault}{\updefault}{\color[rgb]{0,0,0}$\beta$}%
}}}}
\put(2326,-3511){\makebox(0,0)[rb]{\smash{{\SetFigFont{12}{14.4}{\familydefault}{\mddefault}{\updefault}{\color[rgb]{0,0,0}$\beta$}%
}}}}
\put(3676,-2161){\makebox(0,0)[rb]{\smash{{\SetFigFont{12}{14.4}{\familydefault}{\mddefault}{\updefault}{\color[rgb]{0,0,0}$\beta$}%
}}}}
\put(3676,-3511){\makebox(0,0)[rb]{\smash{{\SetFigFont{12}{14.4}{\familydefault}{\mddefault}{\updefault}{\color[rgb]{0,0,0}$\beta$}%
}}}}
\put(1276,-3661){\makebox(0,0)[lb]{\smash{{\SetFigFont{12}{14.4}{\familydefault}{\mddefault}{\updefault}{\color[rgb]{0,0,0}$\zeta$}%
}}}}
\put(1276,-2311){\makebox(0,0)[lb]{\smash{{\SetFigFont{12}{14.4}{\familydefault}{\mddefault}{\updefault}{\color[rgb]{0,0,0}$\zeta$}%
}}}}
\put(2626,-3661){\makebox(0,0)[lb]{\smash{{\SetFigFont{12}{14.4}{\familydefault}{\mddefault}{\updefault}{\color[rgb]{0,0,0}$\zeta$}%
}}}}
\put(2626,-2311){\makebox(0,0)[lb]{\smash{{\SetFigFont{12}{14.4}{\familydefault}{\mddefault}{\updefault}{\color[rgb]{0,0,0}$\zeta$}%
}}}}
\put(2851,-4336){\makebox(0,0)[b]{\smash{{\SetFigFont{12}{14.4}{\familydefault}{\mddefault}{\updefault}{\color[rgb]{0,0,0}$\vdots$}%
}}}}
\put(1501,-4336){\makebox(0,0)[b]{\smash{{\SetFigFont{12}{14.4}{\familydefault}{\mddefault}{\updefault}{\color[rgb]{0,0,0}$\vdots$}%
}}}}
\put(6376,-2161){\makebox(0,0)[rb]{\smash{{\SetFigFont{12}{14.4}{\familydefault}{\mddefault}{\updefault}{\color[rgb]{0,0,0}$\beta+\zeta$}%
}}}}
\put(6376,-3511){\makebox(0,0)[rb]{\smash{{\SetFigFont{12}{14.4}{\familydefault}{\mddefault}{\updefault}{\color[rgb]{0,0,0}$\beta+\zeta$}%
}}}}
\put(5026,-3511){\makebox(0,0)[rb]{\smash{{\SetFigFont{12}{14.4}{\familydefault}{\mddefault}{\updefault}{\color[rgb]{0,0,0}$\beta$}%
}}}}
\put(5026,-2161){\makebox(0,0)[rb]{\smash{{\SetFigFont{12}{14.4}{\familydefault}{\mddefault}{\updefault}{\color[rgb]{0,0,0}$\beta$}%
}}}}
\put(5326,-3661){\makebox(0,0)[lb]{\smash{{\SetFigFont{12}{14.4}{\familydefault}{\mddefault}{\updefault}{\color[rgb]{0,0,0}$\zeta$}%
}}}}
\put(5326,-2311){\makebox(0,0)[lb]{\smash{{\SetFigFont{12}{14.4}{\familydefault}{\mddefault}{\updefault}{\color[rgb]{0,0,0}$\zeta$}%
}}}}
\put(5551,-1186){\makebox(0,0)[b]{\smash{{\SetFigFont{12}{14.4}{\familydefault}{\mddefault}{\updefault}{\color[rgb]{0,0,0}$M-1$}%
}}}}
\put(1051,-2986){\makebox(0,0)[b]{\smash{{\SetFigFont{12}{14.4}{\familydefault}{\mddefault}{\updefault}{\color[rgb]{0,0,0}1}%
}}}}
\put(2851,-1186){\makebox(0,0)[b]{\smash{{\SetFigFont{12}{14.4}{\familydefault}{\mddefault}{\updefault}{\color[rgb]{0,0,0}1}%
}}}}
\put(1051,-1186){\makebox(0,0)[b]{\smash{{\SetFigFont{12}{14.4}{\familydefault}{\mddefault}{\updefault}{\color[rgb]{0,0,0}0}%
}}}}
\put(6901,-1186){\makebox(0,0)[b]{\smash{{\SetFigFont{12}{14.4}{\familydefault}{\mddefault}{\updefault}{\color[rgb]{0,0,0}$M$}%
}}}}
\put(2176,-1486){\makebox(0,0)[b]{\smash{{\SetFigFont{12}{14.4}{\familydefault}{\mddefault}{\updefault}{\color[rgb]{0,0,0}$\alpha$}%
}}}}
\put(4201,-2986){\makebox(0,0)[b]{\smash{{\SetFigFont{12}{14.4}{\familydefault}{\mddefault}{\updefault}{\color[rgb]{0,0,0}$\cdots$}%
}}}}
\put(4201,-1636){\makebox(0,0)[b]{\smash{{\SetFigFont{12}{14.4}{\familydefault}{\mddefault}{\updefault}{\color[rgb]{0,0,0}$\cdots$}%
}}}}
\put(4876,-1486){\makebox(0,0)[b]{\smash{{\SetFigFont{12}{14.4}{\familydefault}{\mddefault}{\updefault}{\color[rgb]{0,0,0}$\alpha$}%
}}}}
\put(6226,-1486){\makebox(0,0)[b]{\smash{{\SetFigFont{12}{14.4}{\familydefault}{\mddefault}{\updefault}{\color[rgb]{0,0,0}$\alpha$}%
}}}}
\put(6226,-2836){\makebox(0,0)[b]{\smash{{\SetFigFont{12}{14.4}{\familydefault}{\mddefault}{\updefault}{\color[rgb]{0,0,0}$\alpha$}%
}}}}
\put(3526,-2836){\makebox(0,0)[b]{\smash{{\SetFigFont{12}{14.4}{\familydefault}{\mddefault}{\updefault}{\color[rgb]{0,0,0}$\alpha$}%
}}}}
\put(3526,-1486){\makebox(0,0)[b]{\smash{{\SetFigFont{12}{14.4}{\familydefault}{\mddefault}{\updefault}{\color[rgb]{0,0,0}$\alpha$}%
}}}}
\put(2176,-2836){\makebox(0,0)[b]{\smash{{\SetFigFont{12}{14.4}{\familydefault}{\mddefault}{\updefault}{\color[rgb]{0,0,0}$\alpha$}%
}}}}
\put(4876,-2836){\makebox(0,0)[b]{\smash{{\SetFigFont{12}{14.4}{\familydefault}{\mddefault}{\updefault}{\color[rgb]{0,0,0}$\alpha$}%
}}}}
\end{picture}%

%% file: chap3.tex
\chapter{Subgraph Selection}
\label{chap:subgraph_selection}

\lettrine{W}{e now} turn to the subgraph selection part of the efficient
operation problem.  This is the problem of determining the coding
subgraph to use given that the network code is decided.  In our case, we
assume that the network code is given by the scheme examined in the
previous chapter.  Since this scheme achieves the capacity of a single
multicast connection in a given subgraph, in using it and determining
the coding subgraph independently, there is no loss of optimality in the
efficient operation problem provided that we are constrained to only
coding packets within a single connection.\footnote{This statement
assumes that no information is conveyed by the timing of packets.  In
general, the timing of packets can be used to convey information, but
the amount of information communicated by timing does not grow in the
size of packets, so the effect of such ``timing channels'' is negligible
for large packet sizes.}
Relaxing this constraint, and allowing coded packets
to be formed using packets from two or more connections, is known to
afford an improvement, but finding capacity-achieving codes is a very
difficult problem---one that, in fact, currently remains open with only
cumbersome bounds on the capability of coding \cite{syc03} and examples
that demonstrate the insufficiency of various classes of linear codes
\cite{dfz05, mek03, ral03, rii04}.  Constraining coding to packets
within a single connection is called superposition coding \cite{yeu95},
and there is evidence to suggest that it may be near-optimal
\cite{lil04}.
We therefore content ourselves with coding only within a single
connection, allowing us to separate network coding from subgraph
selection without loss of optimality.

We formulate the subgraph selection problem in
Section~\ref{sec:problem_formulation}.  The problem we describe is rich
one and the direction we take is simply the one that we believe is most
appropriate.  Certainly, there are many more directions to take, and our
work has lead to follow-on work that extend the problem and explore
other facets of it (see, e.g.,\ \cite{bsg06, brk05, lil05, tam06, wck06,
xiy05, xiy06}).  In Section~\ref{sec:distributed_algorithms}, we discuss
distributed algorithms for solving the problem.  Such algorithms allow
subgraphs to be computed in a distributed manner, with each node making
computations based only on local knowledge and knowledge acquired from
information exchanges.  Perhaps the most well-known distributed
algorithm in networking is the distributed Bellman-Ford algorithm (see,
e.g.,\ \cite[Section 5.2]{beg92}), which is used to find routes in
routed packet networks.  Designing algorithms that can be run in a
distributed manner is not an easy task and, though we do manage to do
so, they apply only in cases where links essentially behave
independently and medium access issues do not pose significant
constraints, either because they are non-existent or because they are 
dealt with
separately (in contrast to the slotted Aloha relay channel of
Section~\ref{sec:example1}, where medium access issues form a large part
of the problem and must be dealt with directly).  In
Section~\ref{sec:dynamic_multicast}, we introduce a dynamic component
into the problem.  Dynamics, such as changes in the membership of the
multicast group or changes in the positions of the nodes, are often
present in problems of interest.  We consider the scenario where
membership of the multicast group changes in time, with nodes joining
and leaving the group, and continuous service to all members of the
group must be maintained---a problem we call dynamic multicast.

\section{Problem formulation}
\label{sec:problem_formulation}

We specify a multicast connection with a triplet $(s, T, \{R_t\}_{t \in
T})$, where $s$ is the source of the connection, $T$ is the set of
sinks, and $\{R_t\}_{t\in T}$ is the set of rates to the sinks (see
Section~\ref{sec:coding_multicast}).
Suppose we wish to establish $C$ multicast connections,
$(s_1, T_1, \{R_{t,1}\}), \ldots, (s_C, T_C, \{R_{t,C}\})$.
Using Theorem~\ref{thm:2100} and the max-flow/min-cut theorem, 
we see that the efficient operation problem
can now be phrased as the following mathematical programming problem:
\begin{equation}
\begin{split}
& \begin{aligned}
\text{minimize }   & f(z) \\
\text{subject to } & z \in Z, 
\end{aligned} \\
&\; \sum_{c=1}^C y_{iJK}^{(c)} \le z_{iJK} ,
  \qquad \text{$\forall$ $(i,J) \in \mathcal{A}$, $K \subset J$} , \\
&\; \sum_{j \in K} x_{iJj}^{(t,c)}
\le \sum_{\{L \subset J | L \cap K \neq \emptyset\}} y_{iJL}^{(c)} ,
  \qquad \text{$\forall$ $(i,J) \in \mathcal{A}$, $K \subset
J$, $t \in T_c$, $c=1,\ldots,C$} , \\
&\; x^{(t,c)} \in F^{(t,c)},
  \qquad \text{$\forall$ $t \in T_c$, $c=1,\ldots,C$} ,
\end{split}
\label{eqn:3110}
\end{equation}
where $x^{(t,c)}$ is the vector consisting of $x_{iJj}^{(t,c)}$,
$(i,J) \in \mathcal{A}$, $j \in J$, and $F^{(t,c)}$ is the bounded
polyhedron of points ${x}^{(t,c)}$ satisfying
the conservation of flow constraints
\[
\sum_{\{J | (i,J) \in \mathcal{A}\}} \sum_{j \in J} x_{iJj}^{(t,c)}
- \sum_{\{j | (j,I) \in \mathcal{A}, i \in I\}} x_{jIi}^{(t,c)}
= 
\begin{cases}
R_{t,c} & \text{if $i = s_c$} , \\
-R_{t,c} & \text{if $i = t$} , \\
0 & \text{otherwise} ,
\end{cases}
\qquad \text{$\forall$ $i \in \mathcal{N}$} ,
\]
and non-negativity constraints
\[
x_{iJj}^{(t,c)} \ge 0 ,
  \qquad \text{$\forall$ $(i,J) \in \mathcal{A}$, $j \in J$} .
\]
In this formulation, $y_{iJK}^{(c)}$ represents the average rate of packets
that are injected on hyperarc $(i,J)$ and received by exactly the set of
nodes $K$ (which occurs with average rate $z_{iJK}$) and that are
allocated to connection $c$.

For simplicity, let us consider the case where $C=1$.  The extension to
$C > 1$ is conceptually straightforward and, moreover, the case where
$C=1$ is interesting in its own right: whenever each multicast group has
a selfish cost objective, or when the network sets link weights to meet
its objective or enforce certain policies and each multicast group is
subject to a minimum-weight objective, we wish to establish single
efficient multicast connections.

Let
\[
b_{iJK} :=
\frac{\sum_{\{L \subset J | L \cap K \neq \emptyset\}} z_{iJL}}{z_{iJ}},
\]
which is the fraction of packets injected on hyperarc $(i,J)$
that are received by a set of nodes that intersects $K$.
Problem (\ref{eqn:3110}) is now
\begin{equation}
\begin{aligned}
\text{minimize }   & f(z) \\
\text{subject to } & z \in Z, \\
& \sum_{j \in K} x_{iJj}^{(t)}
\le z_{iJ}b_{iJK} ,
  \qquad \text{$\forall$ $(i,J) \in \mathcal{A}$, $K \subset
J$, $t \in T$} , \\
& x^{(t)} \in F^{(t)},
  \qquad \text{$\forall$ $t \in T$} .
\end{aligned}
\label{eqn:3210}
\end{equation}

In the lossless case, problem (\ref{eqn:3210}) simplifies to the
following problem:
\begin{equation}
\begin{split}
\text{minimize }   & f(z) \\
\text{subject to }
& z \in Z,  \\
& \sum_{j \in J} x_{iJj}^{(t)} \le z_{iJ},
  \qquad \text{$\forall$ $(i,J) \in \mathcal{A}$, $t \in T$} , \\
& x^{(t)} \in F^{(t)},
  \qquad \text{$\forall$ $t \in T$}.
\end{split}
\label{eqn:3222}
\end{equation}

\begin{figure}
\centering
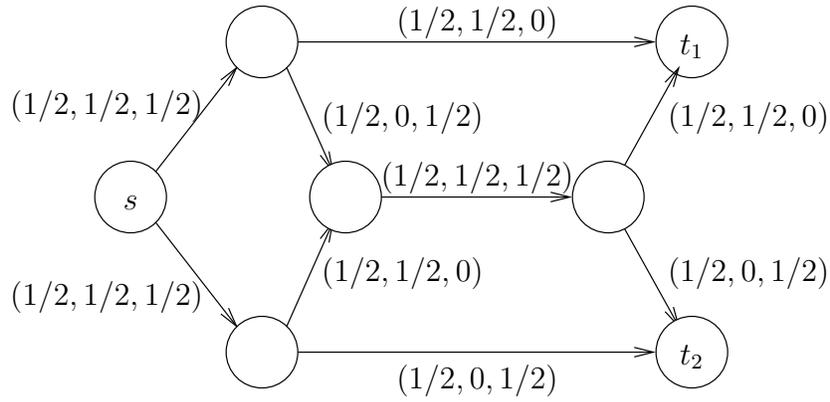 
\caption[A network of lossless point-to-point links with multicast from $s$ to $T=\{t_1, t_2\}$.]
{A network of lossless point-to-point links with multicast from $s$ to $T=\{t_1, t_2\}$.
Each arc is marked with the triple $(z_{ij}, x_{ijj}^{(1)},
x_{ijj}^{(2)})$.}
\label{fig:buttcost}
\end{figure}

\begin{figure}
\centering
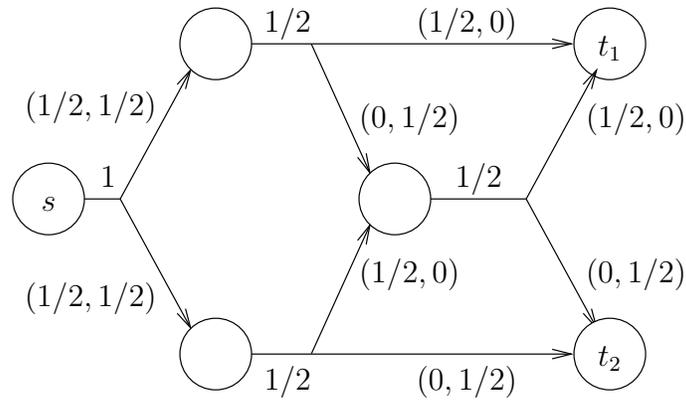 
\caption[A network of lossless broadcast links with multicast from $s$ to $T=\{t_1, t_2\}$.]
{A network of lossless broadcast links with multicast from $s$ to $T=\{t_1, t_2\}$.
Each hyperarc is marked with $z_{iJ}$ at its start and 
the pair $(x_{iJj}^{(1)}, x_{iJj}^{(2)})$ at its ends.}
\label{fig:wbuttcost}
\end{figure}

As an example, consider the network depicted in
Figure~\ref{fig:buttcost}, which consists only of point-to-point links.
Suppose that the network is lossless, that we wish to achieve multicast
of unit rate from $s$ to two sinks, $t_1$ and $t_2$, and that we have $Z =
[0,1]^{|\mathcal{A}|}$ and $f(z) = \sum_{(i,j) \in \mathcal{A}} z_{ij}$.
An optimal solution to problem (\ref{eqn:3222}) is shown in the figure.
We have flows $x^{(1)}$ and $x^{(2)}$ of unit size from $s$ to $t_1$ and
$t_2$, respectively, and, for each arc $(i,j)$, $z_{ij} =
\max(x_{ijj}^{(1)}, x_{ijj}^{(2)})$, as we expect from the optimization.

The same multicast problem in a routed packet network would entail
minimizing the number of arcs used to form a tree that is rooted at $s$
and that reaches $t_1$ and $t_2$---in other words, solving the Steiner
tree problem on directed graphs \cite{ram96}.  The Steiner tree problem
on directed graphs is well-known to be \textsc{np}-complete, but solving
problem (\ref{eqn:3222}) is not.  In this case, problem (\ref{eqn:3222})
is in fact a linear optimization problem.  It is a linear optimization
problem that can be thought of as a fractional relaxation of the Steiner
tree problem \cite{zok02}.  This example illustrates one of the
attractive features of the coded approach: it allows us avoid an
\textsc{np}-complete problem and instead solve its fractional
relaxation.  In Section~\ref{sec:min_weight}, we examine the efficiency
improvements that we can achieve from this feature.

For an example with broadcast links, consider the network depicted in
Figure~\ref{fig:wbuttcost}.  Suppose again that the network is lossless,
that we wish to achieve multicast of unit rate from $s$ to two sinks,
$t_1$ and $t_2$, and that we have $Z = [0,1]^{|\mathcal{A}|}$ and $f(z)
= \sum_{(i,J) \in \mathcal{A}} z_{iJ}$.  An optimal solution to problem
(\ref{eqn:3222}) is shown in the figure.  We still have flows $x^{(1)}$
and $x^{(2)}$ of unit size from $s$ to $t_1$ and $t_2$, respectively,
but now, for each hyperarc $(i,J)$, we determine $z_{iJ}$ from the
various flows passing through hyperarc $(i,J)$, each destined toward a
single node $j$ in $J$, and the optimization gives $z_{iJ} =
\max(\sum_{j \in J} x_{iJj}^{(1)}, \sum_{j \in J} x_{iJj}^{(2)})$.

Neither problem (\ref{eqn:3210}) nor (\ref{eqn:3222}) as it stands is
easy to solve.  But the problems are very general.  Their complexities
improve if we assume that the cost function is separable and possibly
even linear, i.e.,\  if we suppose $f(z) = \sum_{(i,J) \in A}
f_{iJ}(z_{iJ})$, where $f_{iJ}$ is a convex or linear function, which is
a very reasonable assumption in many practical situations.  For example,
packet latency is usually assessed with a separable, convex cost
function and energy, monetary cost, and total weight are usually
assessed with separable, linear cost functions.  The problems examined
in our performance evaluation in
Chapter~\ref{chap:performance_evaluation}, which we believe reflect
problems of practical interest, all involve separable, linear cost
functions.  

The complexities of problems (\ref{eqn:3210}) and (\ref{eqn:3222}) also
improve if we make some assumptions on the form of the constraint set
$Z$, which is the case in most practical situations.

A particular simplification applies if we assume that, when nodes
transmit in a lossless network, they reach all nodes in a certain
region, with cost increasing as this region is expanded.
This applies, for example, if we are interested in minimizing energy
consumption, and the region in which a packet is reliably received
expands as we expend more energy in its transmission.
More
precisely, suppose that we have separable cost, so $f(z) = \sum_{(i,J)
\in \mathcal{A}} f_{iJ}(z_{iJ})$.  Suppose further that
each node $i$ has $M_i$ outgoing hyperarcs
$(i,J_1^{(i)}), (i, J_2^{(i)}), \ldots, (i, J_{M_i}^{(i)})$
with
$J_1^{(i)} \subsetneq J_2^{(i)} \subsetneq \cdots \subsetneq
J_{M_i}^{(i)}$.
(We assume that there are no identical links, as duplicate links
can effectively be treated as a single link.)
Then, we assume that
 $f_{iJ_1^{(i)}}(\zeta) < f_{iJ_2^{(i)}}(\zeta) < \cdots
< f_{iJ_{M_i}^{(i)}}(\zeta)$ for all $\zeta \ge 0$ and
nodes $i$.

Let us introduce,
for $(i,j) \in \mathcal{A}^\prime
:= \{(i,j) | (i,J) \in A, J \ni j\}$,
the variables
\[
\hat{x}_{ij}^{(t)} := \sum_{m=m(i, j)}^{M_i} x_{iJ_m^{(i)}j}^{(t)},
\]
where $m(i,j)$ is the unique $m$ such that $j \in J_m^{(i)} \setminus
J_{m-1}^{(i)}$
(we define $J_0^{(i)} := \emptyset$ for all $i \in \mathcal{N}$
for convenience).
Now, problem (\ref{eqn:3222}) can be reformulated as the following
problem, which has substantially fewer variables:
\begin{equation}
\begin{aligned}
\text{minimize }   & \sum_{(i,J) \in \mathcal{A}} f_{iJ}(z_{iJ}) \\
\text{subject to } & z \in Z \\
&
\sum_{k \in J_{M_i}^{(i)} \setminus J_{m-1}^{(i)}}
  \hat{x}_{ik}^{(t)}
\le
\sum_{n = m}^{M_i} z_{iJ_n^{(i)}},
  \qquad \text{$\forall$ $i \in \mathcal{N}$, $m=1,\ldots,M_i$, $t \in T$} , \\
& \hat{x}^{(t)} \in \hat{F}^{(t)},
  \qquad \text{$\forall$ $t\in T$},
\end{aligned} 
\label{eqn:3250}
\end{equation}
where $\hat{F}^{(t)}$ is the bounded polyhedron of
points $\hat{x}^{(t)}$ satisfying
the conservation of flow constraints
\[
\sum_{\{j | (i,j) \in \mathcal{A}^\prime\}} \hat{x}_{ij}^{(t)}
- \sum_{\{j | (j,i) \in \mathcal{A}^\prime\}} \hat{x}_{ji}^{(t)} = 
\begin{cases}
R_t & \text{if $i = s$} , \\
-R_t & \text{if $i = t$} , \\
0 & \text{otherwise} ,
\end{cases}
\qquad \text{$\forall$ $i \in N$} ,
\]
and non-negativity constraints
\[
0 \le \hat{x}_{ij}^{(t)},
\qquad \text{$\forall$ $(i,j) \in \mathcal{A}^\prime$} .
\]

\begin{Prop}
Suppose that $f(z) = \sum_{(i,J) \in \mathcal{A}} f_{iJ}(z_{iJ})$ and
that
$f_{iJ_1^{(i)}}(\zeta) < f_{iJ_2^{(i)}}(\zeta) < \cdots
< f_{iJ_{M_i}^{(i)}}(\zeta)$ for all $\zeta \ge 0$ and
$i \in \mathcal{N}$.
Then
problem (\ref{eqn:3222}) and problem (\ref{eqn:3250}) are equivalent in
the sense that they have the same optimal cost and $z$ is part of an
optimal solution for (\ref{eqn:3222}) if and only if it is part of an
optimal solution for (\ref{eqn:3250}).
\label{prop:100}
\end{Prop}

\begin{proof}
Suppose $(x, z)$ is a feasible solution to problem (\ref{eqn:3222}).
Then, for all $(i,j) \in \mathcal{A}^\prime$ and $t \in T$,
\[
\begin{split}
\sum_{m = m(i,j)}^{M_i} z_{iJ_m^{(i)}}
&\ge \sum_{m = m(i,j)}^{M_i} \sum_{k \in J_m^{(i)}}
x_{iJ_m^{(i)}k}^{(t)} \\
&= \sum_{k \in J_{M_i}^{(i)}} \sum_{m = \max(m(i,j), m(i,k))}^{M_i}
x_{iJ_m^{(i)}k}^{(t)} \\
&\ge \sum_{k \in J_{M_i}^{(i)} \setminus J^{(i)}_{m(i,j)-1}}
\sum_{m = \max(m(i,j), m(i,k))}^{M_i}
x_{iJ_m^{(i)}k}^{(t)} \\
&= \sum_{k \in J_{M_i}^{(i)} \setminus J^{(i)}_{m(i,j)-1}}
\sum_{m = m(i,k)}^{M_i}
x_{iJ_m^{(i)}k}^{(t)} \\
&= \sum_{k \in J_{M_i}^{(i)} \setminus J^{(i)}_{m(i,j)-1}}
\hat{x}_{ik}^{(t)} .
\end{split}
\]
Hence $(\hat{x},z)$ is a feasible solution of problem (\ref{eqn:3250})
with the same cost.

Now suppose $(\hat{x}, z)$ is an optimal solution of problem
(\ref{eqn:3250}). 
Since $f_{iJ_1^{(i)}}(\zeta) < f_{iJ_2^{(i)}}(\zeta) < \cdots
< f_{iJ_{M_i}^{(i)}}(\zeta)$ for all $\zeta \ge 0$ and
$i \in \mathcal{N}$ by assumption, it
follows that, for all $i \in \mathcal{N}$, the sequence
$z_{iJ_1^{(i)}}, z_{iJ_2^{(i)}}, \ldots, z_{iJ_{M_i}^{(i)}}$ is given
recursively, starting from $m = M_i$, by
\[
z_{iJ_m^{(i)}} = \max_{t \in T}
\left\{
\sum_{k \in J_{M_i}^{(i)} \setminus J_{m-1}^{(i)}} \hat{x}_{ik}^{(t)}
\right\} 
- \sum_{m^\prime=m+1}^{M_i} z_{iJ_{m^\prime}^{(i)}} .
\]
Hence $z_{iJ_m^{(i)}} \ge 0$ for all $i \in \mathcal{N}$ and $m = 1, 2,
\ldots, M_i$.
We then set, starting from $m = M_i$ and $j \in J_{M_i}^{(i)}$,
\[
\begin{split}
x_{iJ_m^{(i)}j}^{(t)}
&:= \min\left(
\hat{x}_{ij}^{(t)} - \sum_{l = m+1}^{M_i} x_{iJ_l^{(i)}j} ,
z_{iJ_m^{(i)}} -
\sum_{k \in J_{M_i}^{(i)} \setminus J_{m(i,j)}^{(i)}}
x_{iJ_m^{(i)}k}^{(t)}
\right) .
\end{split}
\]
It is now not difficult to see that $(x,z)$ is a feasible solution of
problem (\ref{eqn:3222}) with the same cost.

Therefore, the optimal costs of problems (\ref{eqn:3222}) and
(\ref{eqn:3250}) are the same and, since the objective functions for the
two problems are the same, $z$ is part of an optimal solution for
problem (\ref{eqn:3222}) if and only if it is part of an optimal solution
for problem (\ref{eqn:3250}).
\end{proof}

\subsection{An example}
\label{sec:example3}

Let us return again to the slotted Aloha relay channel described in
Section~\ref{sec:example1}.  The relevant optimization problem to solve
in this case is (\ref{eqn:3210}), and it reduces to 
(cf.\  Section~\ref{sec:example2})
\begin{equation*}
\begin{aligned}
\text{minimize }   & z_{1(23)} + z_{23} \\
\text{subject to } & 0 \le z_{1(23)}, z_{23} \le 1, \\
& R \le z_{1(23)}(1-z_{23})(p_{1(23)2} + p_{1(23)3} + p_{1(23)(23)}),
\\
& R \le z_{1(23)}(1-z_{23})(p_{1(23)3} + p_{1(23)(23)}) +
  (1-z_{1(23)})z_{23}p_{233} .
\end{aligned}
\end{equation*}

\begin{figure}
\hspace*{0.25in}
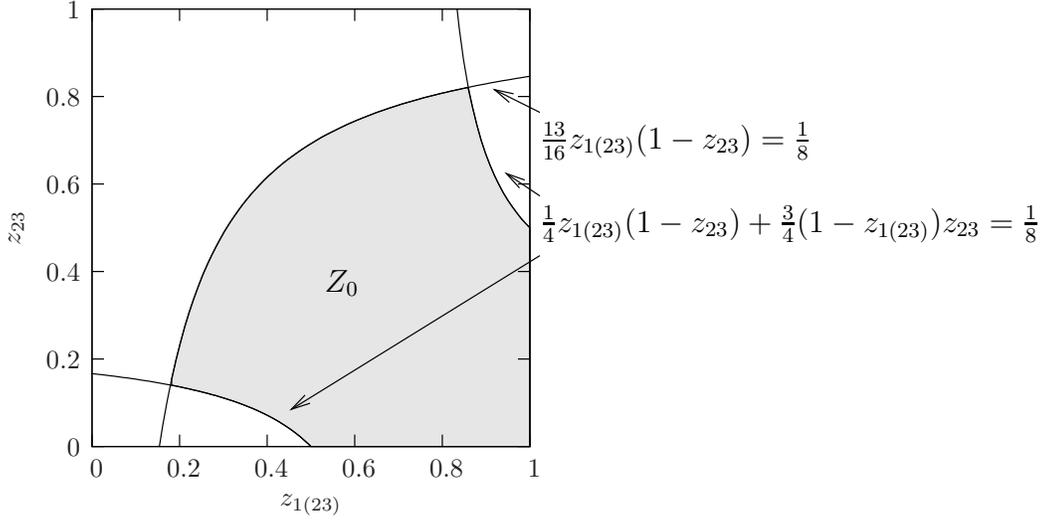
\caption{Feasible set of problem (\ref{eqn:3260}).}
\label{fig:ex_region}
\end{figure}

Let us assume some values for the parameters of the problem and work
through it.  Let $R := 1/8$, $p_{1(23)2} := 9/16$, $p_{1(23)3} := 1/16$,
$p_{1(23)(23)} := 3/16$, and $p_{233} := 3/4$.  Then the optimization
problem we have is
\begin{equation}
\begin{aligned}
\text{minimize }   & z_{1(23)} + z_{23} \\
\text{subject to } & 0 \le z_{1(23)}, z_{23} \le 1, \\
& \frac{1}{8} \le \frac{13}{16}z_{1(23)}(1-z_{23}), \\
& \frac{1}{8} \le \frac{1}{4}z_{1(23)}(1-z_{23}) +
  \frac{3}{4}(1-z_{1(23)})z_{23}.
\end{aligned}
\label{eqn:3260}
\end{equation}
The feasible set of this problem is shown in Figure~\ref{fig:ex_region}.
It is the shaded region labeled $Z_0$.
By inspection, the optimal solution of (\ref{eqn:3260}) is the
lesser of the two intersections between the curves defined by 
\[
\frac{13}{16}z_{1(23)}(1-z_{23}) = \frac{1}{8}
\]
and
\[
\frac{1}{4}z_{1(23)}(1-z_{23}) + \frac{3}{4}(1-z_{1(23)})z_{23}
= \frac{1}{8}.
\]
We obtain $z_{1(23)}^* \simeq 0.179$ and $z_{23}^* \simeq 0.141$.

The problem we have just solved is by no means trivial.  We have taken a
wireless packet network subject to losses that are determined by a
complicated set of conditions---including medium contention---and found
a way of establishing a given unicast connection of fixed throughput
using the minimum number of transmissions per message packet.  
The solution is
that node 1 transmits a packet every time slot with probability 0.179,
and node 2 transmits a packet every time slot independently with
probability 0.141.  Whenever either node transmits a packet, they follow
the coding scheme of Section~\ref{sec:coding_scheme}.

The network we dealt with was, unfortunately, only a small one, and the
solution method we used will not straightforwardly scale to larger
problems.  But the solution method is conceptually simple, and there are
cases where the solution to large problems is computable---and
computable in a distributed manner.  This is the topic of the next
section.

\section{Distributed algorithms}
\label{sec:distributed_algorithms}

In many cases, the optimization problems (\ref{eqn:3210}),
(\ref{eqn:3222}), and (\ref{eqn:3250}) are convex or linear problems and
their solutions can, in theory, be computed.  For practical network
applications, however, it is often important that solutions can be
computed in a distributed manner, with each node making computations
based only on local knowledge and knowledge acquired from information
exchanges.  Thus, we seek distributed algorithms to solve optimization
problems (\ref{eqn:3210}), (\ref{eqn:3222}), and (\ref{eqn:3250}),
which, when paired with the random linear coding scheme of the previous
chapter, yields a distributed approach to efficient operation.  The
algorithms we propose will generally take some time to converge to an
optimal solution, but it is not necessary to wait until the algorithms
have converged before transmission---we can apply the coding scheme to
the coding subgraph we have at any time, optimal or otherwise, and
continue doing so while it converges.  Such an approach is robust to
dynamics such as changes in network topology that cause the optimal
solution to change, because the algorithms will simply converge toward
the changing optimum.

To this end, we simplify the problem by assuming that the objective
function is of the form $f(z) = \sum_{(i,J) \in \mathcal{A}}
f_{iJ}(z_{iJ})$, where $f_{iJ}$ is a monotonically increasing, convex
function, and that, as $z_{iJ}$ is varied, $z_{iJK} / z_{iJ}$ is
constant for all $K \subset J$.  Therefore, $b_{iJK}$ is a constant for
all $(i,J) \in \mathcal{A}$ and $K \subset J$.  We also drop the
constraint set $Z$, noting that separable constraints, at least, can be
handled by making $f_{iJ}$ approach infinity as $z_{iJ}$ approaches its
upper constraint.  These assumptions apply if, at least from the
perspective of the connection we wish to establish, links essentially
behave independently and medium access issues do not pose significant
constraints, either because they are non-existent or because they are
dealt with separately.  The assumptions certainly restrict the range
of applicable cases, but they are not impractical; they apply, in
particular, to all of
the problems examined in our performance evaluation in
Chapter~\ref{chap:performance_evaluation}.

With these assumptions, problem (\ref{eqn:3210}) becomes
\begin{equation}
\begin{split}
\text{minimize }   & \sum_{(i,J) \in \mathcal{A}} f_{iJ}(z_{iJ}) \\
\text{subject to }
& \sum_{j \in K} x_{iJj}^{(t)}
\le z_{iJ} b_{iJK} ,
  \qquad \text{$\forall$ $(i,J) \in \mathcal{A}$, $K \subset
J$, $t \in T$} , \\
& x^{(t)} \in F^{(t)},
  \qquad \text{$\forall$ $t \in T$}.
\end{split}
\label{eqn:3270}
\end{equation}

Since the $f_{iJ}$ are monotonically increasing, the constraint
\begin{equation}
\sum_{j \in K} x_{iJj}^{(t)}
\le z_{iJ} b_{iJK} ,
  \qquad \text{$\forall$ $(i,J) \in \mathcal{A}$, $K \subset
J$, $t \in T$}
\label{eqn:3271}
\end{equation}
gives
\begin{equation}
z_{iJ} = \max_{K \subset J, t \in T}
\left\{
\frac{\sum_{j \in K} x_{iJj}^{(t)}}{b_{iJK}}
\right\}.
\label{eqn:3272}
\end{equation}
Expression (\ref{eqn:3272}) is, unfortunately, not very useful for
algorithm design because the max function is difficult to deal with,
largely as a result of its not being differentiable everywhere.
One way to
overcome this difficulty is to approximate $z_{iJ}$ by
replacing the max in (\ref{eqn:3272})
with an $l^m$-norm (see \cite{des04}), i.e.,\  to approximate $z_{iJ}$
with $z_{iJ}^\prime$, where
\[
z_{iJ}^\prime
:= \left(\sum_{K \subset J, t \in T}
\left(
\frac{\sum_{j \in K} x_{iJj}^{(t)}}{b_{iJK}}
\right)^m\right)^{1/m}.
\]
The approximation becomes exact as $m \rightarrow \infty$.
Moreover, since $z_{iJ}^\prime \ge z_{iJ}$ for all $m > 0$, the coding
subgraph $z^\prime$ admits the desired connection for any feasible
solution.

Now the relevant optimization problem is
\[
\begin{split}
\text{minimize }   & \sum_{(i,J) \in \mathcal{A}} f_{iJ}(z_{iJ}^\prime) \\
\text{subject to }
& x^{(t)} \in F^{(t)},
  \qquad \text{$\forall$ $t \in T$},
\end{split}
\]
which is no more than a convex multicommodity flow problem.
There are many algorithms for convex multicommodity flow problems (see
\cite{omv00} for a survey), some of which (e.g.,\  the algorithms in
\cite{ber80,bgg84}) are well-suited for distributed implementation.  The
primal-dual approach to internet congestion control (see \cite[Section
3.4]{sri04}) can also be used to solve convex multicommodity flow
problems in a distributed manner, and we examine this method in
Section~\ref{sec:primal-dual}.

There exist, therefore, numerous distributed algorithms for the
subgraph selection problem---or, at least, for an approximation of the
problem.  What about distributed algorithms for the true problem?
One clear tactic for finding such algorithms is to eliminate constraint
(\ref{eqn:3271}) using Lagrange multipliers.  Following this tactic, we
obtain a distributed algorithm that we call the subgradient method.  We
describe the subgradient method in Section~\ref{sec:subgradient}.

\subsection{Primal-dual method}
\label{sec:primal-dual}

For the primal-dual method, we assume that the cost functions $f_{iJ}$
are strictly convex and differentiable. 
Hence there is a unique optimal solution to
problem (\ref{eqn:3270}).  We present the
algorithm for the lossless case, with the understanding that it can be
straightforwardly extended to the lossy case.  Thus, the optimization
problem we address is
\begin{equation}
\begin{split}
\text{minimize }   & \sum_{(i,J) \in \mathcal{A}} f_{iJ}(z_{iJ}^\prime) \\
\text{subject to }
& x^{(t)} \in F^{(t)},
  \qquad \text{$\forall$ $t \in T$},
\end{split}
\label{eqn:3340}
\end{equation}
where
\[
z_{iJ}^\prime
:= \left(\sum_{t \in T}
\left(
\sum_{j \in J} x_{iJj}^{(t)}
\right)^m\right)^{1/m}.
\]

Let $(y)_a^+$ denote the following function of $y$:
\begin{equation*}
(y)_a^+ = \begin{cases}
    y &  \text{if $a>0$},\\
    \max\{y,0\} &   \text{if $a \le 0$}.
\end{cases}
\end{equation*}

To solve problem (\ref{eqn:3340}) in a distributed fashion, 
we introduce additional variables $p$ and $\lambda$ and consider varying
$x$, $p$, and $\lambda$ in time $\tau$ according to the following time
derivatives:
\begin{gather}
\dot{x}_{iJj}^{(t)}
=
-k_{iJj}^{(t)}(x_{iJj}^{(t)})\left(\frac{\partial f_{iJ}(z_{iJ}^\prime)}
{\partial x_{iJj}^{(t)}} +
q_{ij}^{(t)}-\lambda_{iJj}^{(t)}\right) , \label{eqn:alg1} \\
\dot{p}_i^{(t)}
=
h_i^{(t)}(p_i^{(t)}) (y_i^{(t)} - \sigma_i^{(t)}) , \\
\dot{\lambda}_{iJj}^{(t)} = m_{iJj}^{(t)}(\lambda_{iJj}^{(t)})
\left(-x_{iJj}^{(t)}\right)_{\lambda_{iJj}^{(t)}}^+ ,
\label{eqn:alg3}
\end{gather}
where
\begin{gather*}
q_{ij}^{(t)} := p_i^{(t)} - p_j^{(t)} , \\
y_i^{(t)} := \sum_{\{J | (i,J) \in \mathcal{A}\}} \sum_{j \in J} x_{iJj}^{(t)}
- \sum_{\{j | (j,I) \in \mathcal{A}, i \in I\}} x_{jIi}^{(t)}
,
\end{gather*}
and
$k_{iJj}^{(t)}(x_{iJj}^{(t)})>0 $, $h_i^{(t)}(p_i^{(t)}) >0 $,
and $m_{iJj}^{(t)}(\lambda_{iJj}^{(t)}) > 0$ are non-decreasing
continuous functions of  $x_{iJj}^{(t)}$, $p_i^{(t)}$,
and $\lambda_{iJj}^{(t)}$ respectively.

\begin{Prop}
The algorithm specified by Equations (\ref{eqn:alg1})--(\ref{eqn:alg3}) 
is globally, asymptotically stable.
\label{prop:stability}
\end{Prop}

\begin{proof}
We prove the stability of the primal-dual algorithm by using the theory
of Lyapunov stability (see, e.g.,\  \cite[Section 3.10]{sri04}).
This proof is based on the proof of Theorem 3.7 of \cite{sri04}.

The Lagrangian for problem (\ref{eqn:3340}) is as follows:
\begin{multline}
L(x,p,\lambda) 
= \sum_{(i,J) \in \mathcal{A}} f_{iJ}(z_{iJ}^\prime) \\
+ \sum_{t \in T} \left\{ \sum_{i \in \mathcal{N}}
p_i^{(t)} \left( \sum_{\{J | (i,J) \in \mathcal{A}\}} 
\sum_{j \in J} x_{iJj}^{(t)} 
  - \sum_{\{j | (j,I) \in \mathcal{A}, i \in I\}} x_{jIi}^{(t)} 
- \sigma_i^{(t)} \right)
\right. \\
\left.
- \sum_{(i,J) \in \mathcal{A}} \sum_{j \in J} \lambda_{iJj}^{(t)}
  x_{iJj}^{(t)} \right\} ,
\label{eqn:3348}
\end{multline}
where
\[
\sigma_i^{(t)} =
\begin{cases}
R_t & \text{if $i=s$} , \\
-R_t & \text{if $i=t$} , \\
0 & \text{otherwise} .
\end{cases}
\]
Since the objective function of problem (\ref{eqn:3340}) is strictly convex,
it has a unique minimizing solution, say $\hat{x}$,
and Lagrange multipliers, say $\hat{p}$ and $\hat{\lambda}$, which
satisfy the following Karush-Kuhn-Tucker conditions:
\begin{gather}
\allowdisplaybreaks
\frac{\partial L(\hat{x},\hat{p},\hat{\lambda})}
{\partial x_{iJj}^{(t)}} = \left(\frac{\partial f_{iJ}(z_{iJ}^\prime)}
{\partial x_{iJj}^{(t)}} + \left(\hat{p}_i^{(t)} - \hat{p}_j^{(t)}\right) 
- \hat{\lambda}_{iJj}^{(t)}\right) = 0 , 
\qquad \text{$\forall$ $(i,J) \in \mathcal{A}$, $j \in J$, $t \in T$} , 
\label{eqn:kkt1} \\
\sum_{\{J | (i,J) \in \mathcal{A}\}} \sum_{j \in J} \hat{x}_{iJj}^{(t)}
  - \sum_{\{j | (j,I) \in \mathcal{A}, i \in I\}} \hat{x}_{jIi}^{(t)} 
= \sigma_i^{(t)}, 
\qquad \text{$\forall$ $i \in \mathcal{N}$, $t \in T$} , \\
\hat{x}_{iJj}^{(t)} \geq 0 \qquad 
\text{$\forall$ $(i,J) \in \mathcal{A}$, $j \in J$, $t \in T$} , \\
\hat{\lambda}_{iJj}^{(t)} \geq 0 \qquad 
\text{$\forall$ $(i,J) \in \mathcal{A}$, $j \in J$, $t \in T$} , \\
\hat{\lambda}_{iJj}^{(t)} \hat{x}_{iJj}^{(t)} = 0 \qquad
\text{$\forall$ $(i,J) \in \mathcal{A}$, $j \in J$, $t \in T$} . 
\label{eqn:kkt4}
\end{gather}

Using equation (\ref{eqn:3348}), we see that $(\hat{x},
\hat{p},\hat{\lambda})$ is an equilibrium point of the
primal-dual algorithm.  We now prove that this point is globally,
asymptotically stable.

Consider the following function as a candidate for the
Lyapunov function:
\begin{multline*}
V(x,p,\lambda) \\
= \sum_{t \in T}
\left\{
\sum_{(i,J) \in \mathcal{A}} \sum_{j \in J}
\left( \int_{\hat{x}_{iJj}^{(t)}}^{x_{iJj}^{(t)}}
\frac{1}{k_{iJj}^{(t)}(\sigma)}(\sigma - \hat{x}_{iJj}^{(t)}) d\sigma
+ \int_{\hat{\lambda}_{iJj}^{(t)}}^{\lambda_{iJj}^{(t)}}
\frac{1}{m_{iJj}^{(t)}(\gamma)}(\gamma - \hat{\lambda}_{iJj}^{(t)})
d\gamma \right)
\right. \\
\left.
+ \sum_{i \in \mathcal{N}} \int_{\hat{p}_i^{(t)}}^{p_i^{(t)}}
\frac{1}{h_i^{(t)}(\beta)}(\beta - \hat{p}_i^{(t)}) d\beta
\right\} .
\end{multline*}
Note that $V(\hat{{x}},\hat{{p}},\hat{\lambda}) = 0$. Since,
$k_{iJj}^{(t)}(\sigma) > 0$, if $x_{iJj}^{(t)} \neq \hat{x}_{iJj}^{(t)}$, 
we have
$\int_{\hat{x_{iJj}^{(t)}}}^{x_{iJj}^{(t)}}\frac{1}{k_{iJj}^{(t)}
(\sigma)}(\sigma - \hat{x}_{iJj}^{(t)})d\sigma >0$. 
This argument can be extended to the other
  terms as well.
Thus, whenever $({x}, {p},{\lambda}) \neq (\hat{x}
,\hat{{p}},\hat{{\lambda}})$, we have $V({x},{p},{\lambda}) > 0$.

Now,
\begin{multline*}
\dot{V}
= \sum_{t \in T}
\left\{
\sum_{(i,J) \in \mathcal{A}} \sum_{j \in J} \left[ 
\left(-x_{iJj}^{(t)}\right)^+_{\lambda_{iJj}^{(t)}}
(\lambda_{iJj}^{(t)} - \hat{\lambda}_{iJj}^{(t)})
\right.\right.
\\
\left.\left.
- \left(\frac{\partial f_{iJ}(z_{iJ}^\prime)}{\partial x_{iJj}^{(t)}}
+ q_{iJj}^{(t)} - \lambda_{iJj}^{(t)} \right) 
\cdot (x_{iJj}^{(t)} - \hat{x}_{iJj}^{(t)})
\right]
\right. \\
\left.
+ \sum_{i \in \mathcal{N}}
(y_i^{(t)} - \sigma_i^{(t)})
(p_i^{(t)} - \hat{p}_i^{(t)}) 
\right\} .
\end{multline*}
Note that
\begin{equation*}
\left(-x_{iJj}^{(t)}\right)^+_{\lambda_{iJj}^{(t)}}
(\lambda_{iJj}^{(t)} - \hat{\lambda}_{iJj}^{(t)})
\le 
-x_{iJj}^{(t)}
(\lambda_{iJj}^{(t)} - \hat{\lambda}_{iJj}^{(t)}) ,
\end{equation*}
since the inequality is an equality if either $x_{iJj}^{(t)} \le 0$ or 
$\lambda_{iJj}^{(t)} \ge 0$; and, in the case when $x_{iJj}^{(t)} > 0$ and
$\lambda_{iJj}^{(t)} < 0$, we have
$(-x_{iJj}^{(t)})^+_{\lambda_{iJj}^{(t)}} = 0$ and, since
$\hat{\lambda}_{iJj}^{(t)} \ge 0$, 
$-x_{iJj}^{(t)}(\lambda_{iJj}^{(t)} - \hat{\lambda}_{iJj}^{(t)}) \ge 0$.
Therefore, 
\begin{equation*}
\begin{split}
\dot{V}
&\le \sum_{t \in T}
\left\{
\sum_{(i,J) \in \mathcal{A}} \sum_{j \in J} \left[ 
-x_{iJj}^{(t)}
(\lambda_{iJj}^{(t)} - \hat{\lambda}_{iJj}^{(t)})
\right.\right.
\\ & \qquad \qquad \qquad \qquad \qquad
\left.\left.
- \left(\frac{\partial f_{iJ}(z_{iJ}^\prime)}{\partial x_{iJj}^{(t)}}
+ q_{iJj}^{(t)} - \lambda_{iJj}^{(t)} \right) 
\cdot (x_{iJj}^{(t)} - \hat{x}_{iJj}^{(t)})
\right]
\right. \\
& \qquad \qquad
 \left.
+ \sum_{i \in \mathcal{N}}
(y_i^{(t)} - \sigma_i^{(t)})
(p_i^{(t)} - \hat{p}_i^{(t)}) 
\right\} \\
&= (\hat{q} - q)^\prime(x - \hat{x})
+ (\hat{p} - p)^\prime(y - \hat{y}) \\
&\qquad
+ \sum_{t \in T}
\left\{
\sum_{(i,J) \in \mathcal{A}} \sum_{j \in J} \left[ 
-\hat{x}_{iJj}^{(t)}
(\lambda_{iJj}^{(t)} - \hat{\lambda}_{iJj}^{(t)})
\right.\right.
\\ & \qquad \qquad \qquad \qquad \qquad \qquad
\left.\left.
- \left(\frac{\partial f_{iJ}(z_{iJ}^\prime)}{\partial x_{iJj}^{(t)}}
+ \hat{q}_{iJj}^{(t)} - \hat{\lambda}_{iJj}^{(t)} \right) 
\cdot (x_{iJj}^{(t)} - \hat{x}_{iJj}^{(t)})
\right]
\right. \\
& \qquad \qquad \qquad
 \left.
+ \sum_{i \in \mathcal{N}}
(\hat{y}_i^{(t)} - \sigma_i^{(t)})
(p_i^{(t)} - \hat{p}_i^{(t)}) 
\right\} \\
&= 
\sum_{t \in T} \sum_{(i,J) \in \mathcal{A}} \sum_{j \in J}
\left(
\frac{\partial f_{iJ}(\hat{z}_{iJ}^\prime)}{\partial \hat{x}_{iJj}^{(t)}}
- \frac{\partial f_{iJ}(z_{iJ}^\prime)}{\partial x_{iJj}^{(t)}}
\right)
(x_{iJj}^{(t)} - \hat{x}_{iJj}^{(t)})
- \lambda^\prime \hat{x} ,
\end{split}
\end{equation*}
where the last line follows from Karush-Kuhn-Tucker conditions
(\ref{eqn:kkt1})--(\ref{eqn:kkt4}) and the fact that
\begin{equation*}
\begin{split}
p^\prime y
&= \sum_{t \in T}
\sum_{i \in \mathcal{N}}
p_{i}^{(t)} 
\left(\sum_{\{J | (i,J) \in \mathcal{A}\}} \sum_{j \in J} {x}_{iJj}^{(t)}
  - \sum_{\{j | (j,I) \in \mathcal{A}, i \in I\}} {x}_{jIi}^{(t)} \right) \\
&= \sum_{t \in T}
\sum_{(i,J) \in \mathcal{A}} \sum_{j \in J}
x_{iJj}^{(t)} (p_i^{(t)} - p_j^{(t)}) = q^\prime x .
\end{split}
\end{equation*}
Thus, owing to the strict convexity of the functions 
$\{f_{iJ}\}$, we have
$\dot{V} \le -\lambda^\prime \hat{x}$,
with equality if and only if $x = \hat{x}$.
So it follows that $\dot{V} \le 0$ for all $\lambda \ge 0$, since
$\hat{x} \ge 0$.

If the initial choice of $\lambda$ is such that $\lambda(0) \geq 0$, we
see from the primal-dual algorithm that $\lambda(\tau) \geq 0$. 
This is
true since $\dot{\lambda} \geq 0$ whenever $\lambda \leq 0$. Thus,
it follows by the theory of Lyapunov stability that the algorithm is
indeed
globally, asymptotically stable.
\end{proof}

The global, asymptotic stability of the algorithm implies that no matter
what the initial choice of $({x},{p})$ is, the primal-dual
algorithm will converge to the unique solution of problem
(\ref{eqn:3340}).  We have to choose ${\lambda}$, however, with
non-negative entries as the initial choice.
Further, there is no guarantee that $x(\tau)$ yields a feasible solution
for any given $\tau$.  Therefore, a start-up time may be required before 
a feasible solution is obtained.

The algorithm that we currently have 
is a continuous time algorithm and,
in practice, an algorithm operating in discrete message exchanges is
required.
To discretize the algorithm, we consider time steps $n = 0,1,\ldots$
and replace the derivatives by differences:
\begin{gather}
x_{iJj}^{(t)}[n+1]
= {x}_{iJj}^{(t)}[n] 
- \alpha_{iJj}^{(t)}[n]\left(\frac{\partial f_{iJ}({z_{iJ}^\prime}[n])}
{\partial x_{iJj}^{(t)}[n]} +
q_{ij}^{(t)}[n]-\lambda_{iJj}^{(t)}[n]\right) , \label{eqn:pd_alg1}\\
p_i^{(t)}[n+1]
= p_i^{(t)}[n] + \beta_i^{(t)}[n] (y_i^{(t)}[n] - \sigma_i^{(t)}) , \\
\lambda_{iJj}^{(t)}[n+1] = \lambda_{iJj}^{(t)}[n] + \gamma_{iJj}^{(t)}[n] \left(-x_{iJj}^{(t)}[n]\right)_{\lambda_{iJj}^{(t)}[n]}^+ ,
\label{eqn:pd_alg3}
\end{gather}
where
\begin{gather*}
q_{ij}^{(t)}[n] := p_i^{(t)}[n] - p_j^{(t)}[n] , \\
y_i^{(t)}[n] := \sum_{\{J | (i,J) \in \mathcal{A}\}} \sum_{j \in J} x_{iJj}^{(t)}[n]
- \sum_{\{j | (j,I) \in \mathcal{A}, i \in I\}} x_{jIi}^{(t)}[n]
,
\end{gather*}
and $\alpha_{iJj}^{(t)}[n] >0$, $ \beta_i^{(t)}[n] > 0$, and $\gamma_{iJj}^{(t)}[n] > 0$ are step sizes.
This discretized algorithm operates in synchronous rounds, with nodes
exchanging information in each round.  We expect that this synchronicity
can be relaxed in practice, but this issue remains to be
investigated.

\begin{figure}
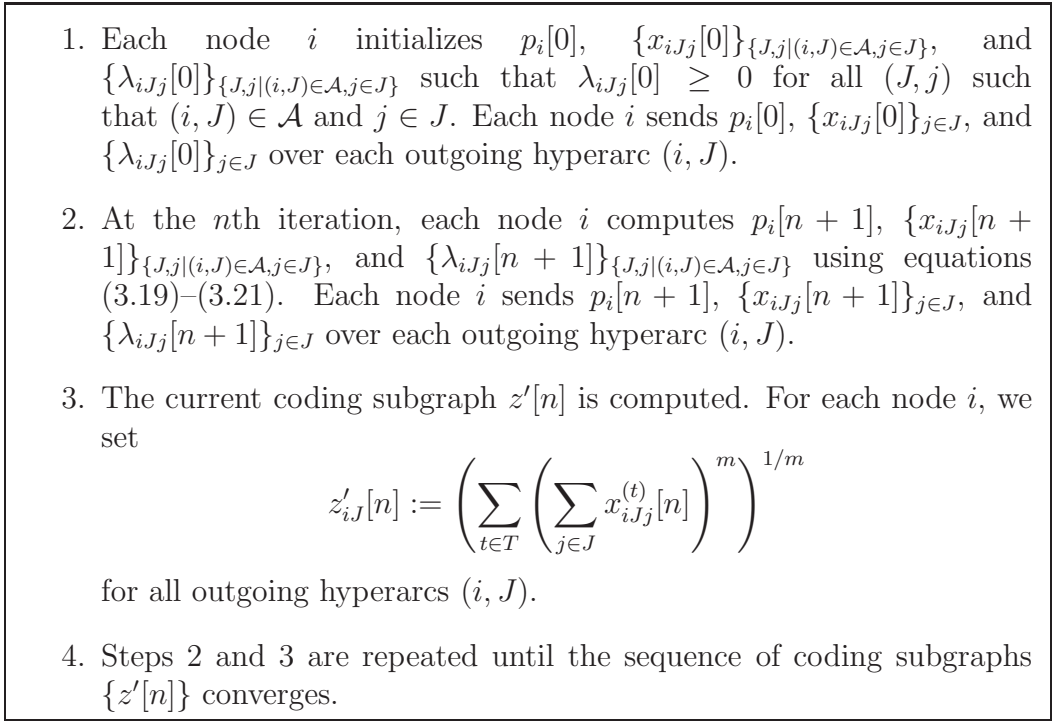

\centering
\framebox{
\begin{minipage}{0.88\textwidth}
\begin{enumerate}

\item Each node $i$ initializes $p_i[0]$, $\{x_{iJj}[0]\}_{\{J,j|(i,J)
\in \mathcal{A}, j \in J\}}$, and $\{\lambda_{iJj}[0]\}_{\{J,j|(i,J) \in
\mathcal{A}, j \in J\}}$ such that $\lambda_{iJj}[0] \ge 0$ for all
$(J,j)$ such that $(i,J) \in \mathcal{A}$ and $j \in J$.  
Each node $i$ sends
$p_i[0]$, $\{x_{iJj}[0]\}_{j \in J}$,
and $\{\lambda_{iJj}[0]\}_{j \in J}$ over each outgoing hyperarc
$(i,J)$.

\item At the $n$th iteration, 
each node $i$ computes $p_i[n+1]$, $\{x_{iJj}[n+1]\}_{\{J,j|(i,J)
\in \mathcal{A}, j \in J\}}$, and $\{\lambda_{iJj}[n+1]\}_{\{J,j|(i,J) \in
\mathcal{A}, j \in J\}}$ using equations
(\ref{eqn:pd_alg1})--(\ref{eqn:pd_alg3}).
Each node $i$ sends
$p_i[n+1]$, $\{x_{iJj}[n+1]\}_{j \in J}$,
and $\{\lambda_{iJj}[n+1]\}_{j \in J}$ over each outgoing hyperarc
$(i,J)$.
\label{pd_loop_beg}

\item The current coding subgraph $z^\prime[n]$ is computed.
For each node $i$, we set
\[
z_{iJ}^\prime[n] := 
\left( \sum_{t \in T} \left( \sum_{j \in J} 
x_{iJj}^{(t)}[n]
\right)^m \right)^{1/m} 
\]
for all outgoing hyperarcs $(i,J)$.
\label{pd_loop_end}

\item Steps \ref{pd_loop_beg} and \ref{pd_loop_end}
are repeated until the sequence of coding subgraphs
$\{z^\prime[n]\}$ converges.
\end{enumerate}
\end{minipage} }
\caption{Summary of the primal-dual method.}
\label{fig:summary_primal-dual}
\end{figure}

We associate a processor with each node. We assume that the processor
for node $i$ keeps track of the variables $p_i$,
$\{x_{iJj}\}_{\{J,j|(i,J) \in \mathcal{A}, j \in J\}}$, and
$\{\lambda_{iJj}\}_{\{J,j|(i,J) \in \mathcal{A}, j \in J\}}$.
With such an assignment of variables to processors, the algorithm is
distributed in the sense that a node exchanges information only with its
neighbors at every iteration of the primal-dual algorithm.
We summarize the primal-dual method in
Figure~\ref{fig:summary_primal-dual}.

\subsection{Subgradient method}
\label{sec:subgradient}

We present the subgradient method for linear cost functions; with some
modifications, it may be made to apply also to convex ones.  Thus, we
assume that the objective function $f$ is of the form
\[
f(z) := \sum_{(i,J) \in \mathcal{A}} a_{iJ}z_{iJ},
\]
where $a_{iJ} > 0$.

Consider the Lagrangian dual of problem (\ref{eqn:3270}):
\begin{equation}
\begin{split}
\text{maximize }   & \sum_{t \in T} q^{(t)}(p^{(t)})  \\
\text{subject to } & \sum_{t \in T}
\sum_{K \subset J} p_{iJK}^{(t)} = a_{iJ}
  \qquad \text{$\forall$ $(i,J) \in \mathcal{A}$} , \\
& p_{iJK}^{(t)} \ge 0 ,
  \qquad \text{$\forall$ $(i,J) \in \mathcal{A}$, $K \subset J$, $t \in T$} ,
\end{split}
\label{eqn:3512}
\end{equation}
where
\begin{equation}
q^{(t)}(p^{(t)}) := \min_{x^{(t)} \in F^{(t)}}
  \sum_{(i,J) \in \mathcal{A}} \sum_{j \in J}
\left(\sum_{\{K \subset J|K \ni j\}}
\frac{p_{iJK}^{(t)}}{b_{iJK}} \right) x_{iJj} .
\label{eqn:3513}
\end{equation}

In the lossless case,
the dual problem defined by equations (\ref{eqn:3512}) and
(\ref{eqn:3513}) simplifies somewhat, and we require only a single dual
variable $p_{iJJ}^{(t)}$ for each hyperarc $(i,J)$.
In the case that relates to optimization problem (\ref{eqn:3250}), the dual
problem simplifies more still, as there are fewer primal variables
associated with it.
Specifically, we obtain, for the Lagrangian dual,
\begin{equation}
\begin{split}
\text{maximize }   & \sum_{t \in T} \hat{q}^{(t)}(p^{(t)})  \\
\text{subject to } & \sum_{t \in T}
p_{iJ_m^{(i)}}^{(t)} = s_{iJ_m^{(i)}} ,
  \qquad \text{$\forall$ $i \in \mathcal{N}$, $m=1,\ldots,M_i$} , \\
& p_{iJ}^{(t)} \ge 0 ,
  \qquad \text{$\forall$ $(i,J) \in \mathcal{A}$, $t \in T$} ,
\end{split}
\label{eqn:3530}
\end{equation}
where
\[
s_{iJ_m^{(i)}} := a_{iJ_m^{(i)}} - a_{iJ_{m-1}^{(i)}},
\]
and
\begin{equation}
\hat{q}^{(t)}(p^{(t)}) := \min_{\hat{x}^{(t)} \in \hat{F}^{(t)}}
  \sum_{(i,j) \in \mathcal{A}^\prime} \left(\sum_{m=1}^{m(i,j)}
p_{iJ_m^{(i)}}^{(t)}\right) \hat{x}_{ij}^{(t)} .
\label{eqn:3533}
\end{equation}
Note that, by the assumptions of the problem, $s_{iJ} > 0$ for all
$(i,J) \in \mathcal{A}$.

In all three cases, the dual problems are very similar, and
essentially the same algorithm can be used to solve them.  We present
the subgradient method for the case that relates to optimization problem
(\ref{eqn:3250})---namely, the primal problem
\begin{equation}
\begin{split}
\text{minimize }   & \sum_{(i,J) \in \mathcal{A}} a_{iJ}z_{iJ} \\
\text{subject to } &
\sum_{k \in J_{M_i}^{(i)} \setminus J_{m-1}^{(i)}}
  \hat{x}_{ik}^{(t)}
\le
\sum_{n = m}^{M_i} z_{iJ_n^{(i)}},
  \qquad \text{$\forall$ $i \in \mathcal{N}$, $m=1,\ldots,M_i$, $t \in T$} , \\
& \hat{x}^{(t)} \in \hat{F}^{(t)},
  \qquad \text{$\forall$ $t\in T$}
\end{split}
\label{eqn:3540}
\end{equation}
with dual (\ref{eqn:3530})---with the
understanding that straightforward modifications can be made for the
other cases.

We first note that problem (\ref{eqn:3533}) is, in fact, a shortest path
problem, which admits a simple, asynchronous distributed solution known
as the distributed asynchronous Bellman-Ford algorithm (see, e.g.,\
\cite[Section 5.2.4]{beg92}).

Now, to solve the dual problem (\ref{eqn:3530}), we employ subgradient
optimization (see, e.g.,\  \cite[Section 6.3.1]{ber95} or 
\cite[Section I.2.4]{new99}).  We start
with an iterate $p[0]$ in the feasible set of
(\ref{eqn:3530}) and, given an iterate $p[n]$ for some
non-negative integer $n$, we solve
problem (\ref{eqn:3533}) for each $t$ in $T$ to obtain 
$x[n]$.  
Let
\[
g_{iJ_m^{(i)}}^{(t)}[n] :=
\sum_{k \in J_{M_i}^{(i)} \setminus J_{m-1}^{(i)}}
\hat{x}_{ik}^{(t)}[n] .
\]
We then assign
\begin{equation}
p_{iJ}[n + 1] :=
\argmin_{v \in P_{iJ}}
\sum_{t \in T} 
(v^{(t)} - (p_{iJ}^{(t)}[n] + \theta[n]g_{iJ}^{(t)}[n]))^2 
\label{eqn:3516}
\end{equation}
for each $(i,J) \in \mathcal{A}$,
where $P_{iJ}$ is the $|T|$-dimensional simplex
\[
P_{iJ} = 
\left\{v \left| 
\sum_{t \in T} v^{(t)} = s_{iJ},\,
v \ge 0 \right.\right\}
\]
and $\theta[n] > 0$ is an appropriate step size.
In other words, $p_{iJ}[n + 1]$ is set to be the Euclidean projection of
$p_{iJ}[n] + \theta[n]g_{iJ}[n]$ onto $P_{iJ}$.

To perform the projection, we use the following proposition.  
\begin{Prop}
Let 
$u := p_{iJ}[n] + \theta[n]g_{iJ}[n]$.  Suppose we index the elements of
$T$ such that $u^{(t_1)} \ge u^{(t_2)} \ge \ldots \ge u^{(t_{|T|})}$.
Take $\hat{k}$ to be the smallest $k$ such that
\[
\frac{1}{k}\left(s_{iJ} - \sum_{r=1}^{t_k} u^{(r)} \right) \le -u^{(t_{k+1})}
\]
or set $\hat{k} = |T|$ if no such $k$ exists.
Then the projection (\ref{eqn:3516}) is achieved by 
\[
p_{iJ}^{(t)}[n+1] 
=  
\begin{cases}
u^{(t)} + \frac{s_{iJ} - \sum_{r=1}^{t_{\hat{k}}} u^{(r)}}{\hat{k}}
  & \text{if $t \in \{t_{1}, \ldots, t_{\hat{k}}\}$}, \\
0 & \text{otherwise} .
\end{cases}
\]
\label{prop:200}
\end{Prop}

\begin{proof}
We wish to solve the following problem.
\begin{equation*}
\begin{aligned}
\text{minimize } & \sum_{t \in T} (v^{(t)} - u^{(t)})^2 \\
\text{subject to } & v \in P_{iJ} .
\end{aligned}
\end{equation*}
First, since the objective function and the constraint set $P_{iJ}$ are
both convex, it is straightforward to establish that a necessary and
sufficient condition for global optimality of $\hat{v}^{(t)}$ in $P_{iJ}$ 
is
\begin{equation}
\hat{v}^{(t)} > 0 \Rightarrow 
(u^{(t)} - \hat{v}^{(t)}) \ge (u^{(r)} - \hat{v}^{(r)}),
\qquad \text{$\forall$ $r \in T$}
\label{eqn:3600}
\end{equation}
(see, e.g.,\  \cite[Section 2.1]{ber95}).
Suppose we index the elements of $T$ such
that $u^{(t_1)} \ge u^{(t_2)} \ge \ldots \ge u^{(t_{|T|})}$.
We then note that there must be an index $k$ in the set 
$\{1, \ldots, |T|\}$ such that $v^{(t_l)} > 0$
for $l = 1, \ldots, k$ and $v^{(t_l)} = 0$ for $l > k+1$, 
for, if not, then a feasible solution with lower cost can be obtained by
swapping around components of the vector.
Therefore, condition (\ref{eqn:3600}) implies that there must exist some
$d$ such that ${\hat{v}^{(t)}} = u^{(t)} + d$ for all $t \in
\{t_1, \ldots, t_k\}$ and that $d \le -u^{(t)}$ for all $t \in
\{t_{k+1}, \ldots, t_{|T|}\}$, which is equivalent to
$d \le -u^{(t_{k+1})}$.  
Since ${\hat{v}^{(t)}}$ is in the simplex $P_{iJ}$, it follows that
\begin{equation*}
kd + \sum_{t = 1}^{t_k} u^{(t)} = s_{iJ},
\end{equation*}
which gives
\begin{equation*}
d = \frac{1}{k}\left(s_{iJ} - \sum_{t = 1}^{t_k} u^{(t)}\right) .
\end{equation*}
By taking $k = \hat{k}$, where $\hat{k}$ is the smallest $k$ such that
\begin{equation*}
\frac{1}{k}\left(s_{iJ} - \sum_{r=1}^{t_k} u^{(r)} \right) 
\le -u^{(t_{k+1})},
\end{equation*}
(or, if no such $k$ exists, then $\hat{k} = |T|$), we see that we have
\begin{equation*}
\frac{1}{\hat{k}-1}\left(s_{iJ} - \sum_{t=1}^{t_{k-1}}u^{(t)}\right)
> -u^{(t_k)},
\end{equation*}
which can be rearranged to give
\begin{equation*}
d = \frac{1}{\hat{k}}\left(s_{iJ} - \sum_{t=1}^{t_k}u^{(t)}\right)
> -u^{(t_k)} .
\end{equation*}
Hence, if $v^{(t)}$ is given by
\begin{equation}
v^{(t)} = 
\begin{cases}
u^{(t)} + \frac{s_{iJ} 
- \sum_{r=1}^{t_{\hat{k}}} u^{(r)}}{\hat{k}}
  & \text{if $t \in \{t_{1}, \ldots, t_{\hat{k}}\}$}, \\
0 & \text{otherwise} ,
\label{eqn:3610}
\end{cases}
\end{equation}
then $v^{(t)}$ is feasible and we see that the optimality condition
(\ref{eqn:3600}) is satisfied.  Note that, since $d \le -u^{(t_{k+1})}$,
equation (\ref{eqn:3610}) can
also be written as
\begin{equation}
v^{(t)} = \max\left(0, u^{(t)} + \frac{1}{\hat{k}}\left(s_{iJ} -
\sum_{r=1}^{t_{\hat{k}}} u^{(r)}\right) \right) .
\label{eqn:3615}
\end{equation}
\end{proof}

The disadvantage of subgradient optimization is that, whilst it yields
good approximations of the 
optimal value of the Lagrangian dual problem (\ref{eqn:3530}) after
sufficient iteration, it does not necessarily yield a primal optimal
solution.  There are, however, methods for recovering primal solutions
in subgradient optimization.  
We employ the following method, which is due to Sherali and
Choi \cite{shc96}.

Let $\{\mu_l[n]\}_{l = 1,\ldots, n}$ be a sequence of 
convex combination weights for each non-negative integer $n$, 
i.e.,\  $\sum_{l=1}^n \mu_l[n] = 1$ and $\mu_l[n] \ge 0$ for all 
$l = 1, \ldots, n$.
Further, let us define
\[
\gamma_{ln} := \frac{\mu_l[n]}{\theta[n]},
\quad\text{$l = 1,\ldots, n$, $n = 0, 1, \ldots$} ,
\]
and
\[
\Delta \gamma_n^{\max} := \max_{l = 2, \ldots, n}
\{\gamma_{ln} - \gamma_{(l-1)n}\} .
\]

\begin{Prop}
If the step sizes $\{\theta[n]\}$ and convex combination weights
$\{\mu_l[n]\}$ are chosen such that
\begin{enumerate}
\item $\gamma_{ln} \ge \gamma_{(l-1)n}$ for all $l = 2, \ldots, n$ and
$n = 0, 1, \ldots$,
\item $\Delta \gamma_n^{\max} \rightarrow 0$ as $n \rightarrow \infty$,
and
\item $\gamma_{1n} \rightarrow 0$ as $n \rightarrow \infty$ and
$\gamma_{nn} \le \delta$ for all $n = 0,1,\ldots$ for some $\delta > 0$,
\end{enumerate}
then 
we obtain an optimal solution to the primal problem from
any accumulation point of the sequence of primal iterates
$\{\tilde{x}[n]\}$ given by
\begin{equation}
\tilde{x}[n] := \sum_{l=1}^n\mu_l[n]\hat{x}[l], 
\quad n = 0,1, \ldots .
\end{equation}
\end{Prop}

\begin{proof}
Suppose that the dual feasible solution that the
subgradient method converges to is $\bar{p}$. 
Then, using (\ref{eqn:3516}),
there exists some $m$ such that for $n \ge m$
\begin{equation*}
p_{iJ}^{(t)}[n+1] = p_{iJ}^{(t)}[n] + \theta[n]g_{iJ}^{(t)}[n]
+ c_{iJ}[n]
\end{equation*}
for all $(i,J) \in \mathcal{A}$ and $t \in T$ such that 
$\bar{p}_{iJ}^{(t)} > 0$.  

Let $\tilde{g}[n] := \sum_{l=1}^n\mu_l[n]g[l]$.
Consider some $(i,J) \in \mathcal{A}$ and $t \in T$.
If $\bar{p}_{iJ}^{(t)} > 0$, then for $n > m$ we have
\begin{equation}
\begin{split}
\tilde{g}_{iJ}^{(t)}[n] &= \sum_{l=1}^m\mu_l[n]g_{iJ}^{(t)}[l] 
    + \sum_{l=m+1}^n\mu_l[n]g_{iJ}^{(t)}[l] \\
  &= \sum_{l=1}^m\mu_l[n]g_{iJ}^{(t)}[l] 
    + \sum_{l=m+1}^n\frac{\mu_l[n]}{\theta[n]}
      (p_{iJ}^{(t)}[n+1] - p_{iJ}^{(t)}[n] - c_{iJ}[n]) \\
  &= \sum_{l=1}^m\mu_l[n]g_{iJ}^{(t)}[l] 
    + \sum_{l=m+1}^n\gamma_{ln}(p_{iJ}^{(t)}[n+1] - p_{iJ}^{(t)}[n]) 
    - \sum_{l=m+1}^n\gamma_{ln}c_{iJ}[n] .
\end{split}
\label{eqn:3620}
\end{equation}
Otherwise, if $\bar{p}_{iJ}^{(t)} = 0$, then from equation
(\ref{eqn:3615}), we have
\begin{equation*}
p_{iJ}^{(t)}[n+1] \ge p_{iJ}^{(t)}[n] + \theta[n]g_{iJ}^{(t)}[n]
+ c_{iJ}[n] ,
\end{equation*}
so
\begin{equation}
\tilde{g}_{iJ}^{(t)}[n] \le \sum_{l=1}^m\mu_l[n]g_{iJ}^{(t)}[l] 
    + \sum_{l=m+1}^n\gamma_{ln}(p_{iJ}^{(t)}[n+1] - p_{iJ}^{(t)}[n]) 
    - \sum_{l=m+1}^n\gamma_{ln}c_{iJ}[n] .
\label{eqn:3630}
\end{equation}

It is straightforward to see that the sequence of iterates
$\{\tilde{x}[n]\}$ is primal feasible, and that we obtain a primal
feasible sequence $\{z[n]\}$ by setting
\[
\begin{split}
z_{iJ_m^{(i)}}[n]
&:= \max_{t \in T} \left\{
\sum_{k \in J_{M_i}^{(i)} \setminus J_{m-1}^{(i)}}
  \tilde{x}_{ik}^{(t)}[n] \right\}
- \sum_{m^\prime = m+1}^{M_i} z_{iJ_{m^\prime}^{(i)}}[n] \\
&= \max_{t \in T} \tilde{g}_{iJ_m^{(i)}} 
- \sum_{m^\prime = m+1}^{M_i} z_{iJ_{m^\prime}^{(i)}}[n]
\end{split}
\]
recursively, starting from $m = M_i$ and proceeding through to $m=1$.
Sherali and Choi \cite{shc96} showed that, if the required conditions on
the step sizes $\{\theta[n]\}$ and convex combination weights
$\{\mu_l[n]\}$ are satisfied, then 
\begin{equation*}
\sum_{l=1}^m\mu_l[n]g_{iJ}^{(t)}[l]
    + \sum_{l=m+1}^n\gamma_{ln}(p_{iJ}^{(t)}[n+1] - p_{iJ}^{(t)}[n])
\rightarrow 0
\end{equation*}
as $k \rightarrow \infty$;
hence we see from equations (\ref{eqn:3620}) and (\ref{eqn:3630}) that, 
for $k$ sufficiently large, 
\begin{equation*}
\sum_{m^\prime=m}^{M_i} z_{iJ_{m^\prime}^{(i)}}[n] = 
- \sum_{l=m+1}^n \gamma_{ln} c_{iJ_m^{(i)}}[n] .
\end{equation*}
Recalling the primal problem (\ref{eqn:3540}), we see that
complementary slackness with $\bar{p}$
holds in the limit of any convergent subsequence
of $\{\tilde{x}[n]\}$.
\end{proof}

The required conditions on the step sizes and convex combination
weights are satisfied by the following choices
(\cite[Corollaries 2--4]{shc96}):
\begin{enumerate}
\item step sizes $\{\theta[n]\}$ such that
$\theta[n] > 0$, $\lim_{n \rightarrow 0} \theta[n] = 0$,
$\sum_{n = 1}^\infty \theta_n = \infty$, and convex combination weights
$\{\mu_l[n]\}$ given by
$\mu_l[n] = \theta[l] / \sum_{k=1}^n\theta[k]$ for all
$l = 1, \ldots, n$, $n = 0,1,\ldots$;
\item step sizes $\{\theta[n]\}$ given by
$\theta[n] = a/(b + cn)$ for all $n = 0,1,\ldots$,
where $a > 0$, $b \ge 0$ and $c > 0$,
and convex combination weights $\{\mu_l[n]\}$ given by
$\mu_l[n] = 1/n$ for all
$l = 1, \ldots, n$, $n = 0,1,\ldots$; and
\item step sizes $\{\theta[n]\}$ given by
$\theta[n] = n^{-\alpha}$ for all $n = 0,1,\ldots$, where
$0 < \alpha < 1$,
and convex combination weights $\{\mu_l[n]\}$ given by
$\mu_l[n] = 1/n$ for all
$l = 1, \ldots, n$, $n = 0,1,\ldots$.
\end{enumerate}
Moreover, for all three choices, we have $\mu_l[n+1]/\mu_l[n]$
independent of $l$ for all $n$, so primal iterates can be computed
iteratively using
\[
\begin{split}
\tilde{x}[n] &= \sum_{l=1}^n\mu_l[n]\hat{x}[l] \\
&= \sum_{l=1}^{n-1}\mu_l[n]\hat{x}[l] + \mu_n[n]\hat{x}[n] \\
&= \phi[n-1]\tilde{x}[n-1] + \mu_n[n]\hat{x}[n] ,
\end{split}
\]
where $\phi[n] := \mu_l[n+1]/\mu_l[n]$.

\begin{figure}
\centering
\framebox{
\begin{minipage}{0.88\textwidth}
\begin{enumerate}
\item Each node $i$ computes $s_{iJ}$ for its outgoing
hyperarcs and initializes $p_{iJ}[0]$ to a point in the
feasible set of (\ref{eqn:3530}).
For example, we take
$p_{iJ}^{(t)}[0] := {s_{iJ}}/{|T|}$.
Each node $i$ sends
$s_{iJ}$ and $p_{iJ}[0]$ over each outgoing hyperarc $(i,J)$.

\item At the $n$th iteration, use $p^{(t)}[n]$ as the hyperarc costs and
run a distributed shortest path algorithm, such as distributed
Bellman-Ford, to determine
$\hat{x}^{(t)}[n]$ for all $t\in T$. 
\label{sg_loop_beg}

\item Each node $i$ computes $p_{iJ}[n+1]$ for its outgoing hyperarcs
using Proposition~\ref{prop:200}.
Each node $i$ sends $p_{iJ}[n+1]$ over each outgoing hyperarc $(i,J)$.

\item Nodes compute the primal
iterate $\tilde{x}[n]$ by setting
\begin{equation*}
\tilde{x}[n] := \sum_{l=1}^n \mu_l[n] \hat{x}[l].
\end{equation*}

\item The current coding subgraph $z[n]$ is computed using the
primal iterate $\tilde{x}[n]$.
For each node $i$, we set
\[
z_{iJ_m^{(i)}}[n]
:= \max_{t \in T} \left\{
\sum_{k \in J_{M_i}^{(i)} \setminus J_{m-1}^{(i)}}
  \tilde{x}_{ik}^{(t)}[n] \right\}
- \sum_{m^\prime = m+1}^{M_i} z_{iJ_{m^\prime}^{(i)}}[n]
\]
recursively, starting from $m = M_i$ and proceeding through to $m=1$.
\label{sg_loop_end}

\item Steps \ref{sg_loop_beg}--\ref{sg_loop_end}
are repeated until the sequence of 
primal iterates $\{\tilde{x}[n]\}$ converges.
\end{enumerate}
\end{minipage} }
\caption{Summary of the subgradient method.}
\label{fig:summary_subgradient}
\end{figure}

This gives us our distributed algorithm.  We summarize the subgradient
method in Figure~\ref{fig:summary_subgradient}.  We see that, although
the method is indeed a distributed algorithm, it again operates in 
synchronous rounds.  
Again, we expect that this synchronicity
can be relaxed in practice, but this issue remains to be
investigated.

\section{Dynamic multicast}
\label{sec:dynamic_multicast}

In many applications, membership of the multicast group
changes in time, with nodes joining and leaving the group, rather than
remaining constant for the duration of the connection, as we have thus
far assumed.  
Under these
dynamic conditions, we often cannot simply re-establish the connection
with every membership change because doing so would cause an
unacceptable disruption in the service being delivered
to those nodes remaining in the group.  A good example of an application
where such issues arise is real-time media distribution.
Thus, we desire to find minimum-cost time-varying 
subgraphs that can deliver continuous service to dynamic multicast
groups.

Although our objective is clear, our description of the problem is
currently vague.  Indeed, one of the principal hurdles to tackling the
problem of dynamic multicast lies in formulating the problem in such a
way that it is suitable for analysis and addresses our objective.  For
routed networks, the problem is generally formulated as the dynamic
Steiner tree problem, which was first proposed in \cite{imw91}.  Under
this formulation, the focus is on worst-case behavior and modifications
of the multicast tree are allowed only when nodes join or leave the
multicast group.  The formulation is adequate, but not
compelling---indeed, there is no compelling reason for the restriction
on when the multicast tree can be modified.

In our formulation for coded networks,
we draw some inspiration from \cite{imw91},
but we focus on expected behavior rather than worst-case behavior, and
we do not restrict modifications of the multicast subgraph to when nodes
join or leave the multicast tree.
We formulate the problem as follows.  

We employ a basic unit of time that
is related to the time that it takes for
changes in the multicast subgraph to settle.  In particular, suppose
that at a given time the multicast subgraph is $z$ and that it is
capable of supporting a multicast connection to
sink nodes $T$.  Then, in one unit
time, we can change the multicast subgraph to $z^\prime$, which is
capable of supporting a multicast connection to 
sink nodes $T^\prime$, without
disrupting the service being delivered to $T \cap T^\prime$ provided
that (componentwise)
$z \ge z^\prime$ or $z \le z^\prime$.  The interpretation of this
assumption is that we allow, in one time unit, only for the subgraph to
increase, meaning that any sink node receiving a particular stream will
continue to receive it (albeit with possible changes in the code,
depending on how the coding is implemented) and therefore facing no
significant disruption to service; or for the subgraph to decrease,
meaning that any sink node receiving a particular stream will be forced to
reduce to a subset of that stream, but one that is sufficient to recover
the source's transmission provided that the sink node is in $T^\prime$, and
therefore again facing no significant disruption to service.  We do not
allow for both operations to take place in a single unit of time (which
would allow for arbitrary changes) because, in that case, sink nodes
may face temporary disruptions to service when decreases to the
multicast subgraph follow too closely to increases.

\begin{figure}
\centering
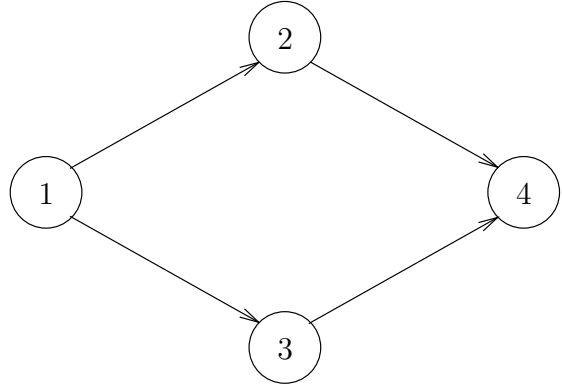
\caption{A four-node lossless network.}
\label{fig:four-node}
\end{figure}

As an example, consider the four-node lossless network shown in
Figure~\ref{fig:four-node}.
Suppose that $s = 1$ and that, at a given time, we have $T = \{2,4\}$.
We support a multicast of unit rate with the subgraph 
\[
(z_{12}, z_{13}, z_{24}, z_{34}) = (1,0,1,0).
\]
Now suppose that the group membership changes, and node
2 leaves while node 3 joins, so $T^\prime = \{3,4\}$.  
As a result, we decide that we wish to
change to the subgraph
\[
(z_{12}, z_{13}, z_{24}, z_{34}) = (0,1,0,1).
\]
If we simply make the change na{\"\i}vely in a single time unit, then
node 4 may face a temporary disruption to its service because packets on
$(2,4)$ may stop arriving before packets on $(3,4)$ start arriving.  The
assumption that we have made on allowed operations ensures that we must
first increase the subgraph to
\[
(z_{12}, z_{13}, z_{24}, z_{34}) = (1,1,1,1),
\]
allow for the change to settle by waiting for one time unit, then
decrease the subgraph to 
\[
(z_{12}, z_{13}, z_{24}, z_{34}) = (0,1,0,1).
\]
With this series of operations, node 4 maintains continuous service
throughout the subgraph change.

We discretize the time axis into time intervals of a single time unit.
We suppose that, at the beginning of each time interval, we receive
zero or more
requests from sink nodes that are not currently part of the multicast
group to join and 
zero or more
requests from sink nodes that are currently part of
the multicast group to leave.  We model these join and leave requests as
a discrete stochastic process and make the assumption that, once all the
members of the multicast group leave, the connection is over and remains
in that state forever.  Let $T_m$ denote the sink nodes in the multicast
group at the end of time interval $m$.  Then, we assume that
\begin{equation}
\lim_{m \rightarrow \infty} \Pr(T_m \neq \emptyset | T_0 = T) = 0
\label{eqn:3750}
\end{equation}
for any initial multicast group $T$.  A possible, simple model of join 
and leave requests is to model $|T_m|$ as a birth-death process with a
single absorbing state at state 0, and to choose a node uniformly from 
$\mathcal{N}^\prime \setminus T_m$, where $\mathcal{N}^\prime :=
\mathcal{N} \setminus \{s\}$,
 at each birth and from $T_m$ at each death.

Let $z^{(m)}$ be the multicast subgraph at the beginning of time
interval $m$, which, by the assumptions made thus far, means that it
supports a multicast connection to sink nodes $T_{m-1}$.  
Let $V_{m-1}$ and $W_{m-1}$ be the join and leave 
requests that arrive at the end of time interval $m-1$,
respectively.  Hence, $V_{m-1} \subset \mathcal{N}^\prime \setminus T_{m-1}$, 
$W_{m-1} \subset T_{m-1}$, and 
$T_m = (T_{m-1} \setminus W_{m-1}) \cup V_{m-1}$.
We choose $z^{(m+1)}$ from $z^{(m)}$ and $T_m$ using the function
$\mu_m$, so $z^{(m+1)} = \mu_m(z^{(m)}, T_m)$, where $z^{(m+1)}$ must
lie in a particular constraint set $U(z^{(m)}, T_m)$.

To characterize the constraint set $U(z, T)$, 
recall the optimization problem for minimum-cost multicast in
Section~\ref{sec:problem_formulation}:
\begin{equation}
\begin{aligned}
\text{minimize }   & f(z) \\
\text{subject to } & z \in Z, \\
& \sum_{j \in K} x_{iJj}^{(t)}
\le z_{iJ}b_{iJK} ,
  \qquad \text{$\forall$ $(i,J) \in \mathcal{A}$, $K \subset
J$, $t \in T$} , \\
& x^{(t)} \in F^{(t)},
  \qquad \text{$\forall$ $t \in T$} .
\end{aligned}
\label{eqn:3790}
\end{equation}
Therefore,
it follows that we can write
$U(z,T) = U_+(z,T) \cup U_-(z,T)$, where 
\begin{gather*}
U_+(z,T) = \{z^\prime \in Z(T) | z^\prime \ge z\}, \\
U_-(z,T) = \{z^\prime \in Z(T) | z^\prime \le z\},
\end{gather*}
and $Z(T)$ is the feasible set of $z$ in 
problem (\ref{eqn:3790}) for a given
$T$, i.e.,\  if we have the subgraph $z$ at the beginning of a time
interval and we must go to a subgraph that supports multicast to $T$,
then the allowable subgraphs are those that support multicast to $T$ and
either increase $z$ (those in
$U_+(z,T)$) or decrease $z$ (those in $U_-(z,T)$).

Note that, 
if we have separable constraints,
then $U(z^{(m)}, T_m) \neq \emptyset$ for all $z^{(m)} \in Z$
provided that $Z(T_m) \neq \emptyset$, i.e.,\  
from any feasible subgraph at stage $m$,
it is possible to go to a feasible subgraph
at stage $m+1$ provided that one exists for the
multicast group $T_m$.  But
while this is the case for coded networks, it is not always the case for
routed networks.  Indeed, if multiple multicast trees are being used (as
discussed in \cite{wcj04}, for example), then it is definitely possible
to find ourselves in a state where we cannot achieve multicast 
at stage $m+1$ even though static multicast to $T_m$ 
is possible using multiple multicast trees.  

\begin{figure}
\centering
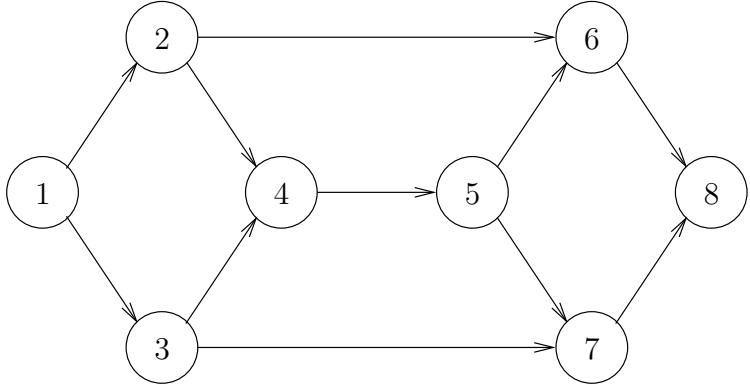
\caption{A lossless network used for dynamic multicast.}
\label{fig:buttvar}
\end{figure}

As an example of this phenomenon, consider the lossless network
depicted in Figure~\ref{fig:buttvar}.  
Suppose that each arc is of unit capacity,
that $s = 1$, and that, at a given time, we have $T =
\{6, 8\}$.  We support a multicast of rate 2 with the two trees 
$\{(1,3), (3,4), (4,5), (5,6), (5,7), (7,8)\}$ and 
$\{(1,2), (2,6), (6,8)\}$, each carrying unit rate.  Now suppose that
the group membership changes, and node 6 leaves while node 7 joins, so
$T^\prime = \{7, 8\}$.  It is clear that static multicast to $T^\prime$
is possible using multiple multicast trees (we simply
reflect the solution for $T$), but we cannot achieve multicast to
$T^\prime$ by only adding edges to the two existing trees.
Our only recourse at this stage is to abandon the existing trees and
establish new ones, which causes a disruption to node 8's service,
or to reconfigure slowly the existing trees, which causes a delay before
node 7 is actually joined to the group.

Returning to the problem at hand,
our objective is to find a policy $\pi = \{\mu_0, \mu_1, \ldots, \}$
that minimizes the cost function
\[
J_\pi(z^{(0)}, T_0)
= \lim_{M \rightarrow \infty} \mathbb{E} \left[
\sum_{m=0}^{M-1} f(z^{(m+1)}) \chi_{2^{\mathcal{N}^\prime} \setminus
\{\emptyset\}}(T_m)
\right] ,
\]
where $\chi_{2^{\mathcal{N}^\prime} \setminus \{\emptyset\}}$ is the characteristic
function for
$2^{\mathcal{N}^\prime} \setminus \{\emptyset\}$ (i.e.,\  
$\chi_{2^{\mathcal{N}^\prime} \setminus \{\emptyset\}}(T) = 1$ if $T \neq \emptyset$,
and $\chi_{2^{\mathcal{N}^\prime} \setminus \{\emptyset\}}(T) = 0$ if $T = \emptyset$).

We impose the assumption that
we have separable constraints and that 
$Z(\mathcal{N}^\prime) \neq \emptyset$,
i.e.,\  we assume that
there exists a subgraph that supports broadcast.  This
assumption ensures that the constraint set $U(z,T)$ is non-empty for all
$z \in Z$ and $T \subset \mathcal{N}^\prime$. 
Thus, from condition (\ref{eqn:3750}), it follows that there exists at
least one policy $\pi$ such that $J_\pi(z^{(0)}, T_0) < \infty$
(namely, one that uses some fixed
$z \in Z(\mathcal{N}^\prime)$ until the multicast group is empty). 

It is now not difficult to see that we are dealing with an undiscounted,
infinite-horizon dynamic programming problem 
(see, e.g.,\  \cite[Chapter 3]{ber01b}), and we can apply the theory
developed for such problems to our problem.  
So doing, 
we first note that the optimal cost function $J^* := \min_\pi J_\pi$
satisfies Bellman's equation, namely, we have
\begin{equation*}
J^*(z, T) 
=
\min_{u \in U(z, T)} \left\{ f(u)
+ \mathbb{E} [ J^*(u, (T \setminus V) \cup W) ]
\right\} 
\end{equation*}
if $T \neq \emptyset$, and $J^*(z,T) = 0$ if $T = \emptyset$.
Moreover, the optimal cost is achieved by the stationary policy $\pi =
\{\mu, \mu, \ldots\}$, where $\mu$ is given by
\begin{equation}
\mu(z, T) 
=
\argmin_{u \in U(z, T)} \left\{ f(u)
+ \mathbb{E} [ J^*(u, (T \setminus V) \cup W) ]
\right\} 
\label{eqn:3795}
\end{equation}
if $T \neq \emptyset$, and $\mu(z,T) = 0$ if $T = \emptyset$.

The fact that the optimal cost can be achieved by a stationary policy
limits the space in which we need to search for optimal policies
significantly, but we are still left with the difficulty that the state
space is uncountably large---it is the space of all possible pairs $(z, T)$,
which is $Z \times 2^{\mathcal{N}^\prime}$.
The size of the state space more or less
eliminates the possibility of using
techniques such as value iteration to obtain $J^*$.

On the other hand, given $J^*$, it does not seem at all implausible that
we can compute the optimal decision at the beginning of each time
interval using (\ref{eqn:3795}).  The constraint set is the union of two
polyhedra, which can simply be handled by optimizing over each
separately.  The objective function can pose a difficulty because, even
if $f$ is convex, it may not necessarily be convex owing to the term
$\mathbb{E} [ J^*(u, (T \setminus V) \cup W) ]$.  But, since we are
unable to obtain $J^*$ precisely on account of the large state space, we
can restrict our attention to approximations that make problem
(\ref{eqn:3795}) tractable.

For dynamic programming problems, there are many approximations that
have been developed to cope with large state spaces (see, e.g.,\ 
\cite[Section 2.3.3]{ber01b}).  In particular, we can approximate
$J^*(z, T)$ by $\tilde{J}(z, T, r)$, where $\tilde{J}(z, T, r)$ is of
some fixed form, and $r$ is a parameter vector that is determined by
some form of optimization, which can be performed offline if the graph
$\mathcal{G}$ is static.  Depending upon the approximation that is used,
we may even be able to solve problem (\ref{eqn:3795}) using the
distributed algorithms described in
Section~\ref{sec:distributed_algorithms} (or simple modifications
thereof).  The specific approximations $\tilde{J}(z, T, r)$ that we can
use and their performance are beyond the scope of this thesis.

%% file: buttcost.pdftex_t
\begin{picture}(0,0)%
\includegraphics{buttcost.eps}%
\end{picture}%
\setlength{\unitlength}{3947sp}%
\begingroup\makeatletter\ifx\SetFigFont\undefined%
\gdef\SetFigFont#1#2#3#4#5{%
  \reset@font\fontsize{#1}{#2pt}%
  \fontfamily{#3}\fontseries{#4}\fontshape{#5}%
  \selectfont}%
\fi\endgroup%
\begin{picture}(5033,2466)(1652,-2819)
\put(5851,-661){\makebox(0,0)[b]{\smash{{\SetFigFont{12}{14.4}{\familydefault}{\mddefault}{\updefault}{\color[rgb]{0,0,0}$t_1$}%
}}}}
\put(5851,-2611){\makebox(0,0)[b]{\smash{{\SetFigFont{12}{14.4}{\familydefault}{\mddefault}{\updefault}{\color[rgb]{0,0,0}$t_2$}%
}}}}
\put(2326,-1636){\makebox(0,0)[b]{\smash{{\SetFigFont{12}{14.4}{\familydefault}{\mddefault}{\updefault}{\color[rgb]{0,0,0}$s$}%
}}}}
\put(5701,-2086){\makebox(0,0)[lb]{\smash{{\SetFigFont{12}{14.4}{\familydefault}{\mddefault}{\updefault}{\color[rgb]{0,0,0}$(1/2,0,1/2)$}%
}}}}
\put(3526,-1111){\makebox(0,0)[lb]{\smash{{\SetFigFont{12}{14.4}{\familydefault}{\mddefault}{\updefault}{\color[rgb]{0,0,0}$(1/2,0,1/2)$}%
}}}}
\put(4501,-511){\makebox(0,0)[b]{\smash{{\SetFigFont{12}{14.4}{\familydefault}{\mddefault}{\updefault}{\color[rgb]{0,0,0}$(1/2,1/2,0)$}%
}}}}
\put(2776,-1036){\makebox(0,0)[rb]{\smash{{\SetFigFont{12}{14.4}{\familydefault}{\mddefault}{\updefault}{\color[rgb]{0,0,0}$(1/2,1/2,1/2)$}%
}}}}
\put(2776,-2236){\makebox(0,0)[rb]{\smash{{\SetFigFont{12}{14.4}{\familydefault}{\mddefault}{\updefault}{\color[rgb]{0,0,0}$(1/2,1/2,1/2)$}%
}}}}
\put(4501,-2761){\makebox(0,0)[b]{\smash{{\SetFigFont{12}{14.4}{\familydefault}{\mddefault}{\updefault}{\color[rgb]{0,0,0}$(1/2,0,1/2)$}%
}}}}
\put(4501,-1486){\makebox(0,0)[b]{\smash{{\SetFigFont{12}{14.4}{\familydefault}{\mddefault}{\updefault}{\color[rgb]{0,0,0}$(1/2,1/2,1/2)$}%
}}}}
\put(5701,-1111){\makebox(0,0)[lb]{\smash{{\SetFigFont{12}{14.4}{\familydefault}{\mddefault}{\updefault}{\color[rgb]{0,0,0}$(1/2,1/2,0)$}%
}}}}
\put(3526,-2086){\makebox(0,0)[lb]{\smash{{\SetFigFont{12}{14.4}{\familydefault}{\mddefault}{\updefault}{\color[rgb]{0,0,0}$(1/2,1/2,0)$}%
}}}}
\end{picture}%

%% file: wbuttcost.pdftex_t
\begin{picture}(0,0)%
\includegraphics{wbuttcost.eps}%
\end{picture}%
\setlength{\unitlength}{3947sp}%
\begingroup\makeatletter\ifx\SetFigFont\undefined%
\gdef\SetFigFont#1#2#3#4#5{%
  \reset@font\fontsize{#1}{#2pt}%
  \fontfamily{#3}\fontseries{#4}\fontshape{#5}%
  \selectfont}%
\fi\endgroup%
\begin{picture}(4315,2466)(2093,-2819)
\put(2326,-1636){\makebox(0,0)[b]{\smash{{\SetFigFont{12}{14.4}{\familydefault}{\mddefault}{\updefault}{\color[rgb]{0,0,0}$s$}%
}}}}
\put(5851,-661){\makebox(0,0)[b]{\smash{{\SetFigFont{12}{14.4}{\familydefault}{\mddefault}{\updefault}{\color[rgb]{0,0,0}$t_1$}%
}}}}
\put(5851,-2611){\makebox(0,0)[b]{\smash{{\SetFigFont{12}{14.4}{\familydefault}{\mddefault}{\updefault}{\color[rgb]{0,0,0}$t_2$}%
}}}}
\put(4276,-1111){\makebox(0,0)[lb]{\smash{{\SetFigFont{12}{14.4}{\familydefault}{\mddefault}{\updefault}{\color[rgb]{0,0,0}$(0,1/2)$}%
}}}}
\put(4276,-2086){\makebox(0,0)[lb]{\smash{{\SetFigFont{12}{14.4}{\familydefault}{\mddefault}{\updefault}{\color[rgb]{0,0,0}$(1/2,0)$}%
}}}}
\put(3826,-2761){\makebox(0,0)[b]{\smash{{\SetFigFont{12}{14.4}{\familydefault}{\mddefault}{\updefault}{\color[rgb]{0,0,0}$1/2$}%
}}}}
\put(3826,-511){\makebox(0,0)[b]{\smash{{\SetFigFont{12}{14.4}{\familydefault}{\mddefault}{\updefault}{\color[rgb]{0,0,0}$1/2$}%
}}}}
\put(3001,-1036){\makebox(0,0)[rb]{\smash{{\SetFigFont{12}{14.4}{\familydefault}{\mddefault}{\updefault}{\color[rgb]{0,0,0}$(1/2,1/2)$}%
}}}}
\put(3001,-2236){\makebox(0,0)[rb]{\smash{{\SetFigFont{12}{14.4}{\familydefault}{\mddefault}{\updefault}{\color[rgb]{0,0,0}$(1/2,1/2)$}%
}}}}
\put(5026,-1486){\makebox(0,0)[b]{\smash{{\SetFigFont{12}{14.4}{\familydefault}{\mddefault}{\updefault}{\color[rgb]{0,0,0}$1/2$}%
}}}}
\put(5701,-1111){\makebox(0,0)[lb]{\smash{{\SetFigFont{12}{14.4}{\familydefault}{\mddefault}{\updefault}{\color[rgb]{0,0,0}$(1/2,0)$}%
}}}}
\put(5701,-2086){\makebox(0,0)[lb]{\smash{{\SetFigFont{12}{14.4}{\familydefault}{\mddefault}{\updefault}{\color[rgb]{0,0,0}$(0,1/2)$}%
}}}}
\put(2701,-1486){\makebox(0,0)[b]{\smash{{\SetFigFont{12}{14.4}{\familydefault}{\mddefault}{\updefault}{\color[rgb]{0,0,0}$1$}%
}}}}
\put(4951,-511){\makebox(0,0)[b]{\smash{{\SetFigFont{12}{14.4}{\familydefault}{\mddefault}{\updefault}{\color[rgb]{0,0,0}$(1/2,0)$}%
}}}}
\put(4951,-2761){\makebox(0,0)[b]{\smash{{\SetFigFont{12}{14.4}{\familydefault}{\mddefault}{\updefault}{\color[rgb]{0,0,0}$(0,1/2)$}%
}}}}
\end{picture}%

%% file: ex_region.pdftex_t
\begin{picture}(0,0)%
\includegraphics{ex_region.eps}%
\end{picture}%
\setlength{\unitlength}{3947sp}%
\begingroup\makeatletter\ifx\SetFigFont\undefined%
\gdef\SetFigFont#1#2#3#4#5{%
  \reset@font\fontsize{#1}{#2pt}%
  \fontfamily{#3}\fontseries{#4}\fontshape{#5}%
  \selectfont}%
\fi\endgroup%
\begin{picture}(3409,3216)(1254,-6409)
\put(4651,-4636){\makebox(0,0)[lb]{\smash{{\SetFigFont{12}{14.4}{\familydefault}{\mddefault}{\updefault}{\color[rgb]{0,0,0}$\frac{1}{4}z_{1(23)}(1-z_{23})+\frac{3}{4}(1-z_{1(23)})z_{23}=\frac{1}{8}$}%
}}}}
\put(4651,-4111){\makebox(0,0)[lb]{\smash{{\SetFigFont{12}{14.4}{\familydefault}{\mddefault}{\updefault}{\color[rgb]{0,0,0}$\frac{13}{16}z_{1(23)}(1-z_{23})=\frac{1}{8}$}%
}}}}
\put(3301,-5011){\makebox(0,0)[lb]{\smash{{\SetFigFont{12}{14.4}{\familydefault}{\mddefault}{\updefault}{\color[rgb]{0,0,0}$Z_0$}%
}}}}
\put(1763,-4948){\makebox(0,0)[rb]{\smash{{\SetFigFont{10}{12.0}{\familydefault}{\mddefault}{\updefault} 0.4}}}}
\put(1763,-3848){\makebox(0,0)[rb]{\smash{{\SetFigFont{10}{12.0}{\familydefault}{\mddefault}{\updefault} 0.8}}}}
\put(1763,-3298){\makebox(0,0)[rb]{\smash{{\SetFigFont{10}{12.0}{\familydefault}{\mddefault}{\updefault} 1}}}}
\put(1838,-6173){\makebox(0,0)[b]{\smash{{\SetFigFont{10}{12.0}{\familydefault}{\mddefault}{\updefault} 0}}}}
\put(2388,-6173){\makebox(0,0)[b]{\smash{{\SetFigFont{10}{12.0}{\familydefault}{\mddefault}{\updefault} 0.2}}}}
\put(2938,-6173){\makebox(0,0)[b]{\smash{{\SetFigFont{10}{12.0}{\familydefault}{\mddefault}{\updefault} 0.4}}}}
\put(3488,-6173){\makebox(0,0)[b]{\smash{{\SetFigFont{10}{12.0}{\familydefault}{\mddefault}{\updefault} 0.6}}}}
\put(4038,-6173){\makebox(0,0)[b]{\smash{{\SetFigFont{10}{12.0}{\familydefault}{\mddefault}{\updefault} 0.8}}}}
\put(1387,-4611){\rotatebox{90.0}{\makebox(0,0)[b]{\smash{{\SetFigFont{10}{12.0}{\familydefault}{\mddefault}{\updefault}$z_{23}$}}}}}
\put(3213,-6360){\makebox(0,0)[b]{\smash{{\SetFigFont{10}{12.0}{\familydefault}{\mddefault}{\updefault}$z_{1(23)}$}}}}
\put(4588,-6173){\makebox(0,0)[b]{\smash{{\SetFigFont{10}{12.0}{\familydefault}{\mddefault}{\updefault} 1}}}}
\put(1763,-6048){\makebox(0,0)[rb]{\smash{{\SetFigFont{10}{12.0}{\familydefault}{\mddefault}{\updefault} 0}}}}
\put(1763,-5498){\makebox(0,0)[rb]{\smash{{\SetFigFont{10}{12.0}{\familydefault}{\mddefault}{\updefault} 0.2}}}}
\put(1763,-4398){\makebox(0,0)[rb]{\smash{{\SetFigFont{10}{12.0}{\familydefault}{\mddefault}{\updefault} 0.6}}}}
\end{picture}%

%% file: four_node.pdftex_t
\begin{picture}(0,0)%
\includegraphics{four_node.eps}%
\end{picture}%
\setlength{\unitlength}{3947sp}%
\begingroup\makeatletter\ifx\SetFigFont\undefined%
\gdef\SetFigFont#1#2#3#4#5{%
  \reset@font\fontsize{#1}{#2pt}%
  \fontfamily{#3}\fontseries{#4}\fontshape{#5}%
  \selectfont}%
\fi\endgroup%
\begin{picture}(3466,2416)(1868,-2769)
\put(3601,-661){\makebox(0,0)[b]{\smash{{\SetFigFont{12}{14.4}{\familydefault}{\mddefault}{\updefault}{\color[rgb]{0,0,0}2}%
}}}}
\put(2101,-1636){\makebox(0,0)[b]{\smash{{\SetFigFont{12}{14.4}{\familydefault}{\mddefault}{\updefault}{\color[rgb]{0,0,0}1}%
}}}}
\put(3601,-2611){\makebox(0,0)[b]{\smash{{\SetFigFont{12}{14.4}{\familydefault}{\mddefault}{\updefault}{\color[rgb]{0,0,0}3}%
}}}}
\put(5101,-1636){\makebox(0,0)[b]{\smash{{\SetFigFont{12}{14.4}{\familydefault}{\mddefault}{\updefault}{\color[rgb]{0,0,0}4}%
}}}}
\end{picture}%

%% file: buttvar.pdftex_t
\begin{picture}(0,0)%
\includegraphics{buttvar.eps}%
\end{picture}%
\setlength{\unitlength}{3947sp}%
\begingroup\makeatletter\ifx\SetFigFont\undefined%
\gdef\SetFigFont#1#2#3#4#5{%
  \reset@font\fontsize{#1}{#2pt}%
  \fontfamily{#3}\fontseries{#4}\fontshape{#5}%
  \selectfont}%
\fi\endgroup%
\begin{picture}(4666,2416)(1868,-2769)
\put(5551,-2611){\makebox(0,0)[b]{\smash{{\SetFigFont{12}{14.4}{\familydefault}{\mddefault}{\updefault}{\color[rgb]{0,0,0}7}%
}}}}
\put(2101,-1636){\makebox(0,0)[b]{\smash{{\SetFigFont{12}{14.4}{\familydefault}{\mddefault}{\updefault}{\color[rgb]{0,0,0}1}%
}}}}
\put(4801,-1636){\makebox(0,0)[b]{\smash{{\SetFigFont{12}{14.4}{\familydefault}{\mddefault}{\updefault}{\color[rgb]{0,0,0}5}%
}}}}
\put(5551,-661){\makebox(0,0)[b]{\smash{{\SetFigFont{12}{14.4}{\familydefault}{\mddefault}{\updefault}{\color[rgb]{0,0,0}6}%
}}}}
\put(6301,-1636){\makebox(0,0)[b]{\smash{{\SetFigFont{12}{14.4}{\familydefault}{\mddefault}{\updefault}{\color[rgb]{0,0,0}8}%
}}}}
\put(2851,-2611){\makebox(0,0)[b]{\smash{{\SetFigFont{12}{14.4}{\familydefault}{\mddefault}{\updefault}{\color[rgb]{0,0,0}3}%
}}}}
\put(3601,-1636){\makebox(0,0)[b]{\smash{{\SetFigFont{12}{14.4}{\familydefault}{\mddefault}{\updefault}{\color[rgb]{0,0,0}4}%
}}}}
\put(2851,-661){\makebox(0,0)[b]{\smash{{\SetFigFont{12}{14.4}{\familydefault}{\mddefault}{\updefault}{\color[rgb]{0,0,0}2}%
}}}}
\end{picture}%

%% file: chap4.tex
\chapter{Performance Evaluation}
\label{chap:performance_evaluation}

\lettrine{I}{n the} preceding two chapters, we laid out a solution to the
efficient operation problem for coded packet networks.  The solution we
described has several attractive properties.  In particular, it can be
computed in a distributed manner and, in many cases, it is possible to
solve the problem, as we have defined it in
Section~\ref{sec:network_model}, optimally for a single multicast
connection.  But the question remains, is it actually useful?  Is there
a compelling reason to abandon the routed approach, with which we have
so much experience, in favor of a new one?

We believe that for some applications the answer to both questions is
yes and, in this chapter, we report on the results of several
simulations that we conducted to assess the performance of the proposed
techniques in situations of interest.  Specifically, we consider three
problems:
\begin{enumerate}
\item minimum-transmission wireless unicast:
the problem of establishing a unicast connection in a lossy 
wireless network using the minimum number of transmissions per message
packet;
\item minimum-weight wireline multicast:
the problem of establishing a multicast connection in a lossless
wireline network using the minimum weight, or artificial cost, per
message packet;
\item minimum-energy wireless multicast:
the problem of establishing a multicast connection in a lossless
wireless network using the minimum amount of energy per message
packet.
\end{enumerate}
We deal with these problems in
Sections~\ref{sec:min_transmission}, \ref{sec:min_weight},
and~\ref{sec:min_energy}, respectively.  
We find that lossy wireless networks generally offer the most potential
for the proposed techniques to improve on existing ones and that
these improvements can indeed be significant.

\section{Minimum-transmission wireless unicast}
\label{sec:min_transmission}

Establishing a unicast connection in a lossy wireless network is not
trivial.  Packets are frequently lost, and some mechanism to ensure
reliable communication is required.  Such a mechanism should not send
packets unnecessarily, and we therefore consider the objective of
minimizing the total number of transmissions per message packet.

There are numerous approaches to
wireless unicast; we consider five, three of which
(approaches \ref{end_retrans}--\ref{link_retrans})
are routed approaches and two of which
(approaches \ref{path_coding} and \ref{end_coding}) are coded
approaches:
\begin{enumerate}

\item \textbf{End-to-end retransmission:} A path is chosen from source
to sink, and packets are acknowledged by the sink, or destination node.
If the acknowledgment for a packet is not received by the source, the
packet is retransmitted.  This represents the situation where
reliability is provided by a retransmission scheme above the link layer,
e.g.,~by the transmission control protocol (\textsc{tcp}) at the
transport layer, and no mechanism for reliability is present at the link
layer.
\label{end_retrans}

\item \textbf{End-to-end coding:} A path is chosen from source to sink,
and an end-to-end forward error correction (\textsc{fec}) code, such as
a Reed-Solomon code, an \textsc{lt} code \cite{lub02}, or a Raptor code
\cite{may02, sho04}, is used to correct for packets lost between source
and sink.  This is the Digital Fountain approach to reliability 
\cite{blm02}.

\item \textbf{Link-by-link retransmission:} A path is chosen from source
to sink, and \textsc{arq} is used at the link layer to
request the retransmission of packets lost on every link in the path.
Thus, on every link, packets are acknowledged by the intended receiver
and, if the acknowledgment for a packet is not received by the sender,
the packet is retransmitted.
\label{link_retrans}

\item \textbf{Path coding:} A path is chosen from source to sink, and
every node on the path employs coding to correct for lost packets.  The
most straightforward way of doing this is for each node to use an
\textsc{fec} code, decoding and re-encoding packets it
receives.  The drawback of such an approach is delay.  Every node
on the path codes and decodes packets in a block.  A way of overcoming
this drawback is to use codes that operate in more of a
``convolutional'' manner, sending out coded packets formed from packets
received thus far, without decoding.  The random linear coding scheme
of Section~\ref{sec:coding_scheme} is such a code.  A variation, with 
lower complexity, is described in \cite{pfs05}.
\label{path_coding}

\item \textbf{Full coding:} In this case, paths are eschewed altogether,
and we use our solution to the efficient operation problem.  
Problem (\ref{eqn:3210}) is solved to find a subgraph,
and the random linear coding scheme of Section~\ref{sec:coding_scheme}
is used.  This represents the limit of
achievability provided that we are restricted from modifying the design
of the physical layer and that we do not exploit the timing of packets
to convey information.
\label{end_coding}

\end{enumerate} 

\subsection{Simulation set-up}

Nodes were placed randomly according
to a uniform distribution over a square region.  The size of the
square was set to achieve unit node density.  We considered a network
where transmissions were subject to distance attenuation and Rayleigh
fading, but not interference (owing to scheduling).  So, when node $i$
transmits, the signal-to-noise ratio (\textsc{snr}) 
of the signal received at node $j$ is 
$\gamma d(i,j)^{-\alpha}$,
where $\gamma$ is an exponentially-distributed random variable with unit
mean, $d(i,j)$ is the distance between node $i$ and node $j$, and
$\alpha$ is an attenuation parameter that we took to be 2.
We assumed that a packet transmitted by node $i$ is successfully
received by node $j$ if the received \textsc{snr} exceeds $\beta$, i.e.,
$\gamma d(i,j)^{-\alpha} \ge \beta$,
where $\beta$ is a threshold that we took to be $1/4$.  If a packet is
not successfully received, then it is completely lost.
If acknowledgments are sent, acknowledgments are subject to loss in
the same way that packets are and follow the reverse path.

\subsection{Simulation results}

\begin{figure}
\begin{center}
\includegraphics{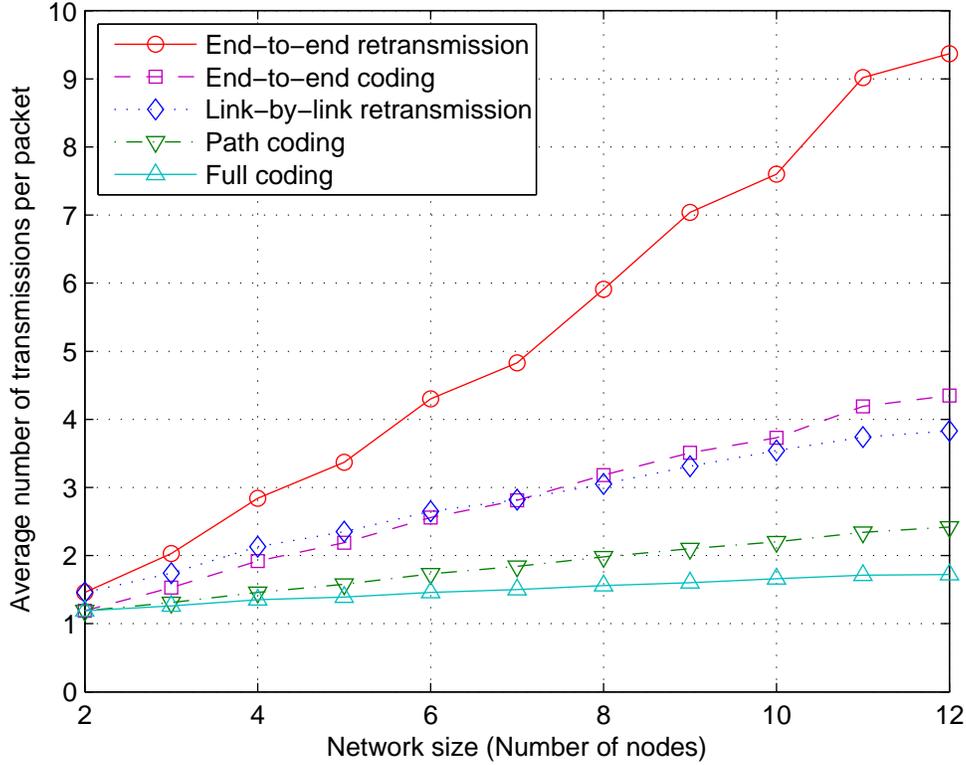}
\end{center}
\caption{Average number of transmissions per packet as a function of
network size for various wireless unicast approaches.}
\label{fig:wucostplot}
\end{figure}

The average number of transmissions required per packet using the
various approaches in random networks of varying size is shown in
Figure~\ref{fig:wucostplot}.    Paths or subgraphs were chosen in each
random instance to minimize the total number of transmissions required,
except in the cases of end-to-end retransmission and end-to-end coding,
where they were chosen to minimize the number of transmissions required
by the source node (the optimization to minimize the total number of
transmissions in these cases cannot be done straightforwardly by a
shortest path algorithm).  We see that, while end-to-end coding and
link-by-link retransmission already represent significant improvements
on end-to-end retransmission, the coded approaches represent
more significant improvements still.  By a network size of nine nodes,
full coding already improves on link-by-link retransmission by a factor
of two.  Moreover, as the network size grows, the performance of the
various schemes diverges.  

Here, we discuss performance simply in terms
of the number of transmissions required per packet; in some cases,
e.g.,~congestion, the performance measure increases super-linearly in
this quantity, and the performance improvement is even greater than that
depicted in Figure~\ref{fig:wucostplot}.  We see, at any rate, that our
prescription for efficient operation promises significant improvements,
particularly for large networks.  

\section{Minimum-weight wireline multicast}
\label{sec:min_weight}

A common networking problem is that of minimizing the weight of a
multicast connection in a lossless wireline network, where the weight of
the connection is determined by weights, or artificial costs, placed on
links to direct the flow of traffic.  Since we consider a wireline
network, the links are all point-to-point and all hyperarcs are simple
arcs.  The cost function is linear and separable, namely, it is $f(z) =
\sum_{(i,j) \in \mathcal{A}} a_{ij}z_{ij}$, where $a_{ij}$ is the weight
of the link represented by arc $(i,j)$.  The constraint set $Z$ is the
entire positive orthant, since it is generally assumed that the rate of
the connection is much smaller than the capacity of the network.

For routed networks, the standard approach to establishing
minimum-weight multicast connections is to find the shortest tree rooted
at the source that reaches all the sinks, which equates to solving the
Steiner tree problem on directed graphs \cite{ram96}.  For coded
networks, we see that optimization problem (\ref{eqn:3222}) is, in this
case, a linear optimization problem and, as such, admits a
polynomial-time solution.  By contrast, the Steiner tree problem on
directed graphs is well-known to be \textsc{np}-complete.  Although
tractable approximation algorithms exist for the Steiner tree problem on
directed graphs (e.g.,\  \cite{ccc99, ram96, zok02}), the solutions
thus obtained are suboptimal relative to minimum-weight multicast without
coding, which in turn is suboptimal relative to when coding is used,
since coding subsumes forwarding and replicating.  Thus, coding promises
potentially significant weight improvements.  

\subsection{Simulation set-up}

We conducted simulations where we took graphs representing various
internet service provider (\textsc{isp}) networks and assessed the
average total weight of random multicast connections using, first, our
network-coding based solution to the efficient operation problem
and, second, routing over the
tree given by the directed Steiner tree (\textsc{dst}) approximation
algorithm described in~\cite{ccc99}.  The graphs, and their associated
link weights, were obtained from the Rocketfuel project of the
University of Washington \cite{msw02}.  The approximation algorithm in
\cite{ccc99} was chosen for comparison as it achieves a poly-logarithmic
approximation ratio (it achieves an approximation ratio of
$O(\log^2|T|)$, where $|T|$ is the number of sink nodes), which is
roughly as good as can be expected from any practical algorithm, since
it has been shown that it is highly unlikely that there exists a
polynomial-time algorithm that can achieve an approximation factor
smaller than logarithmic \cite{ram96}. 

\subsection{Simulation results}
\label{sec:min_weight_sim_results}

\begin{table}
\centering
\begin{tabular}{|c|c|c|c|c|c|} \hline
Network & Approach &
\multicolumn{4}{c|}{Average multicast weight} \\ \cline{3-6}
& & 2 sinks & 4 sinks & 8 sinks & 16 sinks \\ \hline
Telstra (au) & \textsc{dst} approximation &
17.0 & 28.9 & 41.7 & 62.8 \\
& Network coding &
13.5 & 21.5 & 32.8 & 48.0 \\ \hline
Sprint (us) & \textsc{dst} approximation &
30.2 & 46.5 & 71.6 & 127.4 \\ 
& Network coding &
22.3 & 35.5 & 56.4 & 103.6 \\ \hline
Ebone (eu) & \textsc{dst} approximation & 
28.2 & 43.0 & 69.7 & 115.3 \\
& Network coding &
20.7 & 32.4 & 50.4 & 77.8 \\ \hline
Tiscali (eu) & \textsc{dst} approximation &
32.6 & 49.9 & 78.4 & 121.7 \\
& Network coding &
24.5 & 37.7 & 57.7 & 81.7 \\ \hline
Exodus (us) & \textsc{dst} approximation &
43.8 & 62.7 & 91.2 & 116.0 \\
& Network coding &
33.4 & 49.1 &  68.0 &  92.9 \\ \hline 
Abovenet (us) & \textsc{dst} approximation &
27.2 & 42.8 & 67.3 & 75.0 \\
& Network coding &
21.8 & 33.8 & 60.0 & 67.3 \\ \hline
\end{tabular}
\caption{Average weights of random multicast connections of unit rate and
varying size for various approaches
in graphs representing various \textsc{isp} networks.}
\label{tab:1}
\end{table}

The results of the simulations are tabulated in Table~\ref{tab:1}.  We
see that, depending on the network and the size of the multicast group,
the average weight reduction ranges from 10\% to 33\%.  Though these
reductions are modest, it is important to keep in mind that our
solution easily accommodates distributed operation and, by contrast,
computing Steiner trees is generally done at a single point with full
network knowledge.

\section{Minimum-energy wireless multicast} \label{sec:min_energy}

Another problem of interest is that of minimum-energy multicast (see,
e.g.,\  \cite{lia02,wne02b}).  In this problem, we wish to achieve
minimum-energy multicast in a lossless wireless network without explicit
regard for throughput or bandwidth, so the constraint set $Z$ is again
the entire positive orthant.  The cost function is linear and separable,
namely, it is $f(z) = \sum_{(i,J) \in \mathcal{A}} a_{iJ}z_{iJ}$, where
$a_{iJ}$ represents the energy required to transmit a packet to nodes in
$J$ from node $i$.  Hence problem (\ref{eqn:3222}) becomes a linear
optimization problem with a polynomial number of constraints, which can
therefore be solved in polynomial time.  By contrast, the same problem
using traditional routing-based approaches is \textsc{np}-complete---in
fact, the special case of broadcast in itself is \textsc{np}-complete, a
result shown in \cite{lia02, ams02}.  The problem must therefore be
addressed using polynomial-time heuristics such as the Multicast
Incremental Power (\textsc{mip}) algorithm proposed in \cite{wne02b}.

\subsection{Simulation set-up}

We conducted simulations where we placed nodes randomly, according to a
uniform distribution, in a $10 \times 10$ square with a radius of
connectivity of 3 and assessed the average total energy of random
multicast connections using first, our network-coding based
solution to the efficient operation problem
and, second, the routing solution given by the \textsc{mip} 
algorithm.
The energy required to transmit at unit rate to a distance $d$ was taken
to be $d^2$.  

\subsection{Simulation results}

\begin{table*}
\centering
\begin{tabular}{|c|c|c|c|c|c|} \hline
Network size & Approach &
\multicolumn{4}{c|}{Average multicast energy} \\ \cline{3-6}
& & 2 sinks & 4 sinks & 8 sinks & 16 sinks \\ \hline
20 nodes & \textsc{mip} algorithm &
30.6 & 33.8 & 41.6 & 47.4 \\
& Network coding &
15.5 &  23.3 &  29.9 &  38.1 \\ \hline
30 nodes & \textsc{mip} algorithm &
26.8 & 31.9 & 37.7 & 43.3 \\
& Network coding &
15.4 &  21.7 &  28.3 &  37.8 \\ \hline
40 nodes & \textsc{mip} algorithm &
24.4 & 29.3 & 35.1 & 42.3 \\
& Network coding &
14.5 &  20.6 &  25.6 &  30.5 \\ \hline
50 nodes & \textsc{mip} algorithm &
22.6 & 27.3 & 32.8 & 37.3 \\
& Network coding &
12.8 &  17.7 &  25.3 &  30.3 \\ \hline
\end{tabular}
\caption{Average energies of random multicast connections of unit rate
and varying size for various approaches
in random wireless networks of varying size.}
\label{tab:2}
\end{table*}

The results of the simulations are tabulated in Table~\ref{tab:2}.  We
see that, depending on the size of the network and the size of the
multicast group, the average energy reduction ranges from 13\% to 49\%.
These reductions are more substantial than those reported in
Section~\ref{sec:min_weight_sim_results}, but are still somewhat modest.
Again, it is important to keep in mind that our solution easily
accommodates distributed operation.

\begin{sidewaystable}
\centering
\begin{tabular}{|c|c|c|c|c|c|c|} \hline
Network size & Number of sinks & \multicolumn{5}{c|}{Average multicast
energy} \\ \cline{3-7}
& & Optimal & 25 iterations & 50 iterations & 75 iterations & 100
iterations \\ \hline
30 nodes & 2  &               16.2 &                  16.7 &          16.3 &
16.3 &          16.2 \\
& 4 &                21.8 &                  24.0 &          22.7 &
22.3 &          22.1 \\
& 8 &                27.8 &                  31.9 &          29.9 &
29.2 &          28.8 \\ \hline
40 nodes & 2 &                14.4 &                  15.0 & 14.5 &
14.5 &          14.4 \\
& 4 &                18.9 &                  21.8 &          21.2 &
19.6 &          19.4 \\
& 8 &                25.6 &                  31.5 &          29.2 &
28.0 &          27.4 \\ \hline
50 nodes & 2 &                12.4 &                  13.1 & 12.6 &
12.5 &          12.5 \\
& 4 &                17.4 &                  20.7 &          18.9 &
18.2 &          18.0 \\
& 8 &                22.4 &                  29.0 &          26.8 &
25.5 &          24.8 \\ \hline
\end{tabular}
\caption[Average energies of random multicast connections of unit rate
and varying size for the subgradient method 
in random wireless networks of varying size.]
{Average energies of random multicast connections of unit rate
and varying size for the subgradient method 
in random wireless networks of varying size.  The optimal energy was
obtained using a linear program solver.}
\label{tab:3}
\end{sidewaystable}

\begin{figure}
\begin{center}
\includegraphics{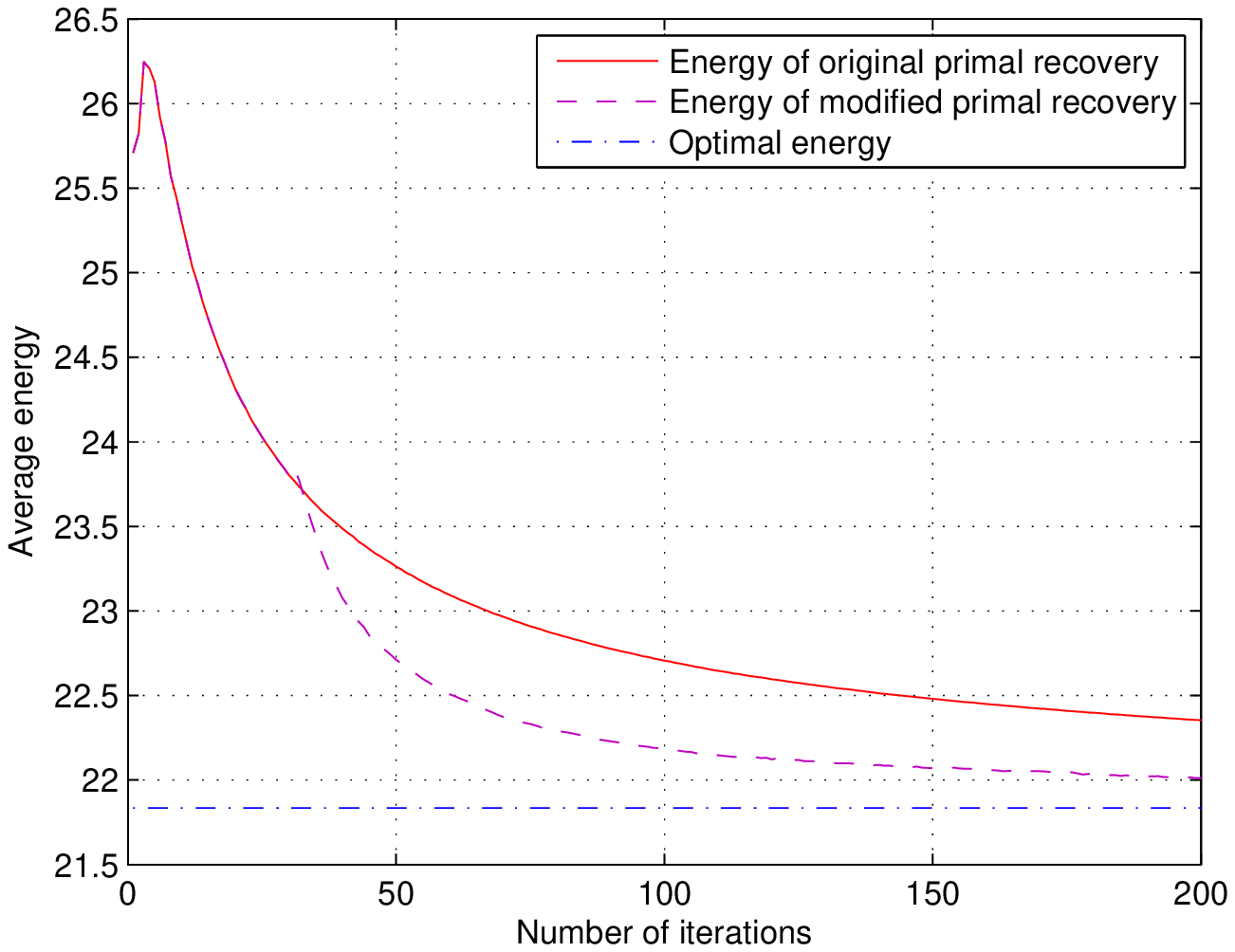}
\end{center}
\caption{Average energy as a function of the number of iterations for
the subgradient method on random 4-sink multicast connections
of unit rate in random 30-node wireless networks.}
\label{fig:fig1a}
\end{figure}

\begin{figure}
\begin{center}
\includegraphics{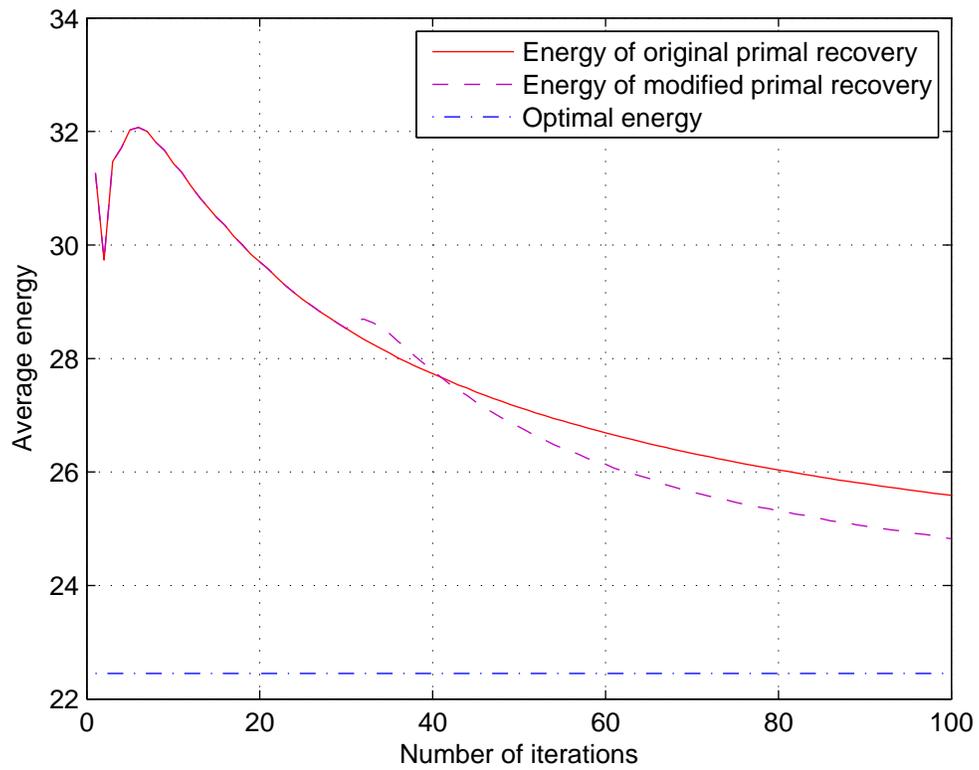}
\end{center}
\caption{Average energy as a function of the number of iterations for
the subgradient method on random 8-sink multicast connections
of unit rate in random 50-node wireless networks.}
\label{fig:fig1b}
\end{figure}

In Table~\ref{tab:3}, we tabulate the behavior of a distributed
approach, specifically, an approach using the subgradient method
(applied to problem (\ref{eqn:3540})).  The algorithm was run under step
sizes given by  $\theta[n] = n^{-0.8}$ and convex combination weights by
$\mu_l[n] = 1/n$, if $n < 30$, and $\mu_l[n] = 1/30$, if $n \ge 30$.  We
refer to this choice of parameters as the case of ``modified primal
recovery''.  Note that, despite our aiming to run sufficiently many
trials to ascertain the true average with high probability, the
simulations reported in Table~\ref{tab:3} do not agree exactly with
those in Table~\ref{tab:2} because they were run on different sets of
random instances.

Our first choice of parameters was step sizes given by $\theta[n] =
n^{-0.8}$ and convex combination weights by $\mu_l[n] = 1/n$.  This
case, which we refer to as ``original primal recovery'', was found to
suffer adversely from the effect of poor primal solutions obtained in
early iterations.  In Figures~\ref{fig:fig1a} and~\ref{fig:fig1b}, we
show the behavior of the subgradient method in the cases of a 4-sink
multicast in a 30-node network and an 8-sink multicast in a 50-node
network, respectively, in detail.  In these figures, we show both
parameter choices, and we see that modified primal recovery performs
substantially better.  For reference, the optimal energy of the problem
is also shown. 


We see that the subgradient method yields solutions that
converge rapidly to an optimal one, and it appears to be a promising
candidate for the basis of a protocol.  

%% file: chap5.tex
\chapter{Conclusion}
\label{chap:conclusion}

\lettrine{R}{outing} is undoubtedly a satisfactory way to operate packet
networks.  It clearly works.  What is not clear is whether it should be
used for all types of networks.  As we mentioned, coding is a
definite alternative at least for application-layer overlay networks and
multi-hop wireless networks.  To actually use coding, however, we must
apply to coding the same considerations that we apply to routing.  This
thesis was motivated by exactly that.  We took the basic premise of
coding and addressed a fundamental problem in packet
networks---efficient operation.  We laid out a solution to the efficient
operation problem, defined as it was to factor in packet loss, packet
broadcast, and asynchronism in packet arrivals.  That, we believe, is
our main contribution.   


From here, there is promising work both in expanding the scope of the
problem and in examining the problem more deeply.  We discuss first the
former.  One way of expanding the scope of the problem is by including
more considerations from networking.  In particular, an important issue
outside the present scope is flow, or congestion, control.  We have
taken, as a starting point, messages admitted into the network at given
rates and left aside the problem of determining which messages to admit
and at what rates.  This problem can be dealt with separately, e.g.,\
using window flow control as in \textsc{tcp}, but it need not be.  In
routed packet networks, flow control can be done jointly with optimal
routing (see \cite[Section 6.5.1]{beg92}), and it may likewise be
possible to extend the subgraph selection techniques that we proposed so
that they jointly perform subgraph selection and flow control.  Indeed,
an extension of the primal-dual method of Section~\ref{sec:primal-dual}
to perform joint subgraph selection and flow control is given in
\cite[Section II-C]{lrm}.  Even if flow control is done separately,
there has not, to our knowledge, been an earnest study of the flow
control problem for coded packet networks.


As for examining the efficient operation problem more deeply, there are
fundamental open questions relating to both network coding and subgraph
selection.  Let us first discuss network coding.  As we mentioned in
Section~\ref{sec:finite-memory}, the random linear coding scheme that we
proposed as a solution to the network coding problem is good in that it
maximizes throughput.  But throughput may not be our principal concern.
Other performance metrics that may be important are memory usage,
computational load, and delay.  Moreover, feedback may be present.  Our
true desire, then, is to optimize over a five-dimensional space whose
five axes are throughput, memory usage, computational load, delay, and
feedback usage.  

Some points in this five-dimensional space are known.  We know, e.g.,\
that random linear coding achieves maximum throughput; we can calculate
or estimate its memory usage, computational load, and delay; and we know
that its feedback usage is minimal or non-existent.  For networks
consisting only of point-to-point links, we have two other useful
points.  We know that, by using a retransmission scheme on each link
(i.e.,\  acknowledging the reception of packets on every link and
retransmitting unacknowledged packets), we achieve maximum throughput
and minimum memory usage, computational load, and delay at the cost of
high feedback usage (we require a reliable feedback message for every
received packet).  We know also that, by using a low-complexity erasure
code on each link (e.g.,\ a Raptor code \cite{may02, sho04} or an LT
code \cite{lub02}), we trade-off, with respect to random linear coding,
computational load for delay.  The challenge is to fill out this space
more.  In the context of channel coding, such a challenge might seem
absurd---an overly ambitious proposition.  But, as the slotted Aloha
relay channel (see Section~\ref{sec:example1}) illustrates, network
coding is different from channel coding, and problems intractable for
the latter may not be for the former.  A preliminary attempt at tackling
this problem is made in \cite{pfs05}.  


Let us discuss, now, open questions relating to subgraph selection.  In
this thesis, we gave distributed algorithms that apply only if the
constraints caused by medium access issues can essentially be
disregarded.  But these issues are important and often must be dealt
with, and it remains to develop distributed algorithms that incorporate
such issues explicitly.  A good starting point would be to develop
distributed algorithms for slotted Aloha networks of the type described
in Section~\ref{sec:example1}.

Much potential for investigation is also present in the cases for which
our algorithms do apply.  Other distributed algorithms are given in
\cite{cxn04, wck06, xiy05, xiy06}, and no doubt more still can be
developed.  For example, our choice to approximate the maximum function
with an $l^m$-norm in Section~\ref{sec:distributed_algorithms} is quite
arbitrary, and it seems likely that there are other approximations that
yield good, and possibly even better, distributed algorithms.

No matter how good the distributed algorithm, however, there will be
some overhead in terms of information exchange and computation.  What we
would like ideally is to perform the optimization instantly without any
overhead.  That goal is impossible, but, failing that, we could content
ourselves with optimization methods that have low overhead and fall
short of the optimal cost.  From such a suboptimal solution, we could
then run a distributed algorithm to bring us to an optimal solution or,
simply, use the suboptimal solution.  A suboptimal, but simple, subgraph
selection method for minimum-energy broadcast in coded wireless networks
is given in \cite{wfl05}.  Little else has been done.  It might seem
contradictory that we started this thesis by lamenting the use of ad hoc
methods and heuristics, yet we now gladly contemplate their use.  There
is a difference, however, between proposing the use of ad hoc methods
when the optimum is unknown or poorly defined and doing so when the
optimum is known but simply cannot be achieved practically.  What we now
call for is the latter.

Another point about the algorithms we have proposed is that they
optimize based on rates---rates of the desired connections and rates of
packet injections.  But we do not necessarily need to optimize based on
rates, and
there is a body of work in networking theory where subgraph selection is
done using queue lengths rather than rates \cite{awl93, tae92}.  This
work generally relates to routed networks, and the first that applies to
coded networks is \cite{hov05}.  Adding such queue-length based
optimization methods to our space of mechanisms for subgraph selection
may prove useful in our search for practical methods.  What we would
like to know, ideally, is the most practical method for network $x$,
given its particular capabilities and constraints.  This might or might
not be one of the methods proposed in this thesis; determining whether
it is, and what is if it is not, is the challenge.


This drive toward practicality fits with the principal motivation of
this thesis: we saw coding as a promising practical technique for packet
networks, so we studied it.  And we believe, on the basis of our
results, that our initial hypothesis has been confirmed.  Realizing
coded packet networks, therefore, is a worthwhile goal, and we see our
work as an integral step toward this goal.  But that is not our only
goal: Gallager's comment on the ``art'' of networking (see
Chapter~\ref{sec:introduction}) is, we believe, indicative of a general
consensus that current understanding of data networks is poor, at least
in relation to current understanding of other engineered systems, such
as communication channels.  There is no clear reason why this disparity
of understanding must exist, and the advances of networking theory have
done much to reduce its extent.  The study of coded networks may reduce
the disparity further---as we have seen in this thesis, we are, in the
context of coded packet networks, able to find optimal solutions to
previously-intractable problems.  This goal, of increasing our general
understanding, is one of the goals of this thesis, and we hope to spawn
more work toward this goal.  Perhaps coding may be the ingredient
necessary to finally put our understanding of data networks on par with
our understanding of communication channels.


%% file: biblio.tex
\bibliographystyle{IEEEtranS}
\bibliography{IEEEabrv,inform_theory}